%% file: paper.tex
\pdfoutput=1
\documentclass[11pt]{article}
\usepackage[margin=0.85in]{geometry}

\usepackage{amsmath,amssymb}
\usepackage{graphicx}
\usepackage[numbers]{natbib}
\usepackage{hyperref}
\usepackage{booktabs}
\usepackage{tikz}
\usetikzlibrary{arrows.meta,positioning}

\definecolor{ConnBlue}{HTML}{0072B2}
\definecolor{ExchVermillion}{HTML}{D55E00}
\definecolor{DoublonGreen}{HTML}{009E73}

\newcommand{\Tr}{\mathrm{Tr}}
\newcommand{\Var}{\mathrm{Var}}

\title{Spectral Moments of the Projected Hubbard Configuration Graph:\\ Connectivity, Exchange, and Virtual Doublons}
\author{D.~Souza}
\date{}

\begin{document}
\maketitle

\begin{abstract}
At strong on-site repulsion, the lowest band of $n$ lattice fermions is
the spectrum of the hopping projected onto configurations without
double occupancy, $H_\infty=P_0H_tP_0$. We study its moments on the
square lattice and, where the arguments allow, on any finite graph.
(1) For spin-independent Hermitian hopping without on-site terms on
$N\geq2$ sites,
$\mu_2=\frac{n(N-n)}{N(N-1)}\sum_{i\neq j}|t_{ij}|^2$ for every spin
composition. (2) On any finite graph, $\mu_4$ is exactly linear in the
same-spin pair count $P_{\rm same}$, because four hops permute at most
three particles. For uniform hopping on a simple graph with $N\geq4$,
$\mu_4=A+\alpha P_{\rm same}$ in closed form: the slope
$\alpha=-8C_4(N-n)(8N-11n-2)/[N(N-1)(N-2)(N-3)]$ is set by the number
$C_4$ of 4-cycles, and the intercept $A$, through the fourth moment of
hard-core bosons, by the numbers of edges, of adjacent and of disjoint
edge pairs, and of 4-cycles. With arbitrary Peierls phases, $\mu_4$ is a
spin-independent baseline plus a coefficient times $\sum_C\cos\Theta_C$
over the Wilson phases of the 4-cycles. (3) At finite $U$, the
second-order Schrieffer--Wolff correction is Hilbert--Schmidt orthogonal
to $H_\infty$ on every triangle-free graph, so the band variance has no
$1/U$ term; its $U^{-2}$ coefficient, an exact trace, is negative at the
low fillings computed. (4) Exchange loops carry the ordinary charge-$e$
harmonic, $\alpha\cos\Phi_\square$ on $L\times L$ tori with $L=3$ or
$L\geq5$, exchange changing the interference parity, not the charge;
the Schrieffer--Wolff-projected doublon hopping acquires a doubled
phase, $t_de^{2iA_{ij}}$, so that the second-order single-doublon
Hamiltonian is exactly $\pi$-periodic ($h/2e$) under a boundary twist,
and the exact two-electron U-band follows up to
$O((2\lambda)^L/U^{L-1})$. The closed forms are cross-checked against
exact sparse traces, the finite-$U$ statements against exact
diagonalization, and the numerical statements are listed as such in the
Discussion. Result (1), the closed form in (2) without phases, the
orthogonality in (3) and the doubled phase are machine-checked in
Lean~4 (Appendix~\ref{app:lean}). Repulsive binding and the pair charge
$2e$ are known; new are the exact moments and the derivation of the
doubled phase. We make no claim about many-pair phase stiffness,
off-diagonal long-range order or superconductivity.
\end{abstract}

\section{Introduction}
\label{sec:intro}

Strong interactions can reorganize a lattice many-body Hilbert space
long before they produce conventional long-range order. A simple
example is a repulsively bound pair, or \emph{doublon}: when the
interaction energy exceeds the available kinetic bandwidth, two
particles occupying the same site form a long-lived composite
excitation whose motion proceeds only through higher-order tunneling.
Repulsively bound pairs were observed experimentally in an optical
lattice~\cite{Winkler2006,Daley2006} and have since become a standard
example of interaction-induced composite dynamics in Hubbard-type
systems. In the two-dimensional electronic Hubbard problem, the
repulsive pair was shown to form a distinct ``U-band'' and to behave
dynamically as a charge-$2e$ composite object~\cite{SouzaClaro2010},
with a doubled Bloch-oscillation frequency as its dynamical
signature. Claro and
Robles~\cite{ClaroRobles2022} solved the one-dimensional bound pair
exactly and showed semiclassically, at leading order in $\lambda/U$,
that its Bloch-oscillation frequency is doubled, compatible with
charge $2e$. More recent work has
extended this picture to tunable doublon--doublon
interactions~\cite{Valmispild2020}, multidoublon Bloch
oscillations~\cite{Zhang2024}, doublon-plus-particle
dynamics~\cite{PaulMishra2024}, doublon/holon attraction from
strong-coupling perturbation theory~\cite{Queisser2023}, QED-analogous
pair creation~\cite{Queisser2025}, non-Hermitian bosonic doublon
dynamics~\cite{BrighiNunnenkamp2024}, and tunable extended-range bound
pairs in a Rydberg-dressed Bose-Hubbard lattice~\cite{Weckesser2025}.
Effective doublon Hamiltonians, with parameters derived perturbatively
from the underlying repulsion, are therefore well established as a
description of strong-coupling dynamics.

There is a complementary, geometric way of viewing this problem. A
wavefunction of $n$ distinguishable particles on a 2D lattice is a
single-particle wavefunction on a $2n$-dimensional configuration
lattice; such mappings of correlated few-particle problems onto
higher-dimensional single-particle tight-binding problems have
precedent~\cite{VallejoNavarro2003,Huipe2024}. Binding and projection
restrict this high-dimensional lattice to lower-dimensional subspaces:
a doublon is the simplest instance, translating with two coordinates
where two independent particles need four. This coordinate-counting
device is \emph{not itself} the novelty of the present work; it is the
starting point for the question we address: once strong repulsion
restricts $n$ fermions to a coordinate skeleton, what determines the
density of states on it (through its moments), and at what order does
fermionic exchange become visible? For the $N_d=0$ band, the subject of
\S\S\ref{sec:Uinfinity}--\ref{sec:SW}, the skeleton is the full
$2n$-dimensional configuration lattice with the doubly occupied points
removed; the reduced doublon sectors enter through
\S\S\ref{sec:benchmark}, \ref{sec:fewbody} and \ref{sec:futurework}.
A second, independent line of work
treats a many-body problem as a hopping problem on a high-dimensional
Fock-space or configuration graph, most influentially for quasiparticle
lifetimes and many-body
localization~\cite{AltshulerGefenKamenevLevitov1997,RoyLogan2025}; we
use the same language below, likewise established rather than novel.

The precedent most conceptually adjacent to \S\ref{sec:mu4exchange} is
the strong-coupling single-hole problem: Brinkman and
Rice~\cite{BrinkmanRice1970} obtained hole-band spectral moments in the
$U\to\infty$ Hubbard model by counting hopping paths that return the
carrier to its origin while restoring or altering the surrounding spin
background, and ten Haaf \emph{et al.}~\cite{tenHaaf1995} made this
explicit in closed-walk/permutation language: a closed walk induces a
permutation of the background spins, decomposable into disjoint cycles,
with the background restored precisely when every spin within a
nontrivial cycle shares the same orientation (see also
\cite{MeshkovBerkov1992}). That closed-walk$\to$permutation-cycle
connection is established, not new here. What differs is the spectral
object: their moments concern a single carrier hopping through a
thermodynamic-limit spin background, evaluated probabilistically
through the background magnetization, whereas we study the eigenvalue
distribution of the entire fixed-particle-number, no-double-occupancy
kinetic operator at finite $N$ and arbitrary $(N_\uparrow,N_\downarrow)$.
Turning that path-counting idea into exact finite-$N$ composition
laws, namely that four projected hops admit only the identity,
transpositions, and three-cycles, forcing
$\mu_4=A(N,n)+\alpha(N,n)P_{\rm same}$ exactly, with $\alpha(N,n)$ and
$A(N,n)$ both derived in closed form, is the content of
\S\ref{sec:mu4exchange}.

Spectral-moment methods have a separate long history in Hubbard-model
theory, from Harris and Lange's moment expansion of the single-particle
spectral function~\cite{HarrisLange1967} through exact strong-coupling
spectral-moment constraints in modern work~\cite{Esterling2023}; both
also use a strong-coupling no-double-occupancy projection, so
projection itself should likewise not be read as novel here. That prior
work characterizes the one-particle addition/removal spectral weight;
the traces considered here instead characterize the eigenvalue
distribution of the projected many-body kinetic operator at fixed
particle number:
\begin{equation}
M_\ell^{\rm sp}(\mathbf k) = \int d\omega\,\omega^\ell\,\mathcal A(\mathbf k,\omega)
\qquad\text{vs.}\qquad
\mu_\ell^{(n)} = \frac{1}{D_0}\Tr_{\mathcal H_n^{(0)}}\big[(P_0H_tP_0)^\ell\big],
\end{equation}
with $\mathcal A$ the single-particle spectral function,
$\mathcal H_n^{(0)}$ the $n$-electron no-double-occupancy space of
one spin composition $(N_\uparrow,N_\downarrow)$, and $D_0$ its
dimension. $M_\ell^{\rm sp}$ probes the frequency distribution of adding or
removing a particle from a many-body state; $\mu_\ell^{(n)}$ is a
moment of the eigenvalue distribution of the projected kinetic operator
itself, restricted to fixed $n$ and doublon occupation, so that
individual closed walks can be classified directly by real-space
occupancy and exchange. Both involve hopping-path combinatorics, but
they concern inequivalent spectral measures.

\paragraph{Principal results.} After establishing the two-particle
doublon problem as a benchmark (\S\ref{sec:benchmark}), we address two
questions, explicitly for $n=2,3,4$ fermions and for every $n$ where
the arguments allow (how many independent translational
coordinates remain after strong-coupling projection, and how does
fermionic exchange modify the spectral moments on the resulting
configuration graph) and then extend the static analysis to a gauge
field. Item 1 below organizes the analysis; items 2--5 are the four
results of the abstract:
\begin{enumerate}
\item A \textbf{cluster-coordinate counting rule} (\S\ref{sec:clusters}):
$n$ fermions in $C$ independently-mobile clusters occupy a
$d_{\rm cl}=2C$ coordinate skeleton, specializing to
$d_{\rm cl}=2(n-N_d)$ for doublons plus unpaired electrons. The sector
structure it implies (which $N_d$ bands exist, and their centres near
$N_d\varepsilon_d$) is verified by exact diagonalization for $n=3,4$
(\S\ref{sec:fewbody}).
\item An \textbf{exact second-moment theorem} (\S\ref{sec:Uinfinity}),
valid for any spin-independent Hermitian hopping amplitudes
($t_{ji}=t_{ij}^*$, no on-site terms) on $N\geq2$ sites,
$\mu_2(U\to\infty)=\tfrac{n(N-n)}{N(N-1)}\sum_{i\ne j}|t_{ij}|^2$,
independent of spin composition; on a $z$-regular lattice,
$\mu_2=\lambda^2nz(N-n)/(N-1)$.
\item \textbf{Where exchange enters the fourth moment}
(\S\ref{sec:mu4exchange}): $\mu_4$ depends on spin composition only
through an exactly linear dependence on the same-spin pair count
$P_{\rm same}$, on any finite graph, because four projected hops admit
only the identity, a transposition, or a 3-cycle (never a four-particle
permutation, shown combinatorially unreachable in four hops). For
uniform hopping the slope is exact in closed form at every $n$ on any
finite simple graph with $N\geq4$,
$\alpha(N,n)=-8C_4(N-n)(8N-11n-2)/[N(N-1)(N-2)(N-3)]$ with $C_4$ the
number of 4-cycles: transpositions around a 4-cycle (on the square
lattice, two geometrically distinct ring-exchange channels) plus
3-cycles of three particles around one. On the $L\times L$ torus with
$L=3$ or $L\geq5$, $\alpha(N,2)=-64/(N-1)$, and on hypercubic tori whose
sides are all $3$ or $\geq5$, $\alpha_d(N,2)=-16(d-1)z/(N-1)$. The
spin-independent part $A(N,n)$ is exact as well. It follows from the
fourth moment of hard-core bosons on the same graph, which depends only
on the numbers of edges, of edge pairs sharing a vertex, of disjoint edge
pairs and of 4-cycles (Eq.~\eqref{eq:Aclosed}). So $\mu_4(\infty)$ is
in closed form on every finite simple graph, at every spin composition.
\item A \textbf{Schrieffer--Wolff expansion} at finite $U$
(\S\ref{sec:SW}): the $1/U$ contribution to $\mu_2$ vanishes
identically on triangle-free graphs, including the torus with $L\geq4$
(a proven cancellation), and the $U^{-2}$
coefficient, which includes a third-order term, is an exact trace
confirmed by exact diagonalization; numerically, at low filling it is
negative (the band narrows) and close to proportional to the
opposite-spin pair count $N_\uparrow N_\downarrow$ (at fixed $n$,
equivalently close to affine in $P_{\rm same}$ with its exact zero at a
single spin species), and near half filling it turns positive.
\item A \textbf{gauge-field hierarchy} (\S\ref{sec:futurework}): the
fourth moment is exact for arbitrary Peierls phases on any finite simple
graph, a spin-independent baseline plus one coefficient times
$\sum_C\cos\Theta_C$ over the 4-cycles; on $L\times L$ tori with $L=3$
or $L\geq5$ the exchange slope is $\alpha(N,n)\cos\Phi_\square$
($-\tfrac{64}{N-1}\cos\Phi_\square$ at $n=2$), sharing one
ordinary charge-$e$ harmonic between orbital and exchange loops
(exchange changes interference \emph{parity}, not charge); the
Schrieffer--Wolff-projected doublon hopping itself carries a
\emph{doubled} phase, $t_de^{2iA_{ij}}$, so that the second-order
effective single-doublon Hamiltonian is exactly $\pi$-periodic ($h/2e$)
under a boundary twist, a periodicity the exact two-electron U-band
follows up to corrections of order $(2\lambda)^L/U^{L-1}$.
\end{enumerate}
A closely related question, whether two doublons bind into a larger
composite, is a self-contained problem with its own subtleties, which
we leave to separate work, in the spirit of~\cite{Queisser2023}. The
single-doublon charge-$2e$ result above is the static half of a longer
program. The \emph{dynamical} extension (driven pair-transport and
current-response observables, and ultimately many-pair phase coherence)
remains future work (\S\ref{sec:discussion}).

\paragraph{Machine-checked results.} The second-moment theorem, the
four-hop permutation bound and the three-cycle identity behind the
linearity of $\mu_4$, the fourth moment of hard-core bosons and the
exact fourth moment $\mu_4(\infty)=A(N,n)+\alpha(N,n)P_{\rm same}$ on
any finite simple graph (without Peierls phases; the form under
arbitrary phases, Eq.~\eqref{eq:mu4flux}, is not formalized), the
vanishing cross term, the doubled doublon
phase, and the finite-population identities are also formalized and
proved in the Lean~4 proof
assistant~\cite{deMouraUllrich2021} with its mathematical library
Mathlib~\cite{Mathlib2020}, with the projected operator $H_\infty$
built there from fermion operators satisfying the canonical
anticommutation relations. Appendix~\ref{app:lean} lists every
machine-checked statement and states precisely what it does not cover;
none of the numerical results depends on it. Derivations too long for
the paper are collected in the Supplemental Material, twelve notes cited
as S1--S12: S1 configuration space and densities of states, S2 the
doublon effective Hamiltonian, S3 the second-moment theorem, S4
finite-population corrections, S5 the exchange topology of $\mu_4$, S6
lattice specificity and finite size, S7 the Schrieffer--Wolff
expansion, S8 the fermion sign rule, S9--S11 the gauge-field
extension, and S12 the Lean statements and definitions.

\section{Model and the two-particle benchmark}
\label{sec:benchmark}

\begin{figure}[htbp]
\centering
\includegraphics[width=0.55\textwidth]{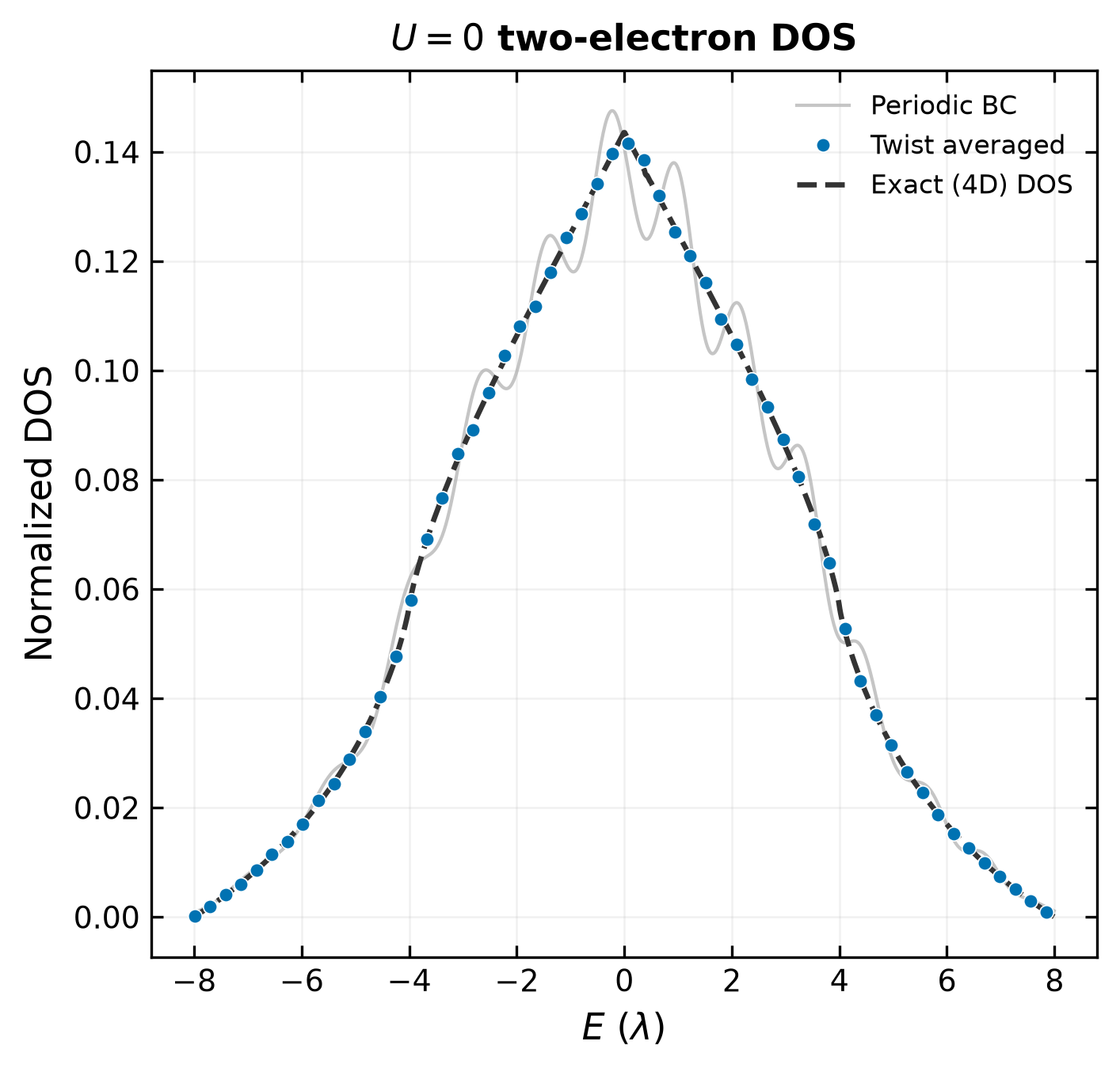}
\caption{Two-electron density of states at $U=0$ (one electron of each spin) for an $L=9$ periodic square lattice. The fixed-boundary spectrum contains $6561$ states; averaging over $300$ independent boundary twists gives $1{,}968{,}300$ eigenvalues. The twist-averaged result approaches the exact continuum $\rho_{4D}(E)$, the free 4D configuration space made explicit.}
\label{fig:tbcU0}
\end{figure}

Two electrons occupy a periodic $L\times L$ square lattice of $N=L^2$
sites with nearest-neighbor hopping $\lambda$ and on-site Hubbard repulsion $U$,
the sum running over nearest-neighbor bonds $\langle ij\rangle$ and both
spin species $\sigma\in\{\uparrow,\downarrow\}$,
\begin{equation}
H = -\lambda\sum_{\langle ij\rangle,\sigma}\big(c^\dagger_{i\sigma}c_{j\sigma}+{\rm h.c.}\big)
+ U\sum_i n_{i\uparrow}n_{i\downarrow}.
\end{equation}
Two \emph{free} ($U=0$) electrons in 2D are exactly a single particle
hopping on a 4-dimensional hypercubic lattice, since
$\Psi(x_1,y_1,x_2,y_2)$ depends on four independent coordinates with
additive dispersion $E=-2\lambda\sum_{\alpha=1}^{4}\cos k_\alpha$. This
gives an exact, parameter-free prediction for the $U=0$ density of
states via the Bessel-integral representation of
Appendix~\ref{app:dos}, confirmed by dense exact diagonalization at
$L=7$ (the lattice of the original density-of-states
figure~\cite{SouzaClaro2010}) and, at $L=9$, sharpened to a
near-exact match by twisted-boundary-condition averaging
(Fig.\,\ref{fig:tbcU0}).

\begin{figure}[htbp]
\centering
\includegraphics[width=0.68\textwidth]{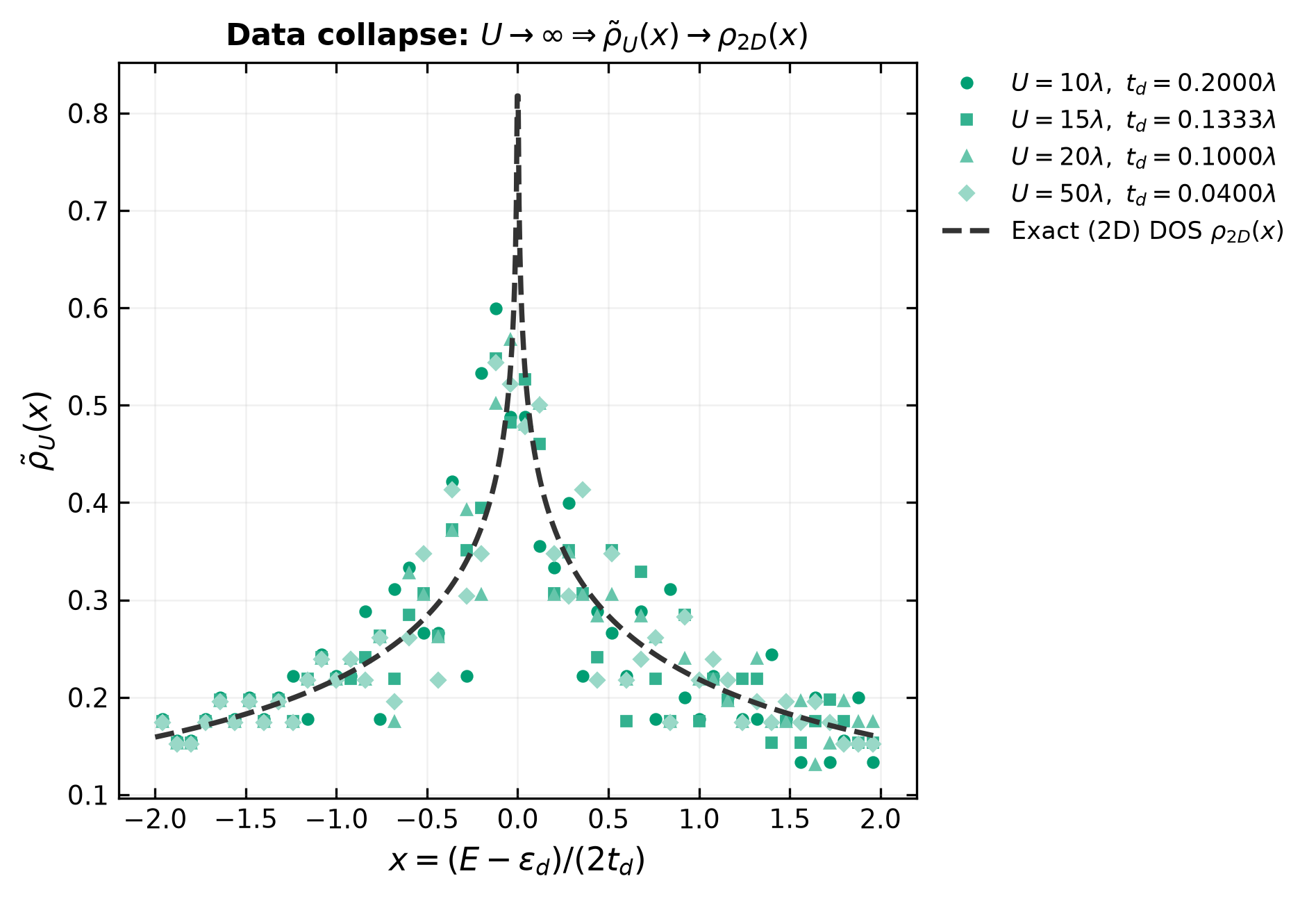}
\caption{The two-electron U-band at $U=10,15,20,50\lambda$, from a relative-coordinate scan (relative lattice $L_{\rm rel}=21$, $48\times48$ pair momenta $K$; Appendix~\ref{app:numerics}), histogrammed (50 bins, unit area) in the reduced variable $x=(E-\varepsilon_d)/(2t_d)$: $\tilde\rho_U(x)$ is the U-band density of states in that variable. It collapses onto the 2D tight-binding density of states of Eq.~(\ref{eq:doublondisp}) in the same variable, with $t_d$ and $\varepsilon_d$ predicted from $U,\lambda$ and no fitted parameters: the $n=2$ benchmark for everything that follows.}
\label{fig:ubandcollapse}
\end{figure}

\paragraph{The doublon effective Hamiltonian.} At fixed total momentum
and every $U$, the pair's reduced (relative-coordinate) equation is
exactly the textbook Koster--Slater bound-state-at-a-repulsive-defect
problem~\cite{KosterSlater1954}. Expanding for large $U$ and converting to the physical pair
momentum $(K_x,K_y)$ gives, to leading order in $\lambda/U$, an
effective 2D tight-binding band,
\begin{equation}
E_{\rm pair}(K_x,K_y) \approx \varepsilon_d + 2t_d(\cos K_x+\cos K_y),
\label{eq:doublondisp}\qquad
\varepsilon_d = U+\frac{8\lambda^2}{U}, \quad t_d=\frac{2\lambda^2}{U},
\end{equation}
i.e.\ a composite boson hopping on the \emph{same} $L\times L$ lattice
with its own 2D dispersion, \emph{inverted} relative to a bare
particle's $-2\lambda(\cos K_x+\cos K_y)$ sign convention: the doublon
band has its maximum, not its minimum, at $K=0$. This sign is fixed,
not a free convention. Direct second-order degenerate perturbation
theory (the $N$-fold-degenerate manifold $\{|r,r\rangle\}$ at energy
$U$ (zeroth order), coupled through the continuum of states with no
doubly occupied site, $N_d=0$, at energy $0$
by two sequential single-electron hops) has exactly two contributing
virtual paths between adjacent doublon sites $r,s$: $\uparrow$ hops
$r\to s$ first, then $\downarrow$; or the reverse order. Each
contributes $(-\lambda)(-\lambda)/(U-0)=+\lambda^2/U$ to the effective
hopping matrix element, with \emph{no} additional fermion sign (a
single particle per spin species carries none, under the sorted-tuple
convention of Appendix~\ref{app:numerics}), giving a net
$+2\lambda^2/U$, the \emph{same} sign as $\lambda^2$, not its
negative. This positive matrix element Fourier-transforms to
$+2t_d(\cos K_x+\cos K_y)$, opposite in sign to the bare kinetic
term, whose convention is $H_t=-\lambda\sum(\cdots)$. The
sign is corroborated independently by gauge covariance: the doublon
operator $d_r=c_{r\downarrow}c_{r\uparrow}$ transforms as
$d_r\to e^{2i\theta_r}d_r$ under a local phase redefinition
$c_{r\sigma}\to e^{i\theta_r}c_{r\sigma}$, twice the phase of a single
fermion, so on a bipartite lattice (even $L$) the staggered
transformation ($\theta=0,\pi$ on the two sublattices) that reverses the bare hopping's sign leaves the
doublon's derived hopping sign \emph{unchanged} ($e^{2i\pi}=+1$),
ruling out the naive expectation that the two signs are gauge-equivalent.
This inverted dispersion is the standard, if underappreciated,
signature of a pair bound by \emph{repulsion}: the doublon descends
from the upper Hubbard band. It is derived in full, together with
both consistency checks above, in S2. The positive sign
is also consistent with the exact one-dimensional repulsively
bound-pair dispersion~\cite{ClaroRobles2022}, whose strong-coupling
expansion gives
$E_{\rm pair}=\varepsilon_d+2t_d\cos K_{\rm CM}+O(\lambda^4/U^3)$ with
$t_d=2\lambda^2/U$ (and $\varepsilon_d=U+4\lambda^2/U$ in one
dimension), an independent, exact
two-body corroboration of the sign correction above, not merely a
consistency check on our own perturbative bookkeeping. For the same
one-dimensional problem, a semiclassical treatment at leading order in
$\lambda/U$ gives a doubled Bloch-oscillation frequency under an
applied field, which that work reads as compatible with the pair's
composite charge $2e$~\cite{ClaroRobles2022};
we return to this, and to its relation to the gauge-covariant result
of \S\ref{sec:futurework}, below. This particular
perturbative form is by now a standard type of result; comparable
effective doublon Hamiltonians
have recently been derived for an extended, mass-imbalanced Hubbard
model~\cite{Zhang2024} and for a periodically driven
one~\cite{Valmispild2020}, and our contribution here is
not the functional form itself but the zero-fitted-parameter connection
between it and the exact two-electron U-band on this specific lattice:
the predicted band centre matches the observed mean to within $0.06\%$
(at most $0.03\,t_d$), and the predicted band edge matches the observed
maximum to within $0.5\%$ (at most $0.26\,t_d$)
($U=10,15,20\lambda$, $L=7$, the top $N$ levels of the spectrum), and the full U-band collapses onto the
predicted 2D dispersion once plotted in the doublon's own reduced
variables (Fig.\,\ref{fig:ubandcollapse}, from a relative-coordinate
momentum scan, Appendix~\ref{app:numerics}). The band-centre test and the
collapse are insensitive to the sign of $t_d$ on any lattice: the grid
average of $\cos K$ vanishes, and the two signs give mirror-image
densities of states about $\varepsilon_d$, which deviate equally from
the symmetric 2D density. The band edge is not: on the odd $L=7$
lattice $(\pi,\pi)$ is not an allowed momentum, so the opposite sign
would place the maximum at $\varepsilon_d+4t_d\cos(\pi/7)$ rather than
$\varepsilon_d+4t_d$.\footnote{For the same reason the zero-flux
doublon bandwidth is $8t_d$ for even $L$ but $4t_d[1+\cos(\pi/L)]$ for
odd $L$; the exact $L=5$ two-electron U-band converges to this odd-$L$
value, not to $8t_d$, as $U$ grows: ordinary momentum-grid
discretization, not a many-body finite-size pathology.} The observed maximum favours the derived sign,
deviating from its prediction by $0.45\%$, $0.11\%$ and $0.04\%$
($0.26$, $0.13$ and $0.08\,t_d$) at
$U=10,15,20\lambda$, against $1.14\%$, $0.44\%$ and $0.23\%$
($0.66$, $0.53$ and $0.47\,t_d$) for the opposite sign. This benchmark is the starting point for the
rest of the paper.

\section{Cluster-coordinate counting}
\label{sec:clusters}

\subsection{A coordinate count, not a spectral claim}

Ignoring interactions and exchange entirely, $n$ particles in 2D,
$\Psi(x_1,y_1,\ldots,x_n,y_n)$, live on a configuration space that is
\emph{exactly} a single particle hopping on a $2n$-dimensional
hypercubic lattice, $\rho_n^{(0)}(E)=\rho_{2nD}(E)$
(Appendix~\ref{app:dos}). This is pure kinematics, a relabeling of coordinates: if
$C$ independently translating clusters move in two spatial
dimensions, then by definition they carry $2C$ centre-of-mass
coordinates. We state this plainly because a natural objection is that
nothing has yet been proved: coordinate counting alone is not a
dynamical result. We use $d_{\rm cl}$ exclusively as a
\textbf{coordinate-counting dimension}: the number of independent
real-space coordinates needed to translate $C$ composite or unpaired
objects in two spatial dimensions. It is not the dimension of the
many-body Hilbert space, the spectral dimension of the configuration
graph, or a scaling dimension, and should not be confused with any of
these. The nontrivial physical content begins only once we ask which
sectors the actual Hubbard spectrum organizes into, and what spectral
measure lives on the resulting skeleton: the subjects of
\S\ref{sec:fewbody} onward.

\subsection{From coordinate count to cluster rule}

Each pairwise on-site interaction imposes $x_i=x_j,y_i=y_j$, two
constraints, reducing the manifold's dimension by 2, consistent with
$4{\rm D}\to2{\rm D}$ at $n=2$. More generally, if $n$ particles organize
into $C$ independently-mobile clusters (a partition
$n=n_1+\cdots+n_C$), each cluster carries exactly one 2D centre-of-mass
coordinate, so
\begin{equation}
d_{\rm cl} = 2C \qquad (\text{codimension } 2(n-C)).
\end{equation}
For $n=4$: four free electrons ($C=4$) occupy 8D; two doublons ($C=2$)
occupy 4D; an (off-site) bound tetramer ($C=1$, if one exists) would
occupy 2D.

Single-band Hubbard physics forbids on-site clusters larger than 2
(only one $\uparrow$ and one $\downarrow$ electron fit per site), so
composites built purely from doublons plus unpaired electrons obey a
testable rule in terms of the doublon occupation $N_d$ alone:
\begin{equation}
d_{\rm cl} = 2(n-N_d).
\end{equation}

\section{Sector structure for \texorpdfstring{$n=3,4$}{n = 3, 4}}
\label{sec:fewbody}

Below, a \emph{composition} is a fixed $(N_\uparrow,N_\downarrow)$; its
states with $N_d=k$ doubly occupied sites form the $N_d=k$
\emph{sector}, whose energies form the $N_d=k$ band at large $U$; and a
momentum block is a fixed total momentum $K$.

Fig.\,\ref{fig:sectors} tests the sector structure implied by the
counting rule for $n=3,4$: which $N_d$ bands exist, and where they are
centred. Each eigenstate of the full exact spectrum is assigned to the
band $N_d=k$ with $|\langle N_d\rangle-k|<1/2$. In every case the band
centres $\bar E_{N_d}$ are confirmed against $N_d\varepsilon_d$, the
\emph{leading isolated-cluster prediction} obtained by simply adding
$N_d$ copies of the single-doublon self-energy: to $\leq0.10\%$ at
$U=40\lambda$ ($0.02\%$ for $n=3$; $0.10\%$ and $0.03\%$ for $N_d=1,2$
at $n=4$), and to $\leq1.1\%$ already at $U=10\lambda$. This residual is not
numerical noise or a higher-order $1/U^2$ tail: exact diagonalization
at $U=10,15,20,30,40,60\lambda$ shows $U\cdot(\bar E_{N_d}-N_d\varepsilon_d)$
converging to a nonzero constant as $U$ grows (not $U^2\cdot(\cdots)$,
which instead diverges), confirming a genuine $O(\lambda^2/U)$
many-body correction: with more than one cluster present, other
particles can block or modify some of the virtual hopping processes
that generate the isolated-pair shift $8\lambda^2/U$, and this
finite-density correction is not yet captured in closed form. For
$N_d=1$ in the $n=3$ ($L=5$) and $n=4$ ($L=4$) sectors this constant
is $\approx-0.33\lambda^2$ and $\approx-1.60\lambda^2$ respectively
(measured at $U=60\lambda$); for $N_d=2$ in the $n=4$ sector it is
$\approx-1.06\lambda^2$, still visibly converging at that $U$. Being
finite-density corrections, these constants depend on $L$.

\subsection{\texorpdfstring{$n=3$}{n = 3}: one doublon plus one electron}
\label{sec:n3test}

Taking $N_\uparrow=2,N_\downarrow=1$ ($L=5$), the energy spectrum
separates cleanly by $N_d$ (Fig.\,\ref{fig:sectors}a): even at
$U=10\lambda$ the $N_d\approx1$ cloud is visually separated from the
$N_d\approx0$ continuum, centred almost exactly at $\varepsilon_d$; by
$U=40\lambda$ the upper-band mean is within $0.02\%$ of
$\varepsilon_d=U+8\lambda^2/U$. The relative coordinate is unbound,
not a genuine three-body composite. Total momentum $K$ is conserved, and
in every momentum block $K$ every $N_d\approx1$ level lies inside the
doublon--electron continuum
$\varepsilon_d+2t_d[\cos(K_x-q_x)+\cos(K_y-q_y)]-2\lambda(\cos q_x+\cos q_y)$
(over all $q$): in exact diagonalization at $U=30\lambda$ for $L\leq9$,
and, with $t_d\to0$, up to $L=32$ in the first-order ($U\to\infty$)
model of a static doublon and a hopping electron, which reproduces the
exact band to $O(\lambda^2/U)$. The only
short-range process at this order, the doublon's $\downarrow$ electron
hopping onto an adjacent electron's site (which exchanges the two),
produces a resonance inside the band near
$E-\varepsilon_d\approx+1.2\lambda$ at $K=0$ (and $-1.2\lambda$ at
$K=(\pi,\pi)$, by the exact symmetry of the first-order model on even
$L$ that maps $K\to K+(\pi,\pi)$ and $E-\varepsilon_d\to-(E-\varepsilon_d)$,
a sublattice sign change), not a bound
state: its weight on the adjacent configurations spreads over more and
more eigenstates as $L$ grows (in the first-order model, computed for
this purpose up to $L=64$, the largest single-state weight falls from
$0.36$ at $L=8$ to $0.034$ at $L=64$). Such doublon-plus-particle configurations have recently been
studied dynamically on quantum hardware, including the same
$N_\uparrow=2,N_\downarrow=1$ sector~\cite{PaulMishra2024}; here we
instead use them as a controlled static test of the counting
rule.

\subsection{\texorpdfstring{$n=4$}{n = 4}: spin composition as a sector-control knob}
\label{sec:n4test}

\begin{figure}[htbp]
\centering
\includegraphics[width=0.98\textwidth]{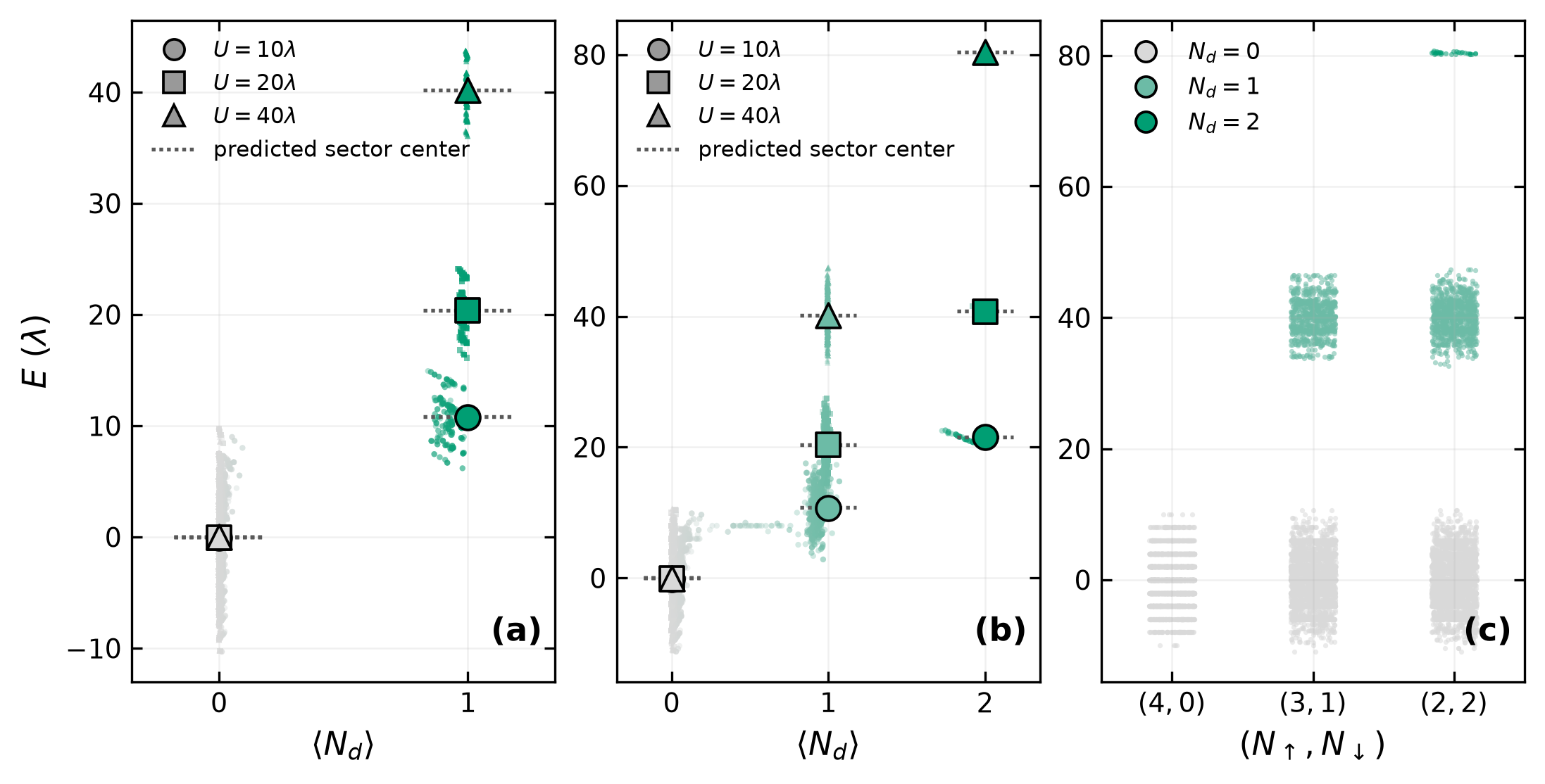}
\caption{Sector structure implied by the counting rule, tested by full
exact diagonalization of every interacting composition. All panels share the energy axis of (a), $E$ in
units of $\lambda$. In (a,b), color encodes the doublon expectation
value $\langle N_d\rangle$ on a continuous white-to-green scale (the
manuscript's virtual-doublon color; darker = more doublons), never a
proxy for $U$. In (a,b) the horizontal axis is the true, unjittered
$\langle N_d\rangle$ of each eigenstate (small pale points: the full
spectrum, a random subset of 4000 levels where it is larger; marker
shape distinguishes $U=10,20,40\lambda$); bold markers with black
edges mark each band's exact mean energy over all its states (a state
belongs to band $N_d=k$ when $|\langle N_d\rangle-k|<1/2$), and the short dotted horizontal ticks mark the
independently predicted sector center ($0$, $\varepsilon_d$, or
$2\varepsilon_d$): coincidence of marker and tick is the closure
test, not a fitted line. (a) $n=3$ ($2\uparrow{+}1\downarrow$, $L=5$):
the $\langle N_d\rangle\approx1$ sector separates cleanly, converging
to $\varepsilon_d$. (b) $n=4$ ($2\uparrow{+}2\downarrow$, $L=4$): the
same test, now with three sectors ($N_d=0,1,2$). (c) $n=4$, three spin
compositions $(N_\uparrow,N_\downarrow)=(4,0),(3,1),(2,2)$ at fixed
$U=40\lambda$; the $(4,0)$ column needs no diagonalization, its levels
being sums of four distinct single-particle energies, of which a random
sample is shown (horizontal position is the categorical composition, not
$N_d$; color is the band each state is assigned to): spin composition alone
controls which sectors are reachable ($(4,0)$: only $N_d=0$; $(3,1)$:
$N_d\in\{0,1\}$; $(2,2)$: $N_d\in\{0,1,2\}$).}
\label{fig:sectors}
\end{figure}

Taking $N_\uparrow=N_\downarrow=2$ ($L=4$), the predicted three-sector
hierarchy sharpens with increasing $U$: three bands at
$E\approx0,\,U,\,2U$ (Fig.\,\ref{fig:sectors}b), with the $N_d\approx1,2$
means converging to $\varepsilon_d$ and $2\varepsilon_d=2U+16\lambda^2/U$
(within $0.10\%$ and $0.03\%$ at $U=40\lambda$, where the bare values
$U$ and $2U$ are still $0.40\%$ and $0.47\%$ away); $N_d=0,1,2$ correspond to
$C=4,3,2$ clusters, the predicted $8{\rm D}\to6{\rm D}\to4{\rm D}$
sequence, and all three bands are present in a single spectrum. Repeating at
three spin compositions, same $U=40\lambda$ (Fig.\,\ref{fig:sectors}c),
shows spin composition directly controls which sectors exist:
$4\uparrow$ is a single band at $N_d=0$ (no interaction is even
possible with one spin species); $3{+}1$ shows exactly two bands
($N_d\in\{0,1\}$); $2{+}2$ shows exactly three ($N_d\in\{0,1,2\}$).

The detailed spectral \emph{shape} within a sector is a separate
matter, addressed from \S\ref{sec:Uinfinity} onward: coordinate
counting alone does not determine the many-body density of states,
since fermionic projection changes the spectral measure on the same
coordinate skeleton. For one electron of each spin at $L=5$, for
example, the $N_d=0$ band has variance $23\lambda^2/3$ against
$8\lambda^2$ for the free four-dimensional lattice with the same
coordinate count, and excess kurtosis $\mu_4/\mu_2^2-3=-204/529\approx-0.386$
against $-3/8$ (Appendix~\ref{app:dos}); the variance is fixed for
every $n$ by the theorem of \S\ref{sec:mu2theorem}
(Fig.\,\ref{fig:naivefpc}a).

We do not extend this sector-by-sector escalation to $n=5$: the results
of \S\ref{sec:Uinfinity}--\S\ref{sec:SW} below are stated for every $n$
(the $\mu_2$ theorem, the exact linearity of $\mu_4$ and its coefficient
$\alpha(N,n)$, the vanishing $1/U$ term), and larger $n$ enter only as
checks of them.

\section{Projected configuration graph at \texorpdfstring{$U\to\infty$}{U → ∞}}
\label{sec:Uinfinity}

\subsection{Setup}

Let $H_t=-\lambda\sum_{\langle ij\rangle,\sigma}(c^\dagger_{i\sigma}c_{j\sigma}+{\rm h.c.})$
and let $P_0$ project onto configurations with no double occupancy of
any kind. Define $H_\infty=P_0H_tP_0$, acting on the $N_d=0$ sector for
$n=N_\uparrow+N_\downarrow$ electrons on $N$ sites of a graph $G$ with
edge set $E$. $H_\infty$ is the adjacency operator of a
\emph{configuration graph}: its vertices are $n$-electron
no-double-occupancy configurations, its edges single allowed hops, with
matrix elements of modulus $\lambda$ whose signs are fixed by the
fermionic ordering of the orbitals (Appendix~\ref{app:numerics}).

\subsection{Exact second-moment theorem}
\label{sec:mu2theorem}

Every basis state in the $N_d=0$ sector has zero diagonal energy, since
a hop always moves a particle to a distinct, previously empty site, so
$\bar E=\Tr(H_\infty)/D_0=0$ and $\mu_2=\Tr(H_\infty^2)/D_0$. For a
configuration $a$ (occupied site set $S$, $|S|=n$),
$(H_\infty^2)_{aa}=\lambda^2M_a$, where $M_a$ counts the allowed
single-particle hops out of $a$: the number of directed (occupied
site, empty neighbor) pairs. So $\mu_2=\lambda^2\langle M\rangle$,
averaged uniformly over the $D_0=\binom{N}{n}\binom{n}{N_\uparrow}$
configurations of the sector. Each $n$-site subset $S$ occurs in exactly
$\binom{n}{N_\uparrow}$ of them (one per spin labelling), and $M_a$
depends only on $S$, so the average is over a uniformly random
$n$-element subset $S$ of the $N$ sites. This reduces to elementary
combinatorics: for \emph{any}
two distinct sites $i\neq j$, $\Pr(i\in S,\,j\notin S)=n(N-n)/[N(N-1)]$,
independent of whether $i,j$ happen to be adjacent. Summing over the
directed edges of $G$, we obtain:

\medskip
\noindent\textbf{Theorem (exact second moment; S3).} \emph{For any finite
simple graph $G=(V,E)$ with $|V|=N\geq2$, hopping amplitude $\lambda$,
and $0\leq n\leq N$, the $U\to\infty$ projected hopping operator in any
spin-composition sector with $N_\uparrow+N_\downarrow=n$ satisfies}
\begin{equation}
\boxed{\mu_2(U\to\infty) = \frac{\lambda^2\cdot 2|E|\cdot n(N-n)}{N(N-1)},}
\end{equation}
\emph{independent of spin composition.}
\medskip

\noindent Here $2|E|=\sum_v\deg(v)$ is the total number of directed edges. This
holds on \emph{any} finite simple graph, regular or not, since the
derivation never used translational invariance, dimensionality, or a
fixed coordination number, only that the $N_d=0$ basis consists of
\emph{all} $n$-site subsets. For a $z$-regular graph, $2|E|=Nz$, giving
the form used throughout the rest of this paper:
\begin{equation}
\mu_2(U\to\infty) = \lambda^2\,n\,z\,\frac{N-n}{N-1}.
\label{eq:mu2regular}
\end{equation}
\textbf{Crucially, $N_\uparrow$ and $N_\downarrow$ never enter}: $M_a$
depends only on which \emph{sites} are occupied, not on how the
occupied sites are split between spins, so every spin composition with
the same total $n$ shares the identical $\mu_2(\infty)$. ($L=4$ square
lattice, $N=16,z=4,n=4$: $\mu_2=(64/5)\lambda^2$, one numerical example,
not the content of the result.)

\paragraph{Generality.} We verified the graph-general form directly by
exhaustive enumeration on three markedly non-regular graphs (an open
chain of degrees 1 and 2, a star graph with one hub of degree $N-1$,
and an irregular graph with mixed degrees 1--4) for $n=2,3,4$; every case
matches $\lambda^2\cdot2|E|\cdot n(N-n)/[N(N-1)]$ to machine precision.
The square-lattice regular-graph form is therefore a special case of a
general graph-combinatorial statement, not a two-dimensional or
lattice-specific coincidence, and the enumeration below
(Table~\ref{tab:mu2verify}) exercises the regular-graph specialization
that the rest of the paper uses. The boxed theorem, and the weighted
form below, are also machine-checked, for $H_\infty$ built from fermion
operators with any spin-independent Hermitian amplitudes (in particular
any Peierls phases) and any ordering of the orbitals, which fixes the
fermionic signs (Appendix~\ref{app:lean}).

The same argument extends immediately to a general (possibly
asymmetric-looking but Hermitian) weighted hopping matrix
$H_t=-\sum_{i\ne j,\sigma}t_{ij}c^\dagger_{i\sigma}c_{j\sigma}$
($t_{ji}=t_{ij}^*$, the same for both spins), since nothing in the derivation used equal
hopping amplitudes either: $(H_\infty^2)_{aa}=\sum_{i\in S,j\notin
S}|t_{ij}|^2$, and averaging over the uniformly random $n$-subset $S$
gives
\begin{equation}
\mu_2(U\to\infty) = \frac{n(N-n)}{N(N-1)}\sum_{i\ne j}|t_{ij}|^2.
\end{equation}
Defining the mean coordination $\bar z\equiv\frac{1}{N}\sum_{i\ne
j}|t_{ij}|^2/\lambda^2$ for some reference scale $\lambda$ recovers
$\mu_2=\lambda^2\bar zn(N-n)/(N-1)$ for \emph{any} weighted graph,
regular or not: the theorem is properly a statement about the squared
Frobenius norm of the off-diagonal part of the one-body hopping matrix
$T=(t_{ij})$, $\sum_{i\ne j}|t_{ij}|^2$, factored against the purely
combinatorial blocking probability
$n(N-n)/[N(N-1)]$; the regular, equal-amplitude square lattice used
throughout this paper is simply the case that makes the factorization
$\bar z\lambda^2$ transparent.

\begin{table}[htbp]
\centering
\begin{tabular}{@{}rrrccc@{}}
\toprule
$L$ & $n$ & $N$ & analytic theorem, Eq.~\eqref{eq:mu2regular} & exact sparse evaluation & error \\
\midrule
3 & 2 & 9  & $7$        & $7$        & $<10^{-13}$ \\
4 & 3 & 16 & $52/5$     & $52/5$     & $<10^{-13}$ \\
4 & 4 & 16 & $64/5$     & $64/5$     & $<10^{-13}$ \\
5 & 4 & 25 & $14$       & $14$       & $<10^{-13}$ \\
6 & 4 & 36 & $512/35$   & $512/35$   & $<10^{-13}$ \\
\bottomrule
\end{tabular}
\caption{Exact $\mu_2(U\to\infty)/\lambda^2$: the analytic column is Eq.~\eqref{eq:mu2regular}; the sparse column is obtained by explicit construction of $H_\infty$ (here in the single-species sector; by the theorem every composition gives the same value) and evaluation of $\Tr H_\infty^2/D_0$, without diagonalization. The two are independent computations, not a formula checked against itself.}
\label{tab:mu2verify}
\end{table}

\subsection{Same-spin exclusion at \texorpdfstring{$U=0$}{U = 0} versus projection at \texorpdfstring{$U=\infty$}{U = ∞}}
\label{sec:fermistats}

\begin{figure}[htbp]
\centering
\includegraphics[width=0.55\textwidth]{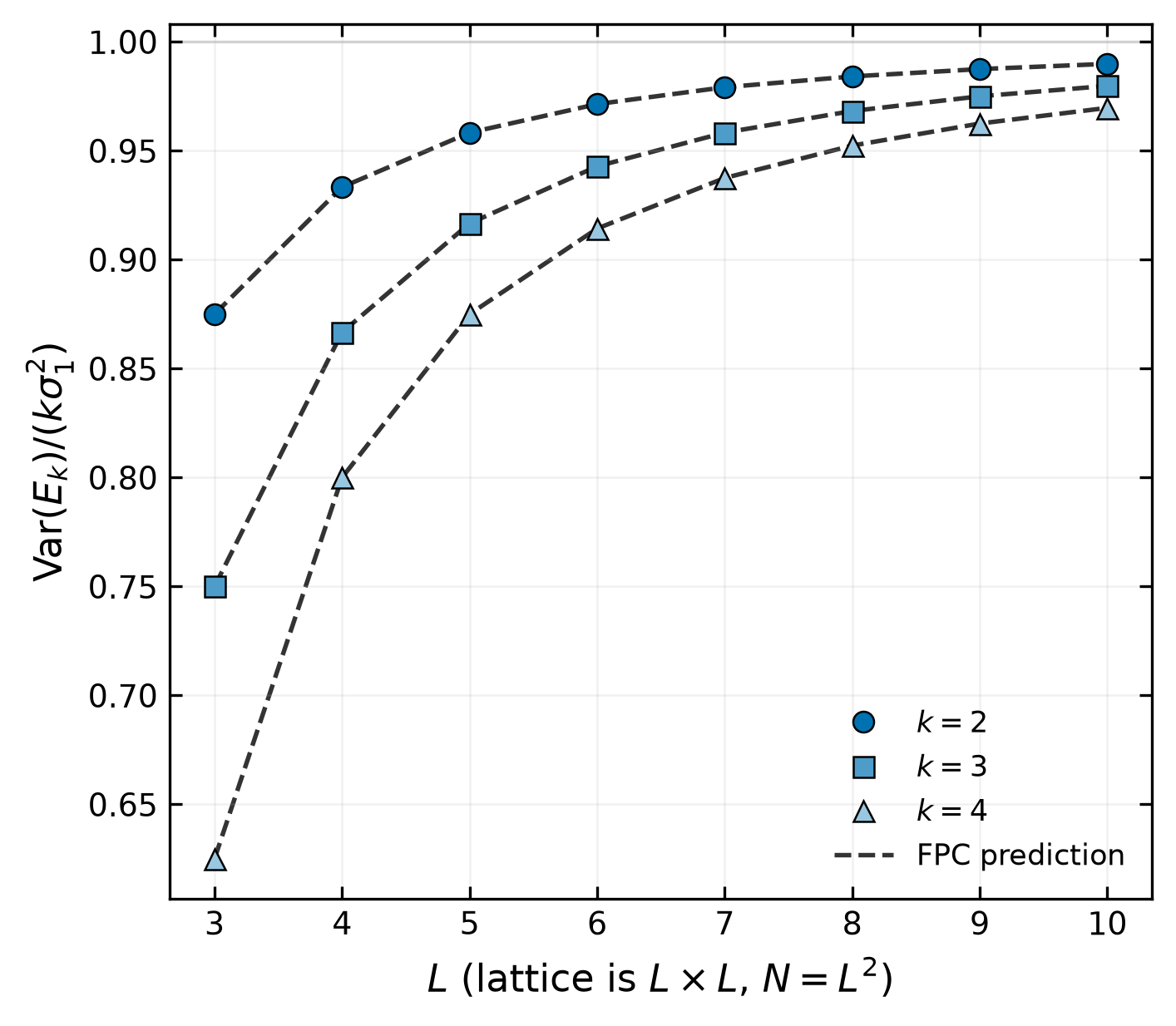}
\caption{Many-fermion variance of a single spin species at $U=0$, in units of $k\sigma_1^2$ ($\sigma_1^2=z\lambda^2$, the variance of the single-particle band), by full enumeration of all $\binom{N}{k}$ many-body energies (markers), against the parameter-free finite-population prediction $(N-k)/(N-1)$ (line), $L=3$--$10$.}
\label{fig:fpcL}
\end{figure}

Does Pauli exclusion among same-spin electrons alone account for the
reduced variance at $U=\infty$? The non-interacting problem answers
this, because its variance is also exact in closed form. Non-interacting identical fermions are exactly Slater
determinants: for a single spin species, their many-body spectrum is
\emph{exactly} every way of choosing $k$ of the $N$ single-particle
levels (distinct labels, possibly degenerate energies), without
replacement (population mean $0$, since the hopping matrix is
traceless, and variance $\sigma_1^2=z\lambda^2$), giving the
classical finite-population-correction (FPC) identity $\Var=k\sigma_1^2(N-k)/(N-1)$ (S4),
identical in \emph{form} to the $U=\infty$ theorem with $k\to n$, but
not in mechanism: at $U=0$ exclusion is momentum-space (same-spin
antisymmetry, so FPC applies only within one spin species); at
$U=\infty$ it is real-space (a spin-blind Gutzwiller constraint on
\emph{all} electrons). Both reduce to the same ``one site blocks one
neighbor'' combinatorics, hence the shared form. An analogous
coincidence-pattern classification of the fourth moment
(Appendix~\ref{app:mu4fpc}) gives the exact $\mu_4$ counterpart. Both
are verified against exact diagonalization (Table~\ref{tab:fpcexact}),
and the variance also against full enumeration across $L=3$--$10$
(Fig.\,\ref{fig:fpcL}): forming all $\binom{N}{k}$ many-body energies of
a single species ($k=2,3,4$, up to $3.9\times10^6$ of them at $L=10$)
and taking their variance reproduces $k\sigma_1^2(N-k)/(N-1)$ to
$10^{-15}$ (relative) at every $L$.
For $n=4$ at $L=4$ this is an exact $20\%$ narrowing relative to the
naive prediction, a finite-\emph{volume} statistical effect (vanishing
as $L\to\infty$), not a finite-$U$ or numerical artifact, since a
single spin species has no Hubbard interaction at all. At $U=0$ the
$n=2$ opposite-spin pair (no mutual exclusion) shows \emph{zero} such
narrowing: precisely zero \emph{Pauli} narrowing, not zero finite-size
deviation of any kind.

For example, at $k=4$, $N=16$, the FPC formula gives
$\mu_2/\lambda^2 = k\,z\,(N-k)/(N-1) = 4\times4\times12/15=64/5$
directly from the sampling argument above, with no diagonalization
required.

\begin{table}[htbp]
\centering
\setlength{\tabcolsep}{4pt}
\begin{tabular}{@{}lccc c@{}}
\toprule
population & {\color{ConnBlue}$\mu_2/\lambda^2$ (FPC formula)} & ED error & {\color{ExchVermillion}$\mu_4/\lambda^4$ (FPC formula)} & ED error \\
\midrule
$k{=}4$ single species ($4\uparrow$, $L{=}4$) & $64/5$   & $<10^{-13}$ & $197632/455$ & $<10^{-13}$ \\
$k{=}2$ (up-pair, $L{=}4$)                    & $112/15$ & $<10^{-13}$ & $2176/15$    & $<10^{-13}$ \\
$k{=}3$ (up-triple, $L{=}4$)                  & $52/5$   & $<10^{-13}$ & $9976/35$    & $<10^{-13}$ \\
\bottomrule
\end{tabular}
\caption{Exact finite-population-correction formulas for $\mu_2,\mu_4$ at $U=0$; ``ED error'' is the relative residual against independent exact diagonalization, confirming the closed form rather than defining it.}
\label{tab:fpcexact}
\end{table}

\begin{figure}[htbp]
\centering
\includegraphics[width=0.98\textwidth]{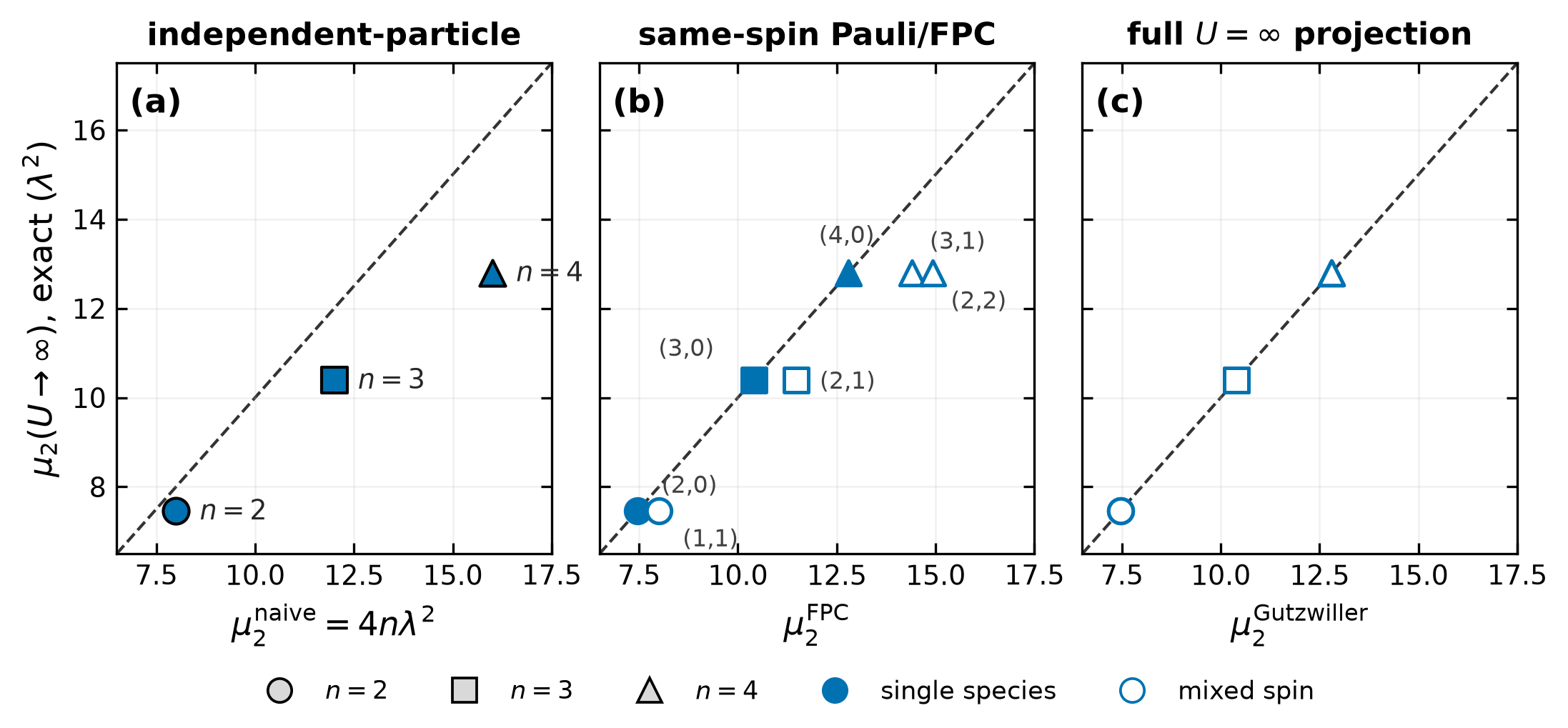}
\caption{$\mu_2$ of the $N_d=0$ sector, $n=2,3,4$ at $L=4$, against
three increasingly complete references, all judged against the same
exact $U\to\infty$ value: (a) independent-particle; (b) same-spin
Pauli/FPC; (c) full $U=\infty$ projection. Single-species compositions
(filled markers) collapse in panel (b); mixed-spin compositions (open
markers) remain off the diagonal there by design, not by failure: the
same-spin FPC correction accounts only for momentum-space exclusion
within each spin species, so a mixed-spin residual is expected wherever
opposite-spin real-space exclusion (the $U=\infty$ Gutzwiller
constraint) also matters. That residual measures exactly that missing
exclusion, and vanishes only once panel (c) restores it.}
\label{fig:naivefpc}
\end{figure}

Does momentum-space exchange alone repair the naive sector prediction?
Fig.\,\ref{fig:naivefpc} tests this at every $N_d=0$ composition for
$n=2,3,4$ ($L=4$), judging three progressively more complete
predictions against the same exact $U\to\infty$ value of $\mu_2$:
(a) the naive coordinate-counting prediction, the variance
$2(2n)\lambda^2=4n\lambda^2$ of the free $2n$-dimensional hypercubic
lattice, is uniformly too large; (b) the exact $U=0$ variance,
$\sum_\sigma N_\sigma z\lambda^2(N-N_\sigma)/(N-1)$ (the FPC identity
applied within each spin species), brings
single-species compositions exactly onto the diagonal (expected, since
FPC \emph{is} the exact result for one species), but leaves mixed-spin
compositions with a residual growing with the degree of spin mixing,
the additional real-space exclusion
between opposite-spin electrons that only exists at strong coupling;
(c) the full $U=\infty$ formula of \S\ref{sec:mu2theorem} brings
\emph{every} composition onto the diagonal, as it must, being exact on
any graph; panels (a) and (b) show that neither the coordinate count
nor same-spin exclusion alone reproduces it.
Doublon-containing ($N_d\geq1$) sectors are a separate question, since
they only exist at finite $U$; their band centres agree with
$N_d\varepsilon_d$ to $\leq0.10\%$ at $U=40\lambda$ for every composition
tested, $(2,1)$, $(3,1)$ and $(2,2)$
(\S\ref{sec:n3test}--\ref{sec:n4test}).

\section{Exchange topology in the fourth moment}
\label{sec:mu4exchange}

Writing $\mu_4=\Tr(H_\infty^4)/D_0=\sum_{a,b}(H_\infty^2)_{ab}^2/D_0$
makes the physical content explicit: $\mu_4$ measures the squared
multiplicity of two-hop connections between configurations, summed over
every pair $(a,b)$: the diagonal terms ($a=b$) are backtracking/local
connectivity, and the off-diagonal terms measure \emph{interference
between distinct two-hop paths} connecting the same two configurations.
Exchange is not one closed-walk type among several but an interference
effect: it arises whenever two \emph{different} particles' two-hop
paths connect the same pair of configurations, since whether the
resulting amplitude constructively or destructively interferes with
the direct path depends on the fermionic sign of the intervening
permutation. The closed-walk classification below organizes the same
sum differently; the two bookkeepings assign the exchange contribution
to different terms but give the same same-spin minus opposite-spin
difference (Appendix~\ref{app:mu4walks}).

A length-4 closed walk in the $N_d=0$ configuration graph returns the
occupied site set to itself and induces a permutation of ``which
original particle ends up where''; we classify walks by the cycle type
of that permutation. Hard-core blocking in this sector does not depend
on spin, so which walks exist from a configuration depends only on its
occupied site set; spin enters only through the requirement that the
permutation map every particle onto one of the same spin, so that the
walk returns to the identical occupation-number state. \textbf{Identity}
walks (a single particle's self-returning excursions, or several
particles' excursions that each return) are therefore spin-blind.
\textbf{Ring exchange}: two occupied sites $i,j$ admit a 4-hop sequence,
through two empty sites, that returns the occupied site set to
$\{i,j\}$ but exchanges which particle sits where; it contributes if and
only if $i,j$ hold the \emph{same spin}. This is the first, and for
$n=2$ the only, spin-dependent closed-walk mechanism; for $n\geq3$,
four-hop three-cycles provide a second. The complete second-quantized
classification, including every fermionic sign, is given in
Appendix~\ref{app:mu4walks}. Throughout, we write
$\mu_4(\infty)=A(N,n)+\alpha(N,n)P_{\rm same}$ in units of $\lambda^4$,
with $P_{\rm same}=\binom{N_\uparrow}{2}+\binom{N_\downarrow}{2}$ the
same-spin pair count; $A$ and $\alpha$ depend on the graph as well as
on $N$ and $n$.

\paragraph{Two geometrically distinct exchange channels.} On the
$L\times L$ torus with $L=3$ or $L\geq5$, ring exchange occurs through
two channels of equal weight: a \emph{diagonal} channel
(next-nearest-neighbor pairs sharing two common neighbors, an
elementary plaquette's opposite corners) and an \emph{edge} channel
(nearest-neighbor pairs, via a $3$-hop$+1$-hop traversal of an adjacent
plaquette). For each eligible pair of sites, the diagonal element
$(H_\infty^4)_{aa}$ of a configuration $a$ occupying that pair is
$8\lambda^4$ lower when the two spins are equal than when they are
opposite (derived in Appendix~\ref{app:mu4walks}, and confirmed by
exact sparse-trace evaluation at $L=5,6,7$); this per-pair value is a
local plaquette-incidence count, not a function of the coordination
number alone (see the hypercubic comparison in
Appendix~\ref{app:mu4walks}). With $2N$ pairs in each channel, the
$n=2$ coefficient follows \textbf{in closed form, not fitted}:
\begin{equation}
\alpha(N,2) = \frac{-8\cdot4N}{\binom N2} = -\frac{64}{N-1},
\end{equation}
matching every lattice of these sizes tested exactly. At $L=4$ the
torus has extra 4-cycles (its axis rings): wraparound adds $16$
diagonal-type pairs and raises the edge channel to $-12$ per pair,
giving $\alpha(16,2)=-32/5$, which the graph-general form below
reproduces (Appendix~\ref{app:mu4walks}). The fermion-sign path-by-path derivation
of both channels is given in Appendix~\ref{app:mu4walks}.

\paragraph{Exact linearity on any graph.} That $\mu_4$ is a symmetric
polynomial of degree at most four in the spin labels does not by itself
exclude a genuine $P_{\rm same}^2$ term: that would
require a four-particle cyclic permutation or a double transposition, both of which force a
$1{+}1{+}1{+}1$ hop distribution among four occupied sites. No such
schedule has a legal first move, since each candidate first hop needs
its target site vacated by another hop of the same schedule, a closed
dependency loop with no entry point (machine-checked on any graph
without self-loops, Appendix~\ref{app:lean}). Four-hop closed walks
therefore induce only the identity, transpositions, and 3-cycles. Now
average over spin labels: the trace over a sector $(N_\uparrow,N_\downarrow)$
runs over site sets $S$ ($|S|=n$) and, for each, over the
$\binom{n}{N_\uparrow}$ spin labellings. A walk inducing $\pi$
contributes $\lambda^4\,{\rm sgn}(\pi)$ whatever the spectators' spins
(the closed-walk amplitude lemma of Appendix~\ref{app:mu4walks}), but only in the labellings that $\pi$
preserves. Two given particles share a spin in a fraction
$P_{\rm same}/\binom n2$ of the labellings, and three given particles
all share a spin in a fraction $P_3/\binom n3$, with
$P_3\equiv\binom{N_\uparrow}{3}+\binom{N_\downarrow}{3}$. Hence
\begin{equation}
\mu_4(\infty)=I-T_2\,\frac{P_{\rm same}}{\binom n2}+T_3\,\frac{P_3}{\binom n3},
\label{eq:mu4decomp}
\end{equation}
where $I$, $T_2$, $T_3$ are the average numbers of identity,
transposition and 3-cycle walks per configuration, averaged over
uniformly random $n$-site sets: they depend on the graph and $(N,n)$,
not on spin. (They are numbers, not to be confused with the hopping
operators $T_0$, $T_1$ of \S\ref{sec:SW}.) Finally $P_3$ is an \emph{exact affine function} of
$P_{\rm same}$ at fixed $n$,
$P_3=\binom{n}{3}-\tfrac{n-2}{2}\big[\binom{n}{2}-P_{\rm same}\big]$
for every $n$ (a combinatorial identity, machine-checked in
Appendix~\ref{app:lean}), despite looking cubic. So $\mu_4(\infty)$ is
exactly affine in $P_{\rm same}$ on any finite graph, with
\begin{equation}
\alpha(N,n)=-\frac{T_2}{\binom n2}+\frac{(n-2)\,T_3}{2\binom n3}.
\label{eq:alphaT}
\end{equation}
This is a first-principles proof of exact linearity, not an empirical
observation, confirmed independently by an over-determined $n=6$ test
(Table~\ref{tab:n6test}: four distinct spin compositions at fixed
$(L,n)=(5,6)$, over-determining a linear fit by two degrees of freedom):
in exact rational arithmetic the four points lie exactly on one line,
$\mu_4=259863/253-(38/23)\,P_{\rm same}$.

\begin{table}[htbp]
\centering
\begin{tabular}{@{}ccrr@{}}
\toprule
composition & $P_{\rm same}$ & {\color{ConnBlue}$\mu_2(\infty)$} & {\color{ExchVermillion}$\mu_4(\infty)$} \\
\midrule
$(6,0)$ & $15$ & \color{ConnBlue!70!black}$19$ & $1002.343874$ \\
$(5,1)$ & $10$ & \color{ConnBlue!70!black}$19$ & $1010.604743$ \\
$(4,2)$ & $7$  & \color{ConnBlue!70!black}$19$ & $1015.561265$ \\
$(3,3)$ & $6$  & \color{ConnBlue!70!black}$19$ & $1017.213439$ \\
\bottomrule
\end{tabular}
\caption{Over-determined $n=6$ linearity test ($L=5$): four distinct spin compositions, all sharing the same spin-blind $\mu_2=19$, against two free parameters ($A,\alpha$) in $\mu_4=A+\alpha P_{\rm same}$. The $\mu_4$ are exact rationals ($253593/253$, $255683/253$, $256937/253$, $257355/253$), and the four points lie exactly on the line $\mu_4=259863/253-(38/23)\,P_{\rm same}$.}
\label{tab:n6test}
\end{table}

\paragraph{The exchange coefficient at every $n$.} On any finite simple
graph, $T_2$ and $T_3$ can be counted exactly, because every
spin-sensitive four-hop walk lives on a 4-cycle of the graph. A
transposition walk exchanges two particles on a 4-cycle (opposite
corners, through the other two; or adjacent corners, three hops around
the cycle and one back) and needs exactly its pair occupied and the
other two corners empty; there are $32C_4$ such walks in all at $n=2$
(per 4-cycle, eight for each of its two diagonal pairs and four for
each of its four edges; S5, Lemma 3),
$C_4$ being the number of 4-cycles of the graph (cycles of length four
counted as subgraphs, chords allowed). A 3-cycle walk is the empty corner
of a 4-cycle going once around it, in either direction, with the other
three corners occupied: $8C_4$ walks in all (four empty corners, two
directions). Averaging the
occupation pattern over uniformly random $n$-site sets (hypergeometric
probabilities, exact) gives
\begin{equation}
\label{eq:alphaapprox}
\frac{T_2}{\binom n2}=\frac{64\,C_4\,(N-n)(N-n-1)}{N(N-1)(N-2)(N-3)},\qquad
\frac{(n-2)\,T_3}{2\binom n3}=\frac{24\,C_4\,(n-2)(N-n)}{N(N-1)(N-2)(N-3)},
\end{equation}
so that, with Eq.~\eqref{eq:alphaT},
\begin{equation}
\alpha(N,n)=-\frac{8\,C_4\,(N-n)(8N-11n-2)}{N(N-1)(N-2)(N-3)}\qquad(N\geq4),
\label{eq:alphaN}
\end{equation}
which reduces to $-64C_4/[N(N-1)]$ at $n=2$. On the $L\times L$ torus
$C_4=N$ for $L=3$ and $L\geq5$ (the plaquettes; at $L=3$ the axis rings
are triangles), giving $\alpha(N,2)=-64/(N-1)$, while at $L=4$ the eight
axis rings of length four make $C_4=24$. In exact rational arithmetic
Eq.~\eqref{eq:alphaN} reproduces every slope we computed: on the square
torus at $L=3$ ($n=3,4,5$), $L=4$ ($n=2,3,4$), $L=5$ ($n=3,4,6$,
including the $-38/23$ of Table~\ref{tab:n6test}), $L=6$ ($n=3,4,5$)
and $L=7$ ($n=3$), and on graphs unlike the square lattice: $K_6$, the
Petersen graph ($C_4=0$, $\alpha=0$), a triangular torus, a random graph
and the kagome lattice ($C_4=0$). The transposition term alone, the
independent blocking of a ring exchange by the $n-2$ other particles,
is therefore not an approximation of $\alpha$ but its exact
transposition part; the 3-cycle term is exactly the remainder that the
hop-class analysis of Appendix~\ref{app:mu4walks} finds in the
$2{+}1{+}1$ class. The walk counts $32C_4$ and $8C_4$ are derived in S5;
the $n=2$ channels are worked out in Appendix~\ref{app:mu4walks}.

The $n=4$ data of Table~\ref{tab:mu4exact} give the same conclusion at $L=4$: the two
independently-determined slopes $(4,0)\to(3,1)$ and $(3,1)\to(2,2)$
agree exactly ($-1968/455$).

\begin{table}[htbp]
\centering
\begin{tabular}{@{}cccrrr@{}}
\toprule
$L$ & composition & $n$ & $P_{\rm same}$ & {\color{ConnBlue}$\mu_2(\infty)$} & {\color{ExchVermillion}$\mu_4(\infty)$} \\
\midrule
4 & $(2,0)$ & 2 & 1 & \color{ConnBlue!70!black}$112/15$ & $2176/15$ \\
4 & $(1,1)$ & 2 & 0 & \color{ConnBlue!70!black}$112/15$ & $2272/15$ \\
\cmidrule(l r){1-6}
4 & $(3,0)$ & 3 & 3 & \color{ConnBlue!70!black}$52/5$ & $9976/35$ \\
4 & $(2,1)$ & 3 & 1 & \color{ConnBlue!70!black}$52/5$ & $10348/35$ \\
\cmidrule(l r){1-6}
4 & $(4,0)$ & 4 & 6 & \color{ConnBlue!70!black}$64/5$ & $197632/455$ \\
4 & $(3,1)$ & 4 & 3 & \color{ConnBlue!70!black}$64/5$ & $203536/455$ \\
4 & $(2,2)$ & 4 & 2 & \color{ConnBlue!70!black}$64/5$ & $15808/35$ \\
\bottomrule
\end{tabular}
\caption{Exact sparse-trace $\mu_2,\mu_4$ at $U=\infty$ (in units of $\lambda^2,\lambda^4$), given as exact rationals ($\mu_4=\Tr(H_\infty^4)/D_0$ with $D_0=\binom{16}{N_\uparrow}\binom{16-N_\uparrow}{N_\downarrow}$ an integer, so $\mu_4$ is exactly rational, no rounding): \textcolor{ConnBlue!70!black}{$\mu_2$ is spin-blind} (identical within each $n$-group, separated by horizontal rules), while \textcolor{ExchVermillion}{$\mu_4$ depends on $P_{\rm same}$}.}
\label{tab:mu4exact}
\end{table}

\paragraph{The spin-independent intercept at every $n$.} The same
counting fixes $A(N,n)$. Let hard-core bosons hop on the same graph.
Their configuration graph is that of one species of projected fermions
($N_\downarrow=0$) with every sign set to $+1$: the $n$-token graph (symmetric $n$-th power) of
$G$~\cite{FabilaMonroy2012}, whose adjacency matrix is the fixed-$n$
block of the hard-core-boson, i.e.\ spin-$\tfrac12$ XX,
Hamiltonian~\cite{MatsubaraMatsuda1956,Audenaert2007}. Every
four-hop closed walk therefore counts once, whatever permutation it
induces, and $\mu_4^{\rm HCB}=I+T_2+T_3$. With
Eq.~\eqref{eq:mu4decomp} and the $P_3$ identity (which gives
$P_3/\binom n3=1-\tfrac32[1-P_{\rm same}/\binom n2]$, since
$\tfrac{n-2}{2}\binom n2/\binom n3=\tfrac32$), the constant term of
$\mu_4(\infty)$ is $I+T_3-\tfrac32T_3$, so
\begin{equation}
A(N,n)=\mu_4^{\rm HCB}-T_2-\tfrac32\,T_3 .
\label{eq:AfromHCB}
\end{equation}
The boson moment is exact on any finite simple graph
(Appendix~\ref{app:mu4walks}). With $m$ edges, $W=\sum_v\binom{d_v}{2}$
pairs of edges sharing a vertex, $D=\binom m2-W$ pairs of vertex-disjoint
edges, and the canonical occupation probabilities
$p_{ab}=(n)_a(N-n)_b/(N)_{a+b}$ (falling factorials: the probability that
$a$ given sites are occupied and $b$ given others empty),
\begin{equation}
\mu_4^{\rm HCB}=(2m+4W+8C_4)\,p_{11}+(24D+16C_4)\,p_{22}.
\label{eq:muHCB}
\end{equation}
The walk counts $32C_4$ and $8C_4$ of Eq.~\eqref{eq:alphaapprox}, with
the occupation probabilities $p_{22}$ and $p_{31}$, give $T_2=32C_4p_{22}$
and $T_3=8C_4p_{31}$ (for $n\leq2$, $T_3=0$, and for $n\leq1$, $T_2=0$,
so that Eq.~\eqref{eq:AfromHCB} holds trivially there), hence
\begin{equation}
A(N,n)=(2m+4W+8C_4)\,p_{11}+(24D-16C_4)\,p_{22}-12C_4\,p_{31}.
\label{eq:Aclosed}
\end{equation}
No step assumes the graph triangle-free (a closed four-hop walk cannot
go around a triangle), so Eq.~\eqref{eq:Aclosed} holds on graphs with
triangles; it is machine-checked on every finite simple graph with
$N\geq4$ (Appendix~\ref{app:lean}). On the $L\times L$ torus with $L=3$ or $L\geq5$
($m=2N$, $W=6N$, $C_4=N$) it reads
$A=36Np_{11}+(48N^2-184N)p_{22}-12Np_{31}$; at $L=4$ ($C_4=24$) it gives
$A=2272/15$, $10534/35$ and $5984/13$ for $n=2,3,4$, the intercepts of
the lines through the values of Table~\ref{tab:mu4exact} (for $n=3,4$ no
composition has $P_{\rm same}=0$). In exact rational arithmetic,
$A+\alpha P_{\rm same}$ reproduces every value of
Tables~\ref{tab:mu4exact} and~\ref{tab:n6test}; every composition of the
square tori with $L=3$ ($n=3,4,5$), $L=4$ ($n=2,3,4$), $L=5$ ($n=3,4$)
and $L=6,7$ ($n=3$), and of $K_6$, the Petersen graph, the $3\times3$
triangular torus and a random graph on eight sites ($n=2,3,4$); and every
sector of every simple graph with
$N\leq6$ (one per isomorphism class) and of 400 random graphs with
$N=7$--$10$.

The graph expansion behind Eq.~\eqref{eq:muHCB} is classical: it is the
fourth-order term of the high-temperature series of the spin-$\tfrac12$
XY model (quantum lattice gas)~\cite{BettsElliottLee1970,Lee1971}, in its
field-dependent, i.e.\ grand-canonical,
form~\cite{RogiersDekeyser1973,DekeyserRogiers1975}; on a finite graph
the canonical moment is a coefficient of that trace in the fugacity
(Appendix~\ref{app:mu4walks}). The induced 4-cycles of a token graph are
classified in Ref.~\cite{FabilaMonroyTrujilloNegrete2022}. Replacing
$p_{ab}$ by $\rho^a(1-\rho)^b$, with $\rho$ the mean filling, gives the
grand-canonical fourth moment
of the hard-core-boson (XY) hopping at infinite temperature,
$(2m+4W+8C_4)\,q+(24D+16C_4)\,q^2$ with $q=\rho(1-\rho)$; subtracting
$3\mu_2^2=12m^2q^2$ and using $24D=12m(m-1)-24W$ gives its fourth
cumulant, $\kappa_4=(2m+4W+8C_4)\,q-(12m+24W-16C_4)\,q^2$.
With the face-centred-cubic counts $m/N=6$, $W/N=66$, $C_4/N=33$ this is
$540q-1128q^2$ per site; divided by $4!$ it is exactly the $K^4$ term,
$\tfrac12(45q-94q^2)$, of $N^{-1}\ln Z$ in the f.c.c. series of
Ref.~\cite{RogiersDekeyser1973}, where $K=\beta J$ for the hopping
$-J\sum_{\langle ij\rangle}(a_i^\dagger a_j+a_j^\dagger a_i)$ and, in
their variables, $x^2y=x^3(y+y^2)=q$, $x^4(y+y^3)=q(1-2q)$ and
$x^4y^2=q^2$. Grand-canonical, thermodynamic-limit counterparts of the
fermionic moment on specific lattices are contained in the
high-temperature series of the $U=\infty$ Hubbard
model~\cite{Plischke1974,KuboTada1983,KuboTada1984} and of the $t$-$J$
model at $J=0$~\cite{Putikka1992}, whose spin sums are organized by the
same permutation cycles; the 3-cycle term is the loop of a hole around a
plaquette of Thouless and Nagaoka~\cite{Thouless1965,Nagaoka1966}. What
is added here is the exact canonical (fixed-$N_\uparrow$, $N_\downarrow$)
form on an arbitrary finite graph and its combination with the exchange
counts into Eq.~\eqref{eq:Aclosed}.

\medskip
\noindent\textbf{Proposition (exact fourth moment).} \emph{On any finite
simple graph with $N\geq4$ sites, $m$ edges, $W$ pairs of edges sharing a
vertex, $D$ pairs of disjoint edges, $C_4$ four-cycles and uniform hopping
$\lambda$ without Peierls phases (for those, see Eq.~\eqref{eq:mu4flux}),
for projected spin-$\tfrac12$ fermions in any sector $(N_\uparrow,N_\downarrow)$
with $n=N_\uparrow+N_\downarrow\leq N$ and
$P_{\rm same}=\binom{N_\uparrow}2+\binom{N_\downarrow}2$, in units of
$\lambda^4$,}
\begin{equation}
\boxed{\begin{aligned}
\mu_4(\infty) &= A(N,n) + \alpha(N,n)\,P_{\rm same},\\
A(N,n) &= (2m+4W+8C_4)\,p_{11}+(24D-16C_4)\,p_{22}-12C_4\,p_{31},\\
\alpha(N,n) &= -\frac{8\,C_4\,(N-n)(8N-11n-2)}{N(N-1)(N-2)(N-3)},
\end{aligned}}
\end{equation}
\emph{with $p_{ab}=(n)_a(N-n)_b/(N)_{a+b}$. For $n\leq1$, $P_{\rm same}=0$ and
$\mu_4(\infty)=A(N,n)$, which at $n=1$ is $\Tr A_G^4/N$, with $A_G$ the
adjacency matrix of the graph. On the square-lattice torus
with $L=3$ or $L\geq5$, $C_4=N$ and $\alpha(N,2)=-64/(N-1)$.}
\medskip

The resulting picture: $\mu_2$ resolves the purely local
\emph{connectivity} of the configuration graph, while on triangle-free graphs,
where no closed three-hop walk exists and so $\mu_3=0$, $\mu_4$ is the
first moment to resolve closed-loop
\emph{geometry}: every area-enclosing four-hop walk, of
which fermionic exchange is one instance and ordinary single-particle
orbital motion (\S\ref{sec:futurework}) is another. Under a magnetic
flux the two are distinguished only by the sign with which their
loops interfere (their interference parity), not by any difference in
the underlying harmonic (\S\ref{sec:futurework}). On graphs with
triangles (the $L=3$ torus, $K_6$, the triangular torus), $\mu_3$ is in
general nonzero and already contains three-hop loops, including a three-hop exchange of two
particles around a triangle.

\begin{figure}[htbp]
\centering
\includegraphics[width=0.98\textwidth]{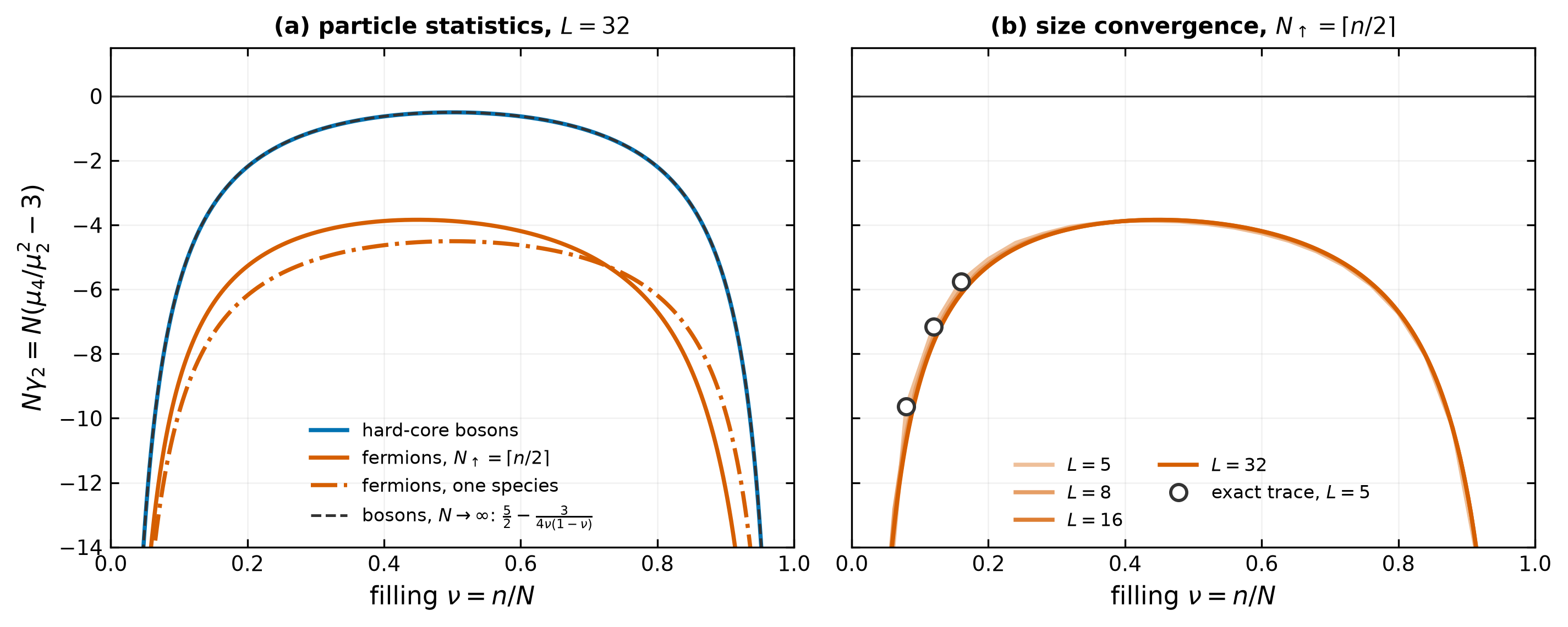}
\caption{Excess kurtosis of the $U=\infty$ spectrum, $N\gamma_2=N(\mu_4/\mu_2^2-3)$, against filling on $L\times L$ square tori, from the Proposition and the second-moment theorem. (a) $L=32$: hard-core bosons, projected fermions with $N_\uparrow=\lceil n/2\rceil$, and one spin species; the dashed line, the boson limit $\tfrac52-3/[4\nu(1-\nu)]$, lies under the hard-core-boson curve at this size. (b) Convergence with $L$ for $N_\uparrow=\lceil n/2\rceil$; open circles are exact sparse traces ($L=5$, $n=2,3,4$). The scaled kurtosis diverges as $\nu\to0$ or $1$, where the fixed-filling scaling does not apply; the axis is clipped.}
\label{fig:mu4kurtosis}
\end{figure}

\paragraph{The shape of the spectrum.} With $\mu_2=2m\,p_{11}$
(\S\ref{sec:mu2theorem}), the Proposition gives the excess kurtosis
$\gamma_2=\mu_4/\mu_2^2-3$ of the $N_d=0$ spectrum in closed form, for
every graph, filling and spin composition (Fig.~\ref{fig:mu4kurtosis}).
At fixed filling $\nu=n/N$ it is of order $1/N$. With $q=\nu(1-\nu)$,
expanding $p_{11}=q\,[1+1/N+O(N^{-2})]$ and
$p_{22}=q^2\,[1+(6-1/q)/N+O(N^{-2})]$ in Eq.~\eqref{eq:muHCB} gives,
for hard-core bosons on the square torus ($L\geq5$),
$N\gamma_2\to\tfrac52-3/(4q)$, which is $-\tfrac12$ at half filling. The fermions lie lower:
their intercept differs from the boson moment by $-T_2-\tfrac32T_3$, and
the exchange term $\alpha P_{\rm same}$ adds to it, both at the same
order and differently for a single species and for balanced spins. The value
$\gamma_2=-204/529$ of the $(1,1)$ sector at $L=5$
(\S\ref{sec:fewbody}) is the $n=2$ point of this family.

\section{Finite \texorpdfstring{$U$}{U}: virtual doublons}
\label{sec:SW}

This section derives how $\mu_2$ of the $N_d{=}0$ band deforms as $U$ is
lowered from infinity, through order $U^{-2}$, and checks the result
against full exact diagonalization of the Hubbard model.

\subsection{Effective Hamiltonian through third order}

Write $H=H_t+U\hat N_d$ ($\hat N_d$ the number of doubly occupied sites). From the
$N_d{=}0$ sector $P_0$, a single hop reaches \emph{only} the $N_d{=}1$
sector $Q_1$, and a path $P_0\to Q_1\to Q_1\to P_0$ cannot pass through
$N_d\geq2$, so through third order $Q_1$ is the only virtual sector.
The Hermitian Schrieffer--Wolff expansion~\cite{ChaoSpalekOles1977,MacDonaldGirvinYoshioka1988}
then gives
\begin{equation}
H_{\rm eff} = T_0 - \frac{1}{U}V + \frac{1}{U^2}H^{(3)} + O(1/U^3),\label{eq:Heff}\qquad
H^{(3)} = T_1^\dagger T_0^{Q}T_1-\tfrac12\{V,T_0\},
\end{equation}
where $T_0=P_0H_tP_0$ is the operator $H_\infty$ of \S\ref{sec:Uinfinity}
(a hop into an empty site; $T_0$ is the usual name in the
strong-coupling expansion),
$T_1=Q_1H_tP_0$ (hop \emph{onto} an opposite-spin occupied site,
creating one doublon), $V\equiv T_1^\dagger T_1$, and $T_0^{Q}=Q_1H_tQ_1$
(a hop that keeps exactly one doublon: a single electron moving into an
empty site, or the doublon moving by an exchange hop); the construction
is given in Appendix~\ref{app:SW}. Writing $\langle X\rangle\equiv\Tr(X)/D_0$
and $\langle X,Y\rangle_{\rm HS}\equiv\Tr(X^\dagger Y)/D_0$ for the
Hilbert--Schmidt inner product, with $\|X\|_{\rm HS}^2=\langle X,X\rangle_{\rm HS}$,
we must be explicit about raw vs.\ central moments here, unlike at
$U=\infty$: $V$ has a nonzero diagonal, so the band's mean energy is
$\bar E(U)=-\langle V\rangle/U+O(1/U^2)$. The physically meaningful
quantity is the \emph{central} second moment
$\mu_{2,c}(U)\equiv\langle(H_{\rm eff}-\bar E(U))^2\rangle$. Since
$\Tr T_0=0$,
\begin{equation}
\mu_{2,c}(U) = \mu_2(\infty) - \frac{2}{U}\langle T_0V\rangle + \frac{c_2}{U^2} + O(1/U^3),
\qquad c_2=\Var(V)+2\langle T_0H^{(3)}\rangle,
\label{eq:mu2cU}
\end{equation}
with $\Var(V)\equiv\langle V^2\rangle-\langle V\rangle^2$. Every term is
an exact sparse trace (an exact rational at $\lambda=1$). Two operators
contribute at order $U^{-2}$: the square of the second-order correction,
$\Var(V)=\|V-\langle V\rangle I\|_{\rm HS}^2\geq0$, and the overlap of the
third-order term with the bare hopping. At the low fillings of
Table~\ref{tab:swamp} the second is larger and of opposite sign, so
$c_2<0$: at large finite $U$ the $N_d{=}0$ band is \emph{narrower} than
at $U=\infty$. This sign is a computed fact, not a theorem, and it
reverses near half filling, where the bare hopping has little room to
act (at half filling $T_0=0$ and $c_2=\Var(V)>0$; at $L=4$, $c_2>0$
already for $(13,1)$ and $(12,2)$).

\subsection{The \texorpdfstring{$1/U$}{1/U} term vanishes on triangle-free graphs}

The $1/U$ coefficient is $-2\langle T_0,V\rangle_{\rm HS}$. Direct computation
gives $\langle T_0,V\rangle_{\rm HS}=0$ exactly (an integer trace) for
$L=4,5,6$, every composition tested. The reason is graph locality:
$T_0$'s edges are single nearest-neighbor hops, while $V$'s off-diagonal
elements connect configurations related either by an exchange of the
spins on a bond (the superexchange piece) or by a hop between two sites
sharing a common mediating neighbor (the correlated three-site-hopping
piece, with its spin-flip part), the standard decomposition of the
strong-coupling
expansion~\cite{ChaoSpalekOles1977,MacDonaldGirvinYoshioka1988}
(Appendix~\ref{app:SW}). Neither shares an off-diagonal matrix element
with the bare projected hopping, unless a mediated pair of sites is itself a
bond, which requires a triangle. On any triangle-free graph
$\Tr(T_0V)=0$, and the $L\times L$ torus ($L\geq3$) is triangle-free
exactly when $L\neq3$; both statements are machine-checked
(Appendix~\ref{app:lean}). The second-order correction operator $V$ is
therefore exactly \emph{Hilbert--Schmidt orthogonal} to the projected
kinetic operator for $L\geq4$, and $\mu_{2,c}(U)$ has no $1/U$ term:
\begin{equation}
\mu_{2,c}(U) - \mu_2(\infty) = \frac{c_2}{U^2} + O(1/U^3)\qquad(L\geq4).
\end{equation}
This is a proven cancellation, not a fit. (The third-order operator is
not orthogonal to $T_0$; its overlap is what enters $c_2$.) At $L=3$
the overlap is nonzero (a small-torus wraparound effect where the
mediated site coincides with a true nearest neighbor), and full exact
diagonalization confirms the resulting $1/U$ term: its coefficient
matches $-2\langle T_0,V\rangle_{\rm HS}$ to better than $10^{-3}$
(relative) for $(1,1)$ and $(2,1)$.

\subsection{The \texorpdfstring{$U^{-2}$}{U⁻²} coefficient and its \texorpdfstring{$N_\uparrow N_\downarrow$}{N↑N↓} dependence}

Exact sparse traces give $c_2$ for every composition
(Table~\ref{tab:swamp} at $L=4$), and full exact diagonalization of the
Hubbard model reproduces it: for $(1,1)$ and $(2,1)$ at $L=4$, the
$U^{-2}$ coefficient of the exact band variance agrees with $c_2$ to
better than $10^{-3}$ (relative). Here the band is the lowest $D_0$
levels of the exact spectrum, and the coefficient is the Richardson
extrapolation of $U^2[\mu_{2,c}(U)-\mu_2(\infty)]$ from $U=400$ and
$800\lambda$. $\Var(V)$ alone has the wrong sign;
for $(2,1)$, $\Var(V)=9.643175$ while $2\langle T_0H^{(3)}\rangle=-816/35$.
Single-species sectors have $c_2=0$: no doublon can form. For two
particles the computed values follow a simple form,
$c_2(1,1)=-96/(N-1)-64/(N-1)^2$ exactly at $L=5$--$8$, which we have not
derived.

At low filling, $c_2/(N_\uparrow N_\downarrow)$ is close to
composition-independent at fixed $(L,n)$, and its relative spread
between two compositions ($|$difference$|$ divided by the mean) shrinks
as $L$ grows:
$0.14\%$, $0.06\%$, $0.03\%$ between $(3,1)$ and $(2,2)$ at $L=4,5,6$,
and $0.34\%$, $0.13\%$ between $(4,1)$ and $(3,2)$ at $L=4,5$. The
near-independence degrades as the filling grows (at $L=4$, $n=6$, the
spread is $0.79\%$ between $(5,1)$ and $(3,3)$), and near half filling
$c_2$ changes sign. At low filling this is numerical evidence that the
leading finite-$U$ correction is governed by the opposite-spin pair
count, $c_2\approx\beta(L,n)\,N_\uparrow N_\downarrow$
with a composition-independent $\beta(L,n)<0$. At fixed $n$ this is
the same statement as $c_2$ being close to affine in $P_{\rm same}$,
since $N_\uparrow N_\downarrow=\binom n2-P_{\rm same}$, with the exact
zero $c_2=0$ of a single species fixing the intercept; with two or
three compositions per $(L,n)$, each test is one ratio. We do not
derive $\beta(L,n)$; at fixed $n$, $|\beta|$ is at most of order $1/N$,
by a bound $|c_2|\leq C(n,z)\lambda^4N_\uparrow N_\downarrow/(N-1)$, with
$C(n,z)$ depending only on $n$ and the coordination $z$,
proved in S7.
The $N_\uparrow N_\downarrow$ law is a numerical trend, epistemically
distinct from the exact $\mu_2(\infty)$ theorem and from the exact
vanishing of the $1/U$ term.

\begin{table}[htbp]
\centering
\begin{tabular}{@{}lrrrrr@{}}
\toprule
composition & $n$ & $N_\uparrow N_\downarrow$ & $\Var(V)$ & $c_2$ (exact) & $c_2$ (decimal) \\
\midrule
$(4,0)$, $(3,0)$ & 4, 3 & 0 & $0$ & $0$ & $0$ \\
$(1,1)$ & 2 & 1 & $5.048889$ & $-1744/225$ & $-7.751111$ \\
$(2,1)$ & 3 & 2 & $9.643175$ & $-3076/225$ & $-13.671111$ \\
$(3,1)$ & 4 & 3 & $13.756484$ & $-40704/2275$ & $-17.891868$ \\
$(2,2)$ & 4 & 4 & $18.376361$ & $-487744/20475$ & $-23.821441$ \\
\bottomrule
\end{tabular}
\caption{The exact $U^{-2}$ coefficient $c_2$ of the $N_d{=}0$ band
variance at $L=4$ (units of $\lambda^4$), Eq.~\eqref{eq:mu2cU}, and its
second-order part $\Var(V)$ alone (to six decimals). For $(1,1)$ and $(2,1)$, $c_2$ agrees
with full exact diagonalization to better than $10^{-3}$. These are
exact numbers, not a closed form: $\beta(L,n)$ is not derived
(\S\ref{sec:SW}).}
\label{tab:swamp}
\end{table}

Both spin-resolved results are therefore statements about the
dependence on $P_{\rm same}$ at fixed $n$: $\mu_4(\infty)$ is exactly
affine in it, with the exact coefficient of \S\ref{sec:mu4exchange};
$c_2$, the leading finite-$U$ correction to $\mu_2$, is close to affine
in it, with its zero at a single species, as a numerical trend. We
keep this asymmetry in confidence explicit rather than presenting both
as theorems.

\section{Gauge response: the flux-resolved fourth moment and the doublon}
\label{sec:futurework}

Promoting the hopping amplitudes to gauge-dependent variables (S9),
$t_{ij}\to t_{ij}e^{iA_{ij}}$ ($A_{ji}=-A_{ij}$), with
$\Phi_\square=\sum_\square A_{ij}$ the flux through an elementary
plaquette (for a uniform flux on the $L\times L$ torus,
$\Phi_\square=2\pi p/L^2$ with integer $p$), is especially natural in the
present formulation because spectral moments are sums over closed
configuration-space paths: at second order the contributing paths
backtrack (occupied site to empty neighbor and back), and are
therefore gauge-blind, $\mu_2(\Phi)=\mu_2(0)$ identically, since a
backtrack's phase is $e^{iA_{ij}}e^{-iA_{ij}}=1$ for any $A_{ij}$; at
fourth order and higher, closed walks enclosing nonzero area can
acquire nontrivial gauge phases. We report the $n=2$ calculation
carried through exactly, then the closed form for every $n$ and
arbitrary phases (Proposition below), together with a machine-precision numerical
cross-check against an explicit Hofstadter-type~\cite{Hofstadter1976}
Peierls Hamiltonian
(Fig.~\ref{fig:mu4exchangeflux}, Appendix~\ref{app:numerics}). Full
derivations are in S10 and S11; the closed form for arbitrary phases,
which also fixes the flux-independent baseline, is the Proposition below
(proof in Appendix~\ref{app:mu4walks}).

\paragraph{The organizing principle.} Since $H_\infty(\Phi)$ is
Hermitian for any $\Phi$,
$\mu_4(\Phi)=\sum_{a,b}|(H_\infty^2(\Phi))_{ab}|^2/D_0$ term by term. A
closed-walk contribution reached by a \emph{single} path is therefore
exactly flux-independent ($|\lambda^2se^{i\theta}|^2=\lambda^4$, with
$s=\pm1$ a fermion sign, regardless of $\theta$); flux dependence requires interference between
\emph{at least two} distinct paths reaching the same intermediate
configuration, with their relative phase equal to the gauge-invariant
Wilson-loop phase $\Theta_C=\sum_{(ij)\in C}A_{ij}$ of the closed loop
they form together (an integer multiple of $\Phi_\square$ for the
contractible loops of the square lattice; on the $4\times4$ torus some
four-hop loops wind around it and pick up the boundary holonomies
instead). Multiple interfering paths are necessary but not sufficient for flux
sensitivity, since two paths could still carry a vanishing relative
phase, but this organizes every closed-walk type classified above into
``single-path, hence flux-blind'' or ``multi-path, hence potentially
flux-sensitive.''

\paragraph{Flux-sensitive four-hop mechanisms.} In the classification
of Appendix~\ref{app:mu4walks}, three closed-walk types are
multi-path: the single-particle plaquette loop (Type I, spin-blind),
the two ring-exchange channels (Type II(b), diagonal and edge), and the
3-cycle (Type III, needed only for $n\geq3$). All of them give the
identical functional dependence on $\Phi_\square$, differing only in
interference \emph{parity}: the
plaquette loop and the 3-cycle are \emph{constructive} at zero flux
(their two contributing routes share a fermion sign, exhaustively
verified for the 3-cycle in Appendix~\ref{app:mu4walks}), giving $\propto(1+\cos\Phi_\square)$;
both ring-exchange channels are \emph{destructive} at zero flux (their
two contributing assignments carry opposite signs, the exact
cancellation derived in Appendix~\ref{app:mu4walks}), giving
$\propto(1-\cos\Phi_\square)$ for the same-spin case that survives,
with the opposite-spin case exactly flux-blind. For $n=2$ on the
$L\times L$ torus with $L=3$ or $L\geq5$ this assembles into an exact
closed form for the exchange coefficient,
\begin{equation}
\alpha(N,2,\Phi) = -\frac{64}{N-1}\cos\Phi_\square,
\label{eq:alphaflux}
\end{equation}
confirmed to machine precision (deviation below $10^{-12}$) against
exact sparse-trace evaluation with Peierls phases at every quantized
flux for $L=3,5,6$ and at every third one for $L=7$. At $L=4$ it fails already at
zero flux ($\alpha(16,2)=-32/5$, \S\ref{sec:mu4exchange}), and under
flux the $L=4$ value also depends on the boundary twists. A
companion check at $L=4,5$ finds no $\cos2\Phi_\square$ component in
the 3-cycle's contribution to the precision of the same sparse-trace
method, as the constructive-interference derivation (S10, Sec.~8)
requires; at $L=4$, loops winding around the torus add further
terms ($\cos4\Phi_\square$ and $\cos8\Phi_\square$, observed
numerically and derived below), in line with the $L=4$ caveat. The
Proposition below gives the fourth moment in closed form for every
graph, every spin composition and arbitrary phases; we call its
$\Theta$-independent term the baseline.

\medskip
\noindent\textbf{Proposition (fourth moment under Peierls phases).}
\emph{On any finite simple graph with $N\geq4$ sites, $m$ edges, $W$
pairs of edges sharing a vertex, $D$ pairs of disjoint edges, uniform
hopping modulus $\lambda$ and arbitrary spin-independent Peierls phases
$A_{ij}=-A_{ji}$, for any number $n\leq N$ of projected spin-$\tfrac12$
fermions (two species), in units of $\lambda^4$,}
\begin{equation}
\mu_4(\Theta)=(2m+4W)\,p_{11}+24D\,p_{22}
+B(N,n,P_{\rm same})\sum_{C}\cos\Theta_C,
\label{eq:mu4flux}
\end{equation}
\begin{equation}
B(N,n,P_{\rm same})=8p_{11}-16p_{22}-12p_{31}
-\frac{8(N-n)(8N-11n-2)}{N(N-1)(N-2)(N-3)}\,P_{\rm same},
\label{eq:Bflux}
\end{equation}
\emph{where the sum runs over every 4-cycle $C$ of the graph (chords
allowed) and $\Theta_C=\sum_{(ij)\in C}A_{ij}$ is its Wilson phase (in
either orientation: only $\cos\Theta_C$ enters); $P_{\rm same}$ is as in
\S\ref{sec:mu4exchange}.}
\medskip

The proof (Appendix~\ref{app:mu4walks}) is the boson count of
Eq.~\eqref{eq:muHCB} with complex amplitudes: every phase-dependent
cross term between two two-hop paths with common endpoints closes exactly
one 4-cycle and carries its Wilson phase (the other cross terms retrace
and carry no phase), and the sector average gives every 4-cycle the same
weight $B$. At $\Theta_C\equiv0$, $\sum_C1=C_4$ and
Eq.~\eqref{eq:mu4flux} is the Proposition of \S\ref{sec:mu4exchange}; at
$n=1$ it is $\Tr A_\Theta^4/N$ with $\Tr A_\Theta^4=2m+4W+8\sum_C\cos\Theta_C$
for the magnetic adjacency matrix $A_\Theta$, the length-4 case of the
closed-walk sum of gain graphs~\cite[Eq.~(4.3)]{CavaleriDonno2022}, which
on the square lattice is the Hofstadter moment $28+8\cos\Phi_\square$ per
site~\cite{Bellissard1997,OuvryWagnerWu2016}.
Three consequences follow. (i) The flux-independent baseline,
$(2m+4W)p_{11}+24Dp_{22}$, is the same for every spin composition: all
spin dependence of $\mu_4(\Theta)$ sits in the coefficient of the Wilson
loops. (ii) On the $L\times L$ torus with $L=3$ or $L\geq5$ every 4-cycle
is a plaquette, so a uniform flux gives
$\sum_C\cos\Theta_C=N\cos\Phi_\square$ and the exchange slope is
$\alpha(N,n)\cos\Phi_\square$ at every $n$; at $n=2$ this is
Eq.~\eqref{eq:alphaflux}. (iii) At $L=4$ the four rows and the four
columns are 4-cycles that wind around the torus, and their Wilson phases
are boundary holonomies, not plaquette fluxes. Wilson phases are gauge
invariant; what the link phases of our numerics fix is the boundary
condition. With those link phases (Landau gauge: a hop along $+\hat x$ at
height $y$ carries
$-y\Phi_\square$, the wrap-around hop along $+\hat y$ in column $x$
carries $4x\Phi_\square$, and every other hop carries none) and with
boundary twists $\theta_x$, $\theta_y$, the row at height $y$ carries
$\theta_x-4y\Phi_\square$ and the column $x$ carries
$\theta_y+4x\Phi_\square$. Without twists, at the quantized fluxes
$\Phi_\square=2\pi p/16$,
$\sum_C\cos\Theta_C=16\cos\Phi_\square+2+4\cos4\Phi_\square+2\cos8\Phi_\square$.
The $\cos4\Phi_\square$ and $\cos8\Phi_\square$ terms and the twist
dependence at $L=4$ are therefore properties of the winding loops'
holonomies, which differ from row to row by the flux $4\Phi_\square$ of
the strip between them and depend on the boundary twists; no
contractible loop produces them. Eq.~\eqref{eq:mu4flux} is checked in exact arithmetic for phases
that are multiples of $\pi/2$, and to $10^{-12}$ for random phases, on
random graphs with triangles and chords in every spin composition; and
against the sparse-trace code of Appendix~\ref{app:numerics} at $L=4$
(every quantized flux, and with twists) and at $L=5$. It is not
machine-checked: the Lean statements of the fourth moment
(Appendix~\ref{app:lean}) are for zero phases.

\paragraph{What this does, and does not, establish.} On $L\times L$
tori with $L=3$ or $L\geq5$, every flux-sensitive mechanism in
$\mu_4(\Phi)$ carries the \emph{same} harmonic, $\cos\Phi_\square$,
the ordinary single-particle Aharonov--Bohm periodicity. At fourth
order there is no $\cos2\Phi_\square$ term (derived in S10, confirmed
numerically for the 3-cycle at $L=4,5$, and implied by
Eq.~\eqref{eq:mu4flux}, in which every 4-cycle enters only through
$\cos\Theta_C$): the bare, projected
fourth moment shows no charge-$2e$ signature. This answers the question
that motivates the section, whether the bare moments carry a
charge-$2e$ signature: fermionic exchange changes the \emph{interference parity} of
a closed walk (constructive vs.\ destructive), not the effective
electromagnetic charge of the object completing it. A clean
consequence is that the two mechanisms trade off maximally at
$\Phi_\square=\pi$: the orbital and 3-cycle contributions
vanish there ($1+\cos\pi=0$) while the same-spin exchange contribution
is maximal ($1-\cos\pi=2$). For $n=2$ the difference
$\mu_4^{(2,0)}(\Phi)-\mu_4^{(1,1)}(\Phi)$ cancels the spin-blind
orbital background \emph{exactly}, isolating pure exchange topology as
a function of flux: a concrete diagnostic of exchange under flux, not
merely a demonstration that $\mu_4(\Phi)\ne\mu_4(0)$.

The doublon operator $d_i=c_{i\downarrow}c_{i\uparrow}$
transforms under a local gauge transformation as $d_i\to
e^{2i\theta_i}d_i$, twice the microscopic phase of a single fermion,
so a genuine charge-$2e$ periodicity, if present at all, must appear in
a \emph{projected}, doublon-sector observable rather than in the bare
$N_d=0$ fourth moment computed above. This question has an exact
answer. Applying the same Peierls
substitution to the microscopic $H_t$ and carrying out the
second-order Schrieffer--Wolff projection onto the doublon (one doubly
occupied site) sector, the counterpart for $N_d=1$ of the projection
onto $N_d=0$ in \S\ref{sec:SW}, every virtual path
contributing to the effective doublon hopping (the two hop orderings
of \S\ref{sec:benchmark}) traverses the \emph{same} bond twice, once
per spin species, so, unlike the ring-exchange channels of
\S\ref{sec:mu4exchange}, no path goes around a loop and there is no
second route with a different phase to interfere against and no possibility of cancellation: the
two-electron process simply accumulates the bond phase twice. In the
two-electron (single-doublon) sector, with energies measured from $U$,
this gives
\begin{equation}
H_d^{(2)}(A) = \varepsilon_d^{(2)}\sum_i n_i^d
+ t_d\sum_{\langle ij\rangle}\big(e^{2iA_{ji}}d_j^\dagger d_i + {\rm h.c.}\big),
\qquad t_d=\frac{2\lambda^2}{U},
\label{eq:doublonSW}
\end{equation}
with $A_{ji}=-A_{ij}$ and the on-site energy shift
$\varepsilon_d^{(2)}=8\lambda^2/U$ (so that $\varepsilon_d=U+\varepsilon_d^{(2)}$)
exactly flux-blind at this order (a backtracking argument, identical to
$\mu_2(\Phi)=\mu_2(0)$), while the inter-site hopping inherits
\emph{twice} the bare link phase, as local gauge covariance requires:
$d_j^\dagger d_i\to e^{2i(\theta_i-\theta_j)}d_j^\dagger d_i$ under a
local gauge transformation must be matched by the transformation of the
coefficient, which depends only on its own bond's phase. The doubled
amplitude $t_de^{2iA_{ji}}$, the absence of longer-range doublon
hopping at this order, and the flux-blind on-site term
($2z\lambda^2/U$ on a site of coordination $z$) are machine-checked on
an arbitrary finite graph for the explicit second-order expression of
the two-electron problem; the Schrieffer--Wolff expansion that leads to
it is not formalized (Appendix~\ref{app:lean}). With several doublons, further
second-order terms appear, which we do not derive: in a sector of
doublons only, with no unpaired electrons and no interaction beyond the
on-site $U$, a nearest-neighbour doublon--doublon interaction. With
unpaired electrons the doublon already moves at first order, by
exchanging places with a singly occupied neighbour, and the second-order
terms include spin exchange and correlated (three-site) hoppings (S11);
nearest-neighbour or longer-range interactions add further terms. Beyond second order the single-doublon Hamiltonian acquires
further flux-dependent terms: a virtual electron can run around a loop
of length $\ell$ at order $\lambda^\ell/U^{\ell-1}$ and contributes that
loop's ordinary charge-$e$ Wilson phase to the doublon's effective
Hamiltonian ($\lambda^3/U^2$ around a
triangle). On the $L\times L$ torus the twist is first felt at length
$L$, when the pair dissociates and its two electrons together wind once
around the torus, which gives the $(2\lambda)^L/U^{L-1}$ deviation from
$\pi$-periodicity found below.
Such terms change the effective Hamiltonian beyond doubled-link hopping;
they do not change the doublon's gauge charge, which is $2e$ at every
order ($d_i\to e^{2i\theta_i}d_i$). This gives two logically distinct, and
equally exact, consequences. Around one elementary plaquette, the
enclosed Wilson phase is doubled,
\begin{equation}
\prod_\square e^{2iA_{ji}} = e^{2i\Phi_\square},
\end{equation}
a \emph{local} statement, the direct analog at doubled phase of the
orbital plaquette-loop mechanism above. Independently, threading a
global boundary twist $\theta$ along $x$, distributed as a phase
$\theta/L$ on every $x$-bond (gauge-equivalent to $\theta$ on one
boundary bond; a bare electron then sees $\cos(k_x+\theta/L)$),
shifts the doublon's own dispersion as
$E_d(K_x,K_y;\theta)=\varepsilon_d+2t_d[\cos(K_x+2\theta/L)+\cos K_y]$,
whose full spectrum is invariant under $\theta\to\theta+\pi$, half the
period of the bare electron's own $2\pi$-periodic twisted-boundary
spectrum,
\begin{equation}
{\rm Spec}\,H_d(\theta+\pi) = {\rm Spec}\,H_d(\theta),
\end{equation}
an exact, genuinely global statement of $h/(2e)$ periodicity for the
effective doublon Hamiltonian, kept
explicitly distinct from the local Wilson-loop result above (derivation
in S11, Sec.~5). The exact two-electron
problem is not $\pi$-periodic as a whole: its $N_d{=}0$ continuum keeps
the bare $2\pi$ period. Its U-band follows the doublon prediction: on
the $6\times6$ torus it is $\pi$-periodic in $\theta$ up to deviations of
about $10^{-4}\lambda$ at $U=20\lambda$ and $3\times10^{-6}\lambda$ at
$U=40\lambda$, falling as $(2\lambda)^L/U^{L-1}$ (breaking the
$\pi$-periodicity requires the pair to dissociate virtually around the
torus), and it matches the doubled-twist dispersion above to
$8\times10^{-3}\lambda$ and $10^{-3}\lambda$, i.e.\ to
$O(\lambda^4/U^3)$. The local, plaquette-flux statement was tested
against the exact flux-threaded two-electron Hubbard problem with no
fitted parameters: on the $6\times6$ torus, over all $36$ quantized
uniform fluxes $\Phi_\square=2\pi p/36$, the doubled-phase prediction for
the U-band outperforms an undoubled control (the same model with phase
$A_{ij}$ instead of $2A_{ij}$) in RMS residual by a factor
of $12.5$ at $U=20\lambda$ and $48$ at $U=40\lambda$. The residual at
one flux is the RMS difference between the $N$ sorted U-band levels of
the exact spectrum and the $N$ sorted eigenvalues of the effective
model; it is averaged over the fluxes. The ratio of the doubled to the
undoubled residual falls steadily with $U$ ($0.28$, $0.080$, $0.021$,
$0.0052$ at $U=10$, $20$, $40$, $80\lambda$), as expected for a
leading-order match.

This single-pair result was anticipated dynamically, if not at the
Hamiltonian level, by two-dimensional
simulations~\cite{SouzaClaro2010}, in which the pair's
Bloch-oscillation frequency is doubled, and by the one-dimensional
treatment of Claro and Robles~\cite{ClaroRobles2022}, which finds the
same doubling semiclassically at leading order in $\lambda/U$,
compatible with charge $2e$. A doubled Peierls phase is what one expects
of a charge-$2e$ pair, and we do not claim it as new. What this paper
adds is the explicit derivation: the same effective charge emerges from
gauge-covariant Schrieffer--Wolff projection of the Peierls-coupled
microscopic Hubbard model on any graph, with its second-order terms
machine-checked for the two-electron problem, giving both the local Wilson-loop doubling and the
global $\pi$-periodicity above as consequences of a single derived
operator, Eq.~\eqref{eq:doublonSW}, tested without fitted parameters
against the exact problem.

The resulting contrast with the flux-sensitive four-hop mechanisms
established earlier in this section is worth stating as one outcome:
the bare projected operator $P_0H_tP_0$'s closed loops carry ordinary
charge-$e$ interference at every order computed here
($\cos\Phi_\square$, with exchange changing interference
\emph{parity} rather than harmonic), while Schrieffer--Wolff-projected
doublon dynamics carries a distinct, phase-preserving charge-$2e$
degree of freedom ($e^{2iA_{ij}}$, $\cos2\Phi_\square$,
$\theta\sim\theta+\pi$). The two are not competing descriptions of the
same physics but successive levels of one hierarchy: exchange-sensitive
interference of charge-$e$ microscopic paths at one order, unitary
propagation of a single charge-$2e$ composite degree of freedom at the
next: not yet the \emph{collective} coherence of many such composites,
which is the open question of \S\ref{sec:discussion}.

\section{Discussion and outlook}
\label{sec:discussion}

\begin{table}[htbp]
\centering
\begin{tabular}{@{}p{0.16\textwidth}p{0.24\textwidth}p{0.22\textwidth}p{0.28\textwidth}@{}}
\toprule
Regime & Geometric object & Spin sensitivity & Gauge response \\
\midrule
{\color{ConnBlue}$\mu_2(\infty)$} & directed edge (single hop) & none & exactly flux-blind, any $\Phi$ \\
{\color{ExchVermillion}$\mu_4(\infty)$} & closed 4-hop loop & $P_{\rm same}$ (exactly affine) & $\cos\Phi_\square$ ($h/e$); exchange changes parity, not charge \\
{\color{DoublonGreen}finite $U$} & virtual doublon excursion ($T_1$, cost $U$) & close to $\propto N_\uparrow N_\downarrow$ (numerical) & not yet examined under flux \\
doublon SW hop & composite two-link hop & one $\uparrow$ and one $\downarrow$ on a site & $e^{2iA_{ij}}$, $\pi$-periodic twist ($h/2e$) \\
\bottomrule
\end{tabular}
\caption{The spectral-moment hierarchy established in this paper: each successive row resolves a more nonlocal structure of the projected configuration graph, and (final row) a distinct gauge-coupled degree of freedom.}
\label{tab:hierarchy}
\end{table}

\begin{figure}[htbp]
\centering
\resizebox{\textwidth}{!}{%
\begin{tikzpicture}[
  site/.style={circle, draw, thick, minimum size=5mm, inner sep=0},
  occ/.style={site, fill=black!75, text=white},
  emp/.style={site, fill=white},
  hop/.style={-{Latex[length=2mm]}, thick},
  vhop/.style={hop, black!45, dashed},
  connhop/.style={hop, ConnBlue},
  exchhop/.style={hop, ExchVermillion},
  doubhop/.style={hop, DoublonGreen},
  vdoubhop/.style={doubhop, dashed, DoublonGreen!55!white},
  lbl/.style={font=\footnotesize},
  panelcap/.style={font=\footnotesize, align=center, text width=3.6cm},
  every node/.style={font=\small}
]

\node[font=\small\bfseries, text=ConnBlue] at (0.5,3.6) {(a) connectivity: $\mu_2$};
\node[emp] (aU) at (0.5,3.0) {};
\node[occ] (a1) at (0.5,2.0) {};
\node[emp] (aD) at (0.5,1.0) {};
\node[emp] (aL) at (-0.5,2.0) {};
\node[emp] (aR) at (1.5,2.0) {};
\node[lbl, left=0.5mm of a1] {$i$};
\node[lbl, below=0.5mm of aD] {$j$};
\draw[connhop] (a1) -- (aU);
\draw[connhop] (a1) -- (aD);
\draw[connhop] (a1) -- (aL);
\draw[connhop] (a1) -- (aR);
\node[panelcap] at (0.5,0.1) {each allowed hop $\to$ one term in $\mu_2$};

\node[font=\small\bfseries, text=ExchVermillion] at (5.6,3.6) {(b) exchange: $\mu_4$};

\node[occ] (i1) at (4.2,3.0) {};
\node[emp] (k1) at (5.2,3.0) {};
\node[emp] (l1) at (4.2,2.0) {};
\node[occ] (j1) at (5.2,2.0) {};
\node[lbl, left=0.5mm of i1] {$i$};
\node[lbl, right=0.5mm of k1] {$k$};
\node[lbl, left=0.5mm of l1] {$l$};
\node[lbl, right=0.5mm of j1] {$j$};
\draw[exchhop] (i1) to[bend left=15] (k1);
\draw[exchhop] (k1) to[bend left=15] (j1);
\draw[exchhop] (j1) to[bend left=15] (l1);
\draw[exchhop] (l1) to[bend left=15] (i1);
\node[panelcap] at (4.7,1.15) {diagonal channel\\($2{+}2$ hops)};

\node[emp] (m2) at (6.6,3.0) {};
\node[emp] (p2) at (7.6,3.0) {};
\node[occ] (i2) at (6.6,2.0) {};
\node[occ] (j2) at (7.6,2.0) {};
\node[lbl, left=0.5mm of m2] {$m$};
\node[lbl, right=0.5mm of p2] {$p$};
\node[lbl, left=0.5mm of i2] {$i$};
\node[lbl, right=0.5mm of j2] {$j$};
\draw[exchhop] (i2) to[bend left=15] (m2);
\draw[exchhop] (m2) to[bend left=15] (p2);
\draw[exchhop] (p2) to[bend left=15] (j2);
\draw[exchhop] (j2) to[bend left=30] (i2);
\node[panelcap] at (7.1,1.15) {edge channel\\($3{+}1$ hops)};

\node[panelcap, text width=7.2cm] at (5.9,0.15) {same spin $\Rightarrow$ both channels return to $|\alpha\rangle$: exchange-sensitive};

\node[font=\small\bfseries, text=DoublonGreen] at (10.9,3.6) {(c) virtual doublon: finite $U$};
\node[lbl] at (10.9,3.15) {$T_1$: cost $U$};
\node[occ] (ci) at (10.3,2.6) {\tiny$\uparrow$};
\node[occ] (cj) at (11.5,2.6) {\tiny$\downarrow$};
\node[lbl, below=0.5mm of ci] {$i$};
\node[lbl, below=0.5mm of cj] {$j$};
\draw[doubhop] (ci) to[bend left=15] (cj);

\node[emp] (ci2) at (10.3,1.5) {};
\node[occ] (cj2) at (11.5,1.5) {\tiny$\uparrow\downarrow$};
\node[lbl, left=0.5mm of ci2] {$i$};
\node[lbl, right=1mm of cj2] {doublon at $j$};
\draw[vdoubhop] (cj2) to[bend left=15] (ci2);
\node[lbl, text=DoublonGreen!55!white] at (10.9,1.0) {$T_1^\dagger$: return$/U$};

\node[panelcap, text width=4.4cm] at (10.9,0.4) {$T_1^\dagger T_1/U$: virtual excursion into the doublon sector};

\end{tikzpicture}%
}
\caption{Microscopic origin of the spectral-moment hierarchy, colored
by the three concepts used consistently throughout this paper's
figures: \textcolor{ConnBlue}{connectivity} ($\mu_2$),
\textcolor{ExchVermillion}{exchange} ($\mu_4$), and
\textcolor{DoublonGreen}{virtual doublon} (finite $U$). (a) $\mu_2$
counts single allowed hops out of an occupied site $i$ to an empty
neighbor $j$, a local, spin-blind connectivity count. (b) $\mu_4$
first becomes exchange-sensitive through two geometrically distinct
four-hop closed paths that return the occupied site set $\{i,j\}$ to
itself: a diagonal ($2{+}2$ hop) channel through shared empty neighbors
$k,l$ (drawn as the closed sequence of both particles' hops, each
particle moving twice), and an edge ($3{+}1$ hop) channel through an adjacent
plaquette's other corners $m,p$. Both return to the identical
occupation-number state $|\alpha\rangle$ (the initial configuration) only when $i,j$ carry the same
spin, making them the exchange-sensitive closed paths of
\S\ref{sec:mu4exchange}. (c) At finite $U$, the virtual hop $T_1$ onto
an opposite-spin occupied site creates a doublon at energy cost $U$;
the return hop $T_1^\dagger/U$ (light green, dashed) generates the
second-order term $V$ of the effective Hamiltonian, which contributes
to $\mu_2$ only at order $U^{-2}$, through $\Var(V)$; with the
third-order term $H^{(3)}$ (a hop within the one-doublon sector between
$T_1$ and $T_1^\dagger$, and the anticommutator $-\tfrac12\{V,T_0\}$),
it fixes the $U^{-2}$ coefficient discussed in \S\ref{sec:SW}.}
\label{fig:conceptual}
\end{figure}
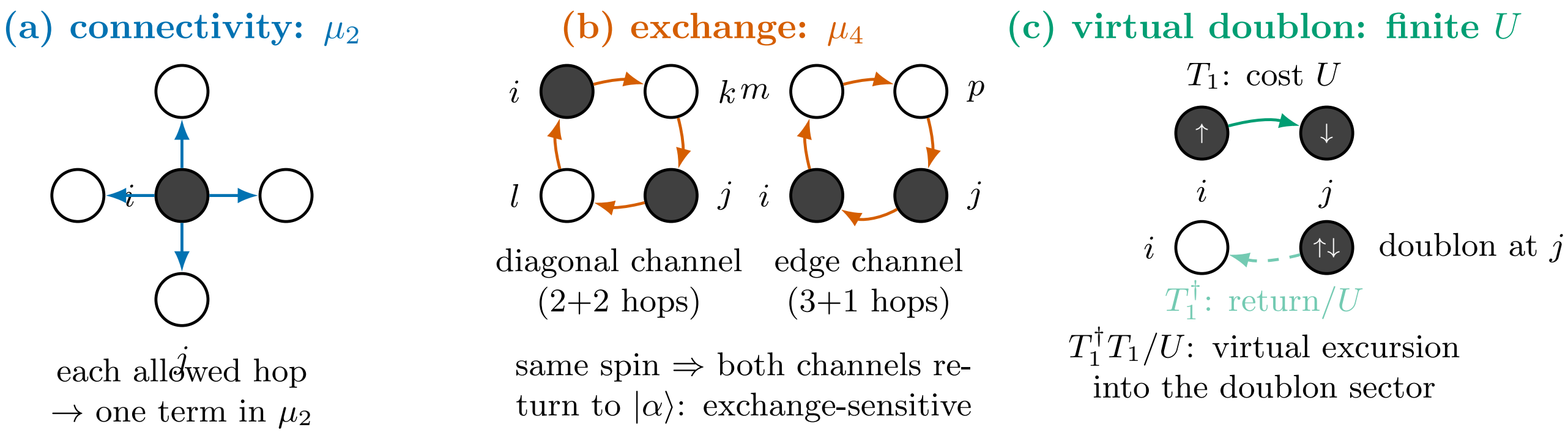

These results form a hierarchy of configuration-space constraints,
spectral signatures, and gauge response
(Table~\ref{tab:hierarchy}). Fig.\,\ref{fig:conceptual} makes the microscopic
origin of that hierarchy explicit, at the level of individual sites and
hops rather than abstract equations: single hops out of one occupied
site fix the spin-blind connectivity moment $\mu_2$; two geometrically
distinct four-hop closed paths (diagonal and edge, of equal weight at
$n=2$) first resolve same-spin exchange in $\mu_4$; and the virtual
doublon-creating hop $T_1$, together with its return $T_1^\dagger/U$
(and, at third order, a hop within the one-doublon sector in between),
generates the leading finite-$U$ correction to $\mu_2$.

\paragraph{Generality and open problems.} Several quantities are
exact traces for which we have no derived closed form: $\beta(L,n)$, the
prefactor of the opposite-spin pair count in the $U^{-2}$ coefficient of
$\mu_2$ (\S\ref{sec:SW}), whose $N_\uparrow N_\downarrow$ law is
established only numerically and only at low filling (exact traces at
$L=4,5,6$ for $n=4$, $L=4,5,6$ for $n=5$ and $L=4$ for $n=6$; for $n=2$,
$c_2$ follows a computed but underived form); and the many-cluster
$O(\lambda^2/U)$ shift of the sector centres (\S\ref{sec:fewbody}). The
fourth moment at $U=\infty$ for uniform hopping, by contrast, is exact at
every $n$ on any finite simple graph: the exchange coefficient
$\alpha(N,n)$ through the number of 4-cycles (Eq.~\eqref{eq:alphaN}), and
the intercept $A(N,n)$ through the numbers of edges, edge pairs and
4-cycles (Eq.~\eqref{eq:Aclosed}), and its dependence on arbitrary
Peierls phases through the Wilson phases of the 4-cycles
(Eq.~\eqref{eq:mu4flux}). Further directions include higher moments ($\mu_6$ and beyond)
of the same projected operators, and extending the cluster-coordinate
rule to non-doublon clusters (e.g.\ a bound tetramer, $C=1$,
$d_{\rm cl}=2$), where the two-doublon binding problem mentioned in
\S\ref{sec:intro} reconnects with this framework.

This program should be pursued in a specific order, because each step
establishes a qualitatively stronger claim than the last, and only the
last begins to justify superconductivity language. (i) \emph{Flux
sensitivity of exchange loops}: established exactly in \S\ref{sec:futurework}, at $h/e$
periodicity. This demonstrates orbital/gauge sensitivity of the
exchange topology already identified in this paper, and nothing more:
a completely normal, non-superconducting system can have flux-sensitive
closed paths. (ii) \emph{A charge-$2e$ signature at the single-pair
level}: established in \S\ref{sec:futurework} for the spectrum (a static
property; transport is not computed here), Eq.~\eqref{eq:doublonSW};
the projected pair sector does carry the doubled Peierls phase,
confirmed against the exact microscopic problem with no fitted
parameters. (iii) \emph{Collective stiffness}: whether any such
doubling \emph{survives, or grows, with particle number and system
size}, rather than remaining the dynamics of one isolated doublon,
e.g.\ via the finite-system charge stiffness
$D_s\sim\partial^2E_0(\theta)/\partial\theta^2|_{\theta=0}$ compared
across sectors with unpaired fermions, one doublon, and multiple
pairs, and whether any collective response is genuinely cooperative
rather than merely additive across independently-mobile doublons.
Only this last step addresses coherence rather than local bound-pair
motion, and remains entirely future work, together with the
doublon--doublon interaction (and, with unpaired electrons, the
first-order doublon--singlon exchange, spin exchange and correlated
hoppings) that a multi-doublon Schrieffer--Wolff projection would need to
derive first,
so that a stiffness calculation tests the repulsive Hubbard model
itself rather than an assumed hard-core-boson model.

The present paper's two spin-resolved results suggest a specific
question to ask of this program. The fourth moment
found here is controlled by the same-spin pair count $P_{\rm same}$
(\S\ref{sec:mu4exchange}), while the leading finite-$U$ correction to
$\mu_2$ is, numerically, close to proportional to the opposite-spin
pair count $N_\uparrow N_\downarrow$ (\S\ref{sec:SW}). At fixed $n$ the
two pair counts are affinely related, so the distinction is not
between two variables but between two mechanisms: same-spin exchange,
exact at $U=\infty$, and opposite-spin virtual doublon formation, a
numerical trend at finite $U$. Under flux, only the first has been
examined so far: it
appears as an interference \emph{parity} of the ordinary $h/e$
harmonic (Table~\ref{tab:hierarchy}). We phrase the resulting question deliberately
short of superconductivity: can virtual opposite-spin pairing and
same-spin exchange topology combine to produce anomalously collective
(non-additive) charge-$2e$ transport in the repulsive Hubbard model? This paper
establishes an explicit ladder toward that question without yet
answering it: repulsive binding into a mobile composite
(\S\ref{sec:benchmark}, \cite{SouzaClaro2010,ClaroRobles2022}), a
microscopic charge-$2e$ gauge coupling and single-pair $h/(2e)$ flux
periodicity at leading order in $1/U$ (\S\ref{sec:futurework}), but not yet many-pair phase
stiffness, off-diagonal long-range order, or condensation. Only these
collective properties, step (iii) above and beyond, would justify
superconductivity language, and they are left, deliberately, entirely
to follow-on work.

\paragraph{Numerical statements.} The following results are computed,
not derived, and are presented as such: the two-electron benchmark
numbers, the U-band collapse and the odd-$L$ bandwidth convergence
(\S\ref{sec:benchmark}); the agreement
of the sector centres with $N_d\varepsilon_d$, the many-cluster
constants, and the $n=3$ resonance inside the doublon--electron
continuum (\S\ref{sec:fewbody}); the sign of $c_2$, its closeness to
$\beta(L,n)N_\uparrow N_\downarrow$ at low filling, and the form of
$c_2(1,1)$ at $L=5$--$8$ (\S\ref{sec:SW}); and, in
\S\ref{sec:futurework}, the $\pi$-periodicity of the exact U-band up to
corrections of order $(2\lambda)^L/U^{L-1}$, its agreement with the
doubled-twist doublon dispersion to $O(\lambda^4/U^3)$, and the RMS
agreement of the effective doublon model with the exact flux-threaded
U-band. Every other result is derived, and every exact closed form is
also checked against exact evaluation; the local values quoted in
Appendix~\ref{app:mu4walks} as cross-checks are exact enumerations at
the sizes stated.

\paragraph{What is not claimed.} We collect here the limits stated
throughout. (i) $\beta(L,n)$ is not derived: the sign of the $U^{-2}$
coefficient and its opposite-spin pair-count law at low filling are
numerical findings (\S\ref{sec:SW}); nor is the many-cluster shift of
the sector centres given in closed form. The fourth moment under
Peierls phases, Eq.~\eqref{eq:mu4flux}, is derived but not
machine-checked. The graph expansion behind the boson
moment, Eq.~\eqref{eq:muHCB}, is not new (\S\ref{sec:mu4exchange}).
(ii) For uniform hopping, the exchange coefficient depends on the graph
only through its number of 4-cycles (Eq.~\eqref{eq:alphaN}, finite
simple graphs, $N\geq4$); with bond-dependent amplitudes (spin-independent,
Hermitian, no on-site terms) $\mu_4$ is still linear in $P_{\rm same}$ (the argument of \S\ref{sec:mu4exchange}
goes through unchanged, S5) but the slope differs. The
square-lattice value $-64/(N-1)$, written $-16z/(N-1)$, is not a law in
the coordination number $z$ (Appendix~\ref{app:mu4walks}), and the
hypercubic form $\alpha_d(N,2)$ holds only on tori whose sides are all
$3$ or $\geq5$.
(iii) The doublon Hamiltonian, Eq.~\eqref{eq:doublonSW}, is derived for
a single doublon; its $\pi$-periodicity under a twist is exact for that
effective model, the exact two-electron U-band follows it only up to
corrections of order $(2\lambda)^L/U^{L-1}$, and the full two-electron
spectrum is not $\pi$-periodic. (iv) Nothing here concerns many-pair
phase stiffness, off-diagonal long-range order, condensation,
collective transport, or superconductivity; the single-pair flux
response is a standard consequence of the effective doublon Hamiltonian.
(v) Neither the no-double-occupancy projection, nor the moment method,
nor the functional form of the doublon Hamiltonian is claimed as new
(\S\ref{sec:intro}, \S\ref{sec:benchmark}). (vi) What is and is not
machine-checked is listed in Appendix~\ref{app:lean}.

\section*{Acknowledgments}
The formal proofs build on Mathlib, and we thank its contributors. This
work received no external funding. Generative AI tools were used in
this work. Claude (Anthropic) was used under the author's direction to
draft and revise text, code and Lean proofs, and as independent
reviewers of the statements, proofs and prose, following the review
procedure documented in the accompanying repository; GPT models
(OpenAI) also served as independent reviewers of the statements. Qwen
models (Alibaba Cloud) were used to test and review the scientific
code, DeepSeek models in writing the numerical solvers, and the
open-source coding agent opencode in the development of the code.
The author checked the results and takes full responsibility for the
content.

\makeatletter\immediate\write\@auxout{\string\citation{apsrev42Control}}\makeatother
\bibliographystyle{apsrev4-2}
\bibliography{references}

\appendix

\section{Exact hypercubic-lattice density of states}
\label{app:dos}

For a hypercubic lattice of dimension $d$, hopping $\lambda$, dispersion
$E(\mathbf{k})=-2\lambda\sum_i\cos k_i$, the density of states has the
exact integral representation~\cite{Economou2006} (S1)
\begin{equation}
\rho_d(E) = \frac{1}{\pi}\int_0^\infty \cos(Et)\,[J_0(2\lambda t)]^d\,dt,
\end{equation}
following from $E$ being a sum of $d$ independent ``arcsine'' random
variables. For $d=1$ the density of states is elementary
(inverse-square-root band edges); for $d=2$ it is a complete elliptic
integral, logarithmically divergent at the band
centre~\cite{Morita1971}; for $d\geq3$ we evaluate the integral
numerically.

\begin{figure}[htbp]
\centering
\includegraphics[width=0.98\textwidth]{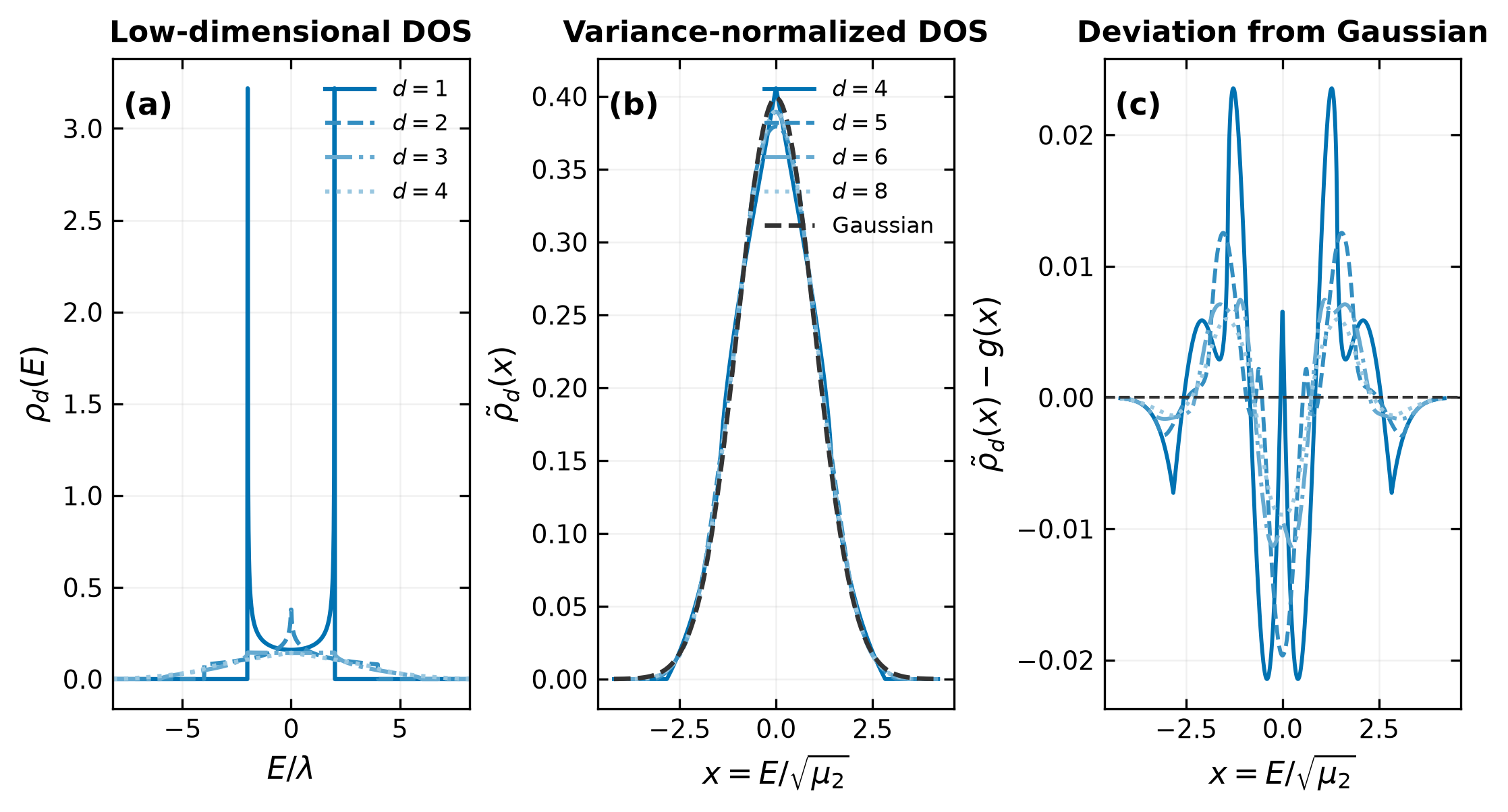}
\caption{Dimensional evolution of the hypercubic tight-binding density
of states. (a) Low-dimensional DOS, showing the progressive smoothing
of van Hove structure. (b) Variance-normalized DOS for $d=4,5,6,8$,
compared with the unit Gaussian approached as $d\to\infty$. (c)
Difference from the Gaussian reference. Although the high-$d$ curves
are nearly indistinguishable on the scale of the DOS itself, their
non-Gaussianity remains exactly resolved by the excess kurtosis
$\gamma_2=-3/(2d)$.}
\label{fig:dosfamily}
\end{figure}

Rather than display $d=1$--$8$ side by side, where the high-$d$ curves
become visually redundant once the van Hove structure has washed out,
Fig.\,\ref{fig:dosfamily} makes the approach to the central-limit
Gaussian quantitative. Panel (a) shows $d=1$--$4$ on their native energy
scale, where the van Hove structure (inverse-square-root divergences at
$d=1$; a logarithmic divergence at $d=2$; square-root cusps, of finite
value but divergent slope, at $d=3$; a smooth maximum at $d=4$) is
visibly disappearing. Panel (b)
variance-normalizes $d=4,5,6,8$, $x=E/\sqrt{\mu_2}$,
$\tilde\rho_d(x)=\sqrt{\mu_2}\,\rho_d(x\sqrt{\mu_2})$, against the unit
Gaussian $g(x)=e^{-x^2/2}/\sqrt{2\pi}$: on this scale the curves are
nearly indistinguishable by eye. Panel (c) plots the residual
$\tilde\rho_d(x)-g(x)$, which remains clearly resolved and shrinks
systematically with $d$.

This is quantified exactly by the excess kurtosis. For isotropic
hopping ($c_i=\lambda$, $i=1,\ldots,d$), $\mu_2=2d\lambda^2$ and
$\mu_4=12d^2\lambda^4-6d\lambda^4$ (from the additive moment formulas
below), so
\begin{equation}
\gamma_2 \equiv \frac{\mu_4}{\mu_2^2}-3 = -\frac{3}{2d}.
\end{equation}
The non-Gaussianity of $\rho_d$ is therefore always present and exactly
known at any $d$, even though it becomes visually imperceptible in the
DOS itself for $d\gtrsim5$: $\gamma_2\to0$ as $1/d$, not as a step
function.

The representation generalizes immediately to anisotropic hopping
amplitudes $c_i$ (needed for mixed doublon/electron clusters, e.g.\
\S\ref{sec:n3test}), since the characteristic function of a sum of
independent cosines factorizes regardless of amplitude:
\begin{equation}
\rho(E) = \frac1\pi\int_0^\infty\cos(Et)\prod_iJ_0(2c_it)\,dt.
\end{equation}
For a sum of independent cosine terms $E=-2\sum_ic_i\cos k_i$, the exact
moments follow without numerical integration: per-term variance
$2c_i^2$ and per-term 4th cumulant $-6c_i^4$, both additive, giving
$\Var=2\sum c_i^2$, $\mu_4=-6\sum c_i^4+3\Var^2$, used for the naive
reference variance of \S\ref{sec:fermistats} and the
excess-kurtosis comparison of \S\ref{sec:fewbody}.

\section{The exact finite-population \texorpdfstring{$\mu_4$}{μ4} formula}
\label{app:mu4fpc}

For $k$ distinct labels (single-particle states, possibly degenerate) sampled without replacement from a population
$\{\varepsilon_1,\ldots,\varepsilon_N\}$ of zero mean
($\sum_i\varepsilon_i=0$, as for any lattice band without on-site
energy), classifying the 4-index
expansion of $S^4$ ($S=$ sum of sampled values) by coincidence pattern
(all equal; 3+1; 2+2; 2+1+1; all distinct) and writing
$p_m=k(k-1)\cdots(k-m+1)/[N(N-1)\cdots(N-m+1)]$ for the probability
that $m$ given distinct labels all belong to the sample gives
\begin{equation}
\mu_4 = P_4(p_1-7p_2+12p_3-6p_4) + 3P_2^2(p_2-2p_3+p_4),
\end{equation}
with the power sums $P_2=\sum\varepsilon_i^2$, $P_4=\sum\varepsilon_i^4$
(not to be confused with the pair counts $P_{\rm same}$, $P_3$ of
\S\ref{sec:mu4exchange}). Validated
against brute-force enumeration on a random centred toy population
(exactly, in rational arithmetic, $k=2,3,4$) and against the real $L=4$
single-particle spectrum (Table \ref{tab:fpcexact}); for a $U{=}0$
spectrum with both spin species, the independent pieces are combined
via the sum-of-independent-variables identity
$\mu_4(X{+}Y)=\mu_4(X)+\mu_4(Y)+6\Var(X)\Var(Y)$. The formula, valid for every
$k\le N$ with $p_m=0$ for $m>k$, the finite-population variance, the
additivity identity, and the six exact values of
Table~\ref{tab:fpcexact} are machine-checked (Appendix~\ref{app:lean}).

\section{The \texorpdfstring{$\mu_4(U\to\infty)$}{μ4(U → ∞)} closed-walk classification}
\label{app:mu4walks}

Four-step closed walks in the $N_d=0$ configuration graph classify by
how many original particles participate:

\begin{description}
\item[Type I --- single particle, self-returning.] Four sub-types:
same-neighbor out-and-back twice; two different neighbors,
sequentially; ``there, one step further, and back''; and a genuine
4-cycle/plaquette loop. Each depends only on joint ``these $m$
additional sites are simultaneously empty, given the source occupied''
probabilities: a symmetric hypergeometric factor in $(n,N)$ only, with
no dependence on how $n$ splits into $(N_\uparrow,N_\downarrow)$, by
the identical argument as $\mu_2$ (\S\ref{sec:mu2theorem}).
\item[Type II(a) --- two different particles, each returning.] Two
particles each make a two-hop excursion and return (the excursions may
share an intermediate site); the walk induces the identity and depends
only on which sites are empty, so it is spin-blind.
\item[Type II(b) --- ring exchange (transposition).] Two occupied sites
$i,j$ on a common 4-cycle admit a 4-hop sequence, through the cycle's
two other (empty) corners, that returns the occupied \emph{site set} to
$\{i,j\}$ but permutes which original particle occupies which site.
\item[Type III --- 3-cycle.] Three occupied sites $i,j,k$ and one
auxiliary empty site $m$ (adjacent to two of the three, forming an
L-shaped or plaquette-corner arrangement) admit the $2{+}1{+}1$-hop
sequence $i\to m$, $j\to i$, $k\to j$, $m\to k$: the particle originally
at $i$ ends at $k$ (via $m$, two hops), the particle at $j$ ends at $i$
(one hop), the particle at $k$ ends at $j$ (one hop): a genuine 3-cycle
$(i\,k\,j)$ using exactly 4 hops. Explicit check: on the square
lattice, taking $i=(0,0)$, $j=(1,0)$, $k=(1,1)$, $m=(0,1)$ realizes
this, since $m$ is a common neighbor of both $i$ and $k$.
\end{description}

\begin{figure}[htbp]
\centering
\input{fig_walktypes.tex}
\caption{The four-hop closed-walk types of $\mu_4(U\to\infty)$, drawn on one plaquette.
The number of routes that reach the same two-hop intermediate decides the response to a
plaquette flux $\Phi_\square$ (\S\ref{sec:futurework}): a single route is flux-blind, while two or
more interfere with relative Wilson phase $\Phi_\square$. The spin-sensitive types (vermillion)
return to the initial configuration only when the particles involved have equal spins.}
\label{fig:walktypes}
\end{figure}
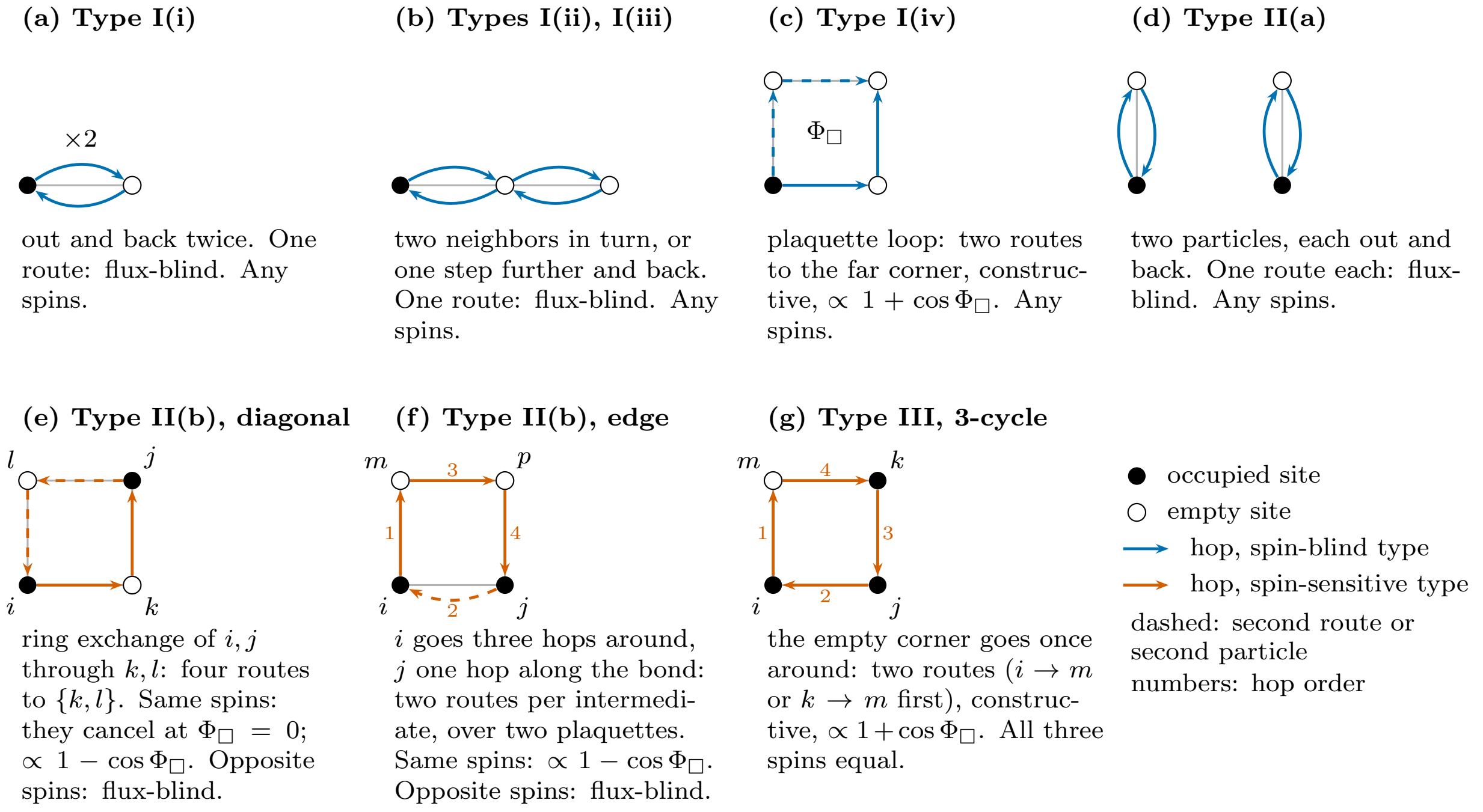

Figure~\ref{fig:walktypes} draws each type, with the interference that sets its response to a
magnetic flux (\S\ref{sec:futurework}).

For Type II(b), the correct statement is in the occupation-number
basis, not in labelled-particle language: writing the initial state as
$|\alpha\rangle=c^\dagger_{i\sigma}c^\dagger_{j\sigma'}|\cdots\rangle$
and the representative diagonal contribution to $H_\infty^4$ as the
normal-ordered hop sequence around the plaquette $i,k,j,l$,
\begin{equation}
\langle\alpha|\,c^\dagger_{j\sigma'}c_{k\sigma'}\;c^\dagger_{i\sigma}c_{l\sigma}\;c^\dagger_{l\sigma'}c_{j\sigma'}\;c^\dagger_{k\sigma}c_{i\sigma}\,|\alpha\rangle,
\end{equation}
the operator string returns every mode's occupation number to its
initial value (hence contributes to the diagonal) \emph{only} when
$\sigma=\sigma'$. For $\sigma\ne\sigma'$ it vanishes identically: its
third hop (the second factor as written),
$c^\dagger_{i\sigma}c_{l\sigma}$, must remove a $\sigma$
electron from $l$, which holds $\sigma'$. Written instead with the spin
labels following the particles ($\sigma'$ from $l$ to $i$, $\sigma$ from
$k$ to $j$), the string returns the occupied sites but leaves $\sigma'$
at $i$ and $\sigma$ at $j$, the opposite of $|\alpha\rangle$, so
$\langle\alpha|\cdots|\alpha\rangle=0$ again, independent of any sign. For $\sigma=\sigma'$ the occupation numbers do return to
$|\alpha\rangle$ and the matrix element is nonzero; its sign and
magnitude are fixed by the fermionic operator ordering above (the
closed-walk amplitude lemma below).

The identical argument applies to the Type III 3-cycle: the final
occupation-number state returns to $|\alpha\rangle$ if and only if the
spin at each site after the cycle matches the spin before it, which
for the cycle $(i\,k\,j)$ requires $\sigma_i=\sigma_j=\sigma_k$ (all
three participating particles the same spin), contributing a term
proportional to $P_3=\binom{N_\uparrow}{3}+\binom{N_\downarrow}{3}$
when summed over eligible triples.

\paragraph{Lemma (amplitude of a closed walk).} \emph{Let $w$ be a site
walk of four hops from an occupied site set $S$ that returns the
occupied set to $S$, inducing the permutation $\pi_w$ of $S$, and let
$\sigma$ be a spin labelling of $S$. Applying the four hops of $w$ to
the basis state $|S,\sigma\rangle$, each as its term
$-\lambda\,c^\dagger_{y\tau}c_{x\tau}$ of $H_\infty$ ($\tau$ the spin of
the moving particle), gives a basis state; it is $|S,\sigma\rangle$
itself if and only if $\sigma$ is constant on the cycles of $\pi_w$,
and then the product of the four matrix elements is
$(-\lambda)^4{\rm sgn}(\pi_w)=\lambda^4{\rm sgn}(\pi_w)$, whatever the
spins of the other particles and whatever ordering of the $2N$ modes
defines the basis.}

\emph{Proof.} Write $|S,\sigma\rangle=\epsilon\,c^\dagger_{s_1\sigma(s_1)}
\cdots c^\dagger_{s_n\sigma(s_n)}|0\rangle$, with $s_1,\ldots,s_n$ the
sites of $S$ in any fixed order and $\epsilon=\pm1$ the sign relating
this product to the basis vector of the chosen mode ordering. Let a hop
$c^\dagger_{y\tau}c_{x\tau}$, with $y$ empty, act on such a product in
which $c^\dagger_{x\tau}$ stands in slot $k$. Moving $c_{x\tau}$ to slot
$k$ gives $(-1)^{k-1}$; there $c_{x\tau}c^\dagger_{x\tau}$ acts as the
identity, since no other factor creates the mode $(x,\tau)$; and moving
$c^\dagger_{y\tau}$ into slot $k$ gives $(-1)^{k-1}$ again. So a hop
replaces $c^\dagger_{x\tau}$ by $c^\dagger_{y\tau}$ in its slot, with no
sign. After the four hops, slot $k$ holds
$c^\dagger_{\pi_w(s_k)\,\sigma(s_k)}$ and the prefactor is
$\epsilon(-\lambda)^4$: site $\pi_w(s_k)$ carries spin $\sigma(s_k)$.
This is $|S,\sigma\rangle$ if and only if
$\sigma(\pi_w(s))=\sigma(s)$ for all $s$, and otherwise an orthogonal
basis state. In the first case, sorting the slots back into the order
$s_1,\ldots,s_n$ permutes $n$ anticommuting factors by $\pi_w$, giving
${\rm sgn}(\pi_w)$, and $\epsilon^2=1$. $\square$

The individual hop signs depend on the mode ordering and on where the
spectators sit; only the product around a closed walk is the
convention-free ${\rm sgn}(\pi_w)$. Counting the labellings that a
given $\pi$ preserves gives the fractions used in
\S\ref{sec:mu4exchange}: for a transposition $(ab)$,
$[\binom{n-2}{N_\uparrow-2}+\binom{n-2}{N_\downarrow-2}]/\binom{n}{N_\uparrow}
=P_{\rm same}/\binom n2$; for a 3-cycle,
$[\binom{n-3}{N_\uparrow-3}+\binom{n-3}{N_\downarrow-3}]/\binom{n}{N_\uparrow}
=P_3/\binom n3$. With bond-dependent amplitudes $t_{xy}$ (the same for
both spins, Hermitian, no on-site terms) the same proof gives the product of the walk's four amplitudes in place of
$\lambda^4$, a factor that depends on the walk but not on the spins, so
Eq.~\eqref{eq:mu4decomp} holds with $I$, $T_2$, $T_3$ replaced by
amplitude-weighted walk sums, and $\mu_4$ stays affine in $P_{\rm same}$
(S5).

\paragraph{The 3-cycle's local path multiplicity, exhaustively.} The
construction above exhibits \emph{one} explicit 4-hop realization of
the 3-cycle; exhaustively enumerating
$(H_\infty^2)_{ab}=\sum_e(H_\infty)_{ae}(H_\infty)_{eb}$ over every
locally allowed intermediate $e$, for $a=\{i,j,k\}$ and
$b=\{i,k,m\}$, shows this is incomplete. Restricted to the four local
sites, only two first hops are ever legal from $a$ ($i\to m$ or
$k\to m$; every other local pair is occupied-occupied), and each
completes to $b$ uniquely ($j\to i$ or $j\to k$ respectively): an
exhaustive, not illustrative, count. Both routes carry the
\emph{same} fermion sign under the sorted-tuple rule (direct
computation), so they interfere constructively:
$(H_\infty^2)_{ab}=-2\lambda^2$, twice the magnitude
($|(H_\infty^2)_{ab}|^2=4\lambda^4$) of a single route. The exact
sparse-trace numbers of this paper sum over every second-quantized path
and so include both routes.

A four-particle cyclic permutation's diagonal spin condition is $\sigma_i=\sigma_j=\sigma_k=\sigma_l$
(all four the same spin), contributing a term $\propto
P_4^{\rm same}=\binom{N_\uparrow}{4}+\binom{N_\downarrow}{4}$, which is
\emph{not} affine in $P_{\rm same}$ (unlike $P_3$). A double
transposition $(ij)(kl)$ has a different and more dangerous condition:
only $\sigma_i=\sigma_j$ \emph{and independently} $\sigma_k=\sigma_l$,
with no requirement relating the two pairs to each other: an
$\uparrow\uparrow\downarrow\downarrow$ configuration survives it as well
as an all-$\uparrow$ one. Its spin trace is therefore a \emph{product}
of two pair-equality indicators, exactly the kind of structure capable
of generating a genuine $P_{\rm same}^2$ term. Both cycle types must
therefore be excluded for exact linearity to hold.

\paragraph{No four-particle permutation is reachable in four hops.} A
permutation with no fixed points among four sites $\{i,j,k,l\}$ (either
cycle type above) requires all four participating
particles to move at least once; with exactly four hops available,
this forces a $1{+}1{+}1{+}1$ distribution, since any particle
receiving zero hops would remain fixed. But a $1{+}1{+}1{+}1$
realization requires each particle's single hop to land directly on
its final site among the four, and every one of those four sites is
occupied at the outset (the four sites are exactly the support of the
permutation). The hop into a site $i$ can proceed only once $i$'s
occupant has left, but that occupant's hop is itself one of the four
hops being scheduled, and by the same argument it must wait for the
occupant of its own target site to leave first, and so on around the
cycle back to $i$: a closed dependency loop with no legal first move. (A double transposition faces
the identical obstruction independently on each of its two 2-cycles.)
Four-particle permutations are therefore impossible within a four-hop
closed walk on any graph without self-loops; only $e$, transpositions,
and 3-cycles are accessible, matching the closed-walk types enumerated
above. The same argument proves more: on any graph, a closed walk of
$h$ hard-core hops moves either no particle or fewer than $h$ of them
(if $h$ particles moved, each would move once, onto a site whose
occupant had to leave first, and the hop times would decrease around
every cycle). This general statement, its $h=4$ case, and the
realization of both a transposition and a 3-cycle in four hops on every
$L\times L$ torus with $L\geq3$ are machine-checked
(Appendix~\ref{app:lean}).

We cross-checked this directly on a fully occupied plaquette at
$n=4$: a diagonal pair $(i,j)$ with both mediators $k,l$ also occupied
($k,l$ are themselves a diagonal pair, with $i,j$ as their mediators).
With all four corners occupied, the diagonal exchanges of this
plaquette are blocked; what remains spin-dependent is the edge channel,
in which a same-spin nearest-neighbor pair of the plaquette exchanges
by three hops around its outer neighboring plaquette and one back, four
walks of sign $-1$. Writing $b_{\rm s}$ for the number of same-spin
bonds of the plaquette, the measured values follow exactly
$(H_\infty^4)_{aa}=208-4b_{\rm s}$: $208$ when $i,j$ and $k,l$ carry opposite spins (no
same-spin bond), $200$ when one of $k,l$ shares the spin of $i,j$ (two),
and $192$ when all four spins are equal (four), identical at $L=5$ and
$L=6$. No four-particle process contributes.

\subsection*{Two geometrically distinct channels, and their exact $n=2$ coefficient}

Evaluating the analogous strings for every distinct pair type on the
square lattice (verified against exact sparse traces at $n=2$, where
Type II(b) is isolated exactly since no other particles can occupy the
intermediate sites) reveals \textbf{two}, not one, contributing
geometries:
\begin{itemize}
\item \textbf{Diagonal pairs} ($i,j$ next-nearest neighbors, sharing
two common neighbors $k,l$ directly, an elementary plaquette's
opposite corners): the cyclic sequence $i\to k\to j\to l\to i$ (either
orientation) contributes.
\item \textbf{Edge pairs} ($i,j$ nearest neighbors): a $3$-hop$+1$-hop
traversal of an adjacent plaquette also contributes, e.g.\ one
particle traverses $i\to m\to p\to j$ (three sides of the plaquette,
$m,p$ its other two corners) while the other simply crosses the shared
edge $j\to i$ once the direct site is vacated. This channel has no
counterpart in a count of opposite plaquette corners alone.
\end{itemize}
At $n=2$, where the two occupied sites are the entire system and the
required intermediate sites are guaranteed empty, exact evaluation
gives \textbf{identical magnitude for both channels}: $-8\lambda^4$ per
eligible pair, verified at $L=5,6,7$ (at $L=3$ as well; $L=4$ differs,
see below). Each channel has $2N$ eligible pairs, which gives
$\alpha(N,2)=(-8)\cdot2\cdot2N/\binom N2=-64/(N-1)$
(\S\ref{sec:mu4exchange}), matching the exact evaluation at every
$L\neq4$ tested.

\paragraph{The sign mechanism, derived completely.} Writing
$\mu_4=\sum_{a,b}(H_\infty^2)_{ab}^2/D_0$, the square-lattice $-8\lambda^4$
per-pair magnitude
common to both channels follows from a direct, complete fermion-sign
count of every second-quantized hopping path connecting the reference
state $a$ to each relevant two-hop intermediate $b$, not merely a
qualitative interference argument.

\emph{Diagonal channel.} Take a same-spin pair $i,j$ with common
neighbors $k,l$. The single relevant intermediate is the ``fully
swapped'' configuration $c=\{k,l\}$, reached in exactly two hops by
four distinct paths: two choices of which particle moves to which
mediator ($i\to k,j\to l$ or $i\to l,j\to k$), each with two
hop-orderings. Direct evaluation of the fermion sign (using the
sorted-tuple rule of App.~\ref{app:numerics}) for a concrete instance ($i,j$ diagonal,
$k,l$ the shared neighbors) gives the four path signs $-1,+1,+1,-1$,
which \emph{sum to zero}: $(H_\infty^2)_{a,c}=0$ exactly for same spin,
an exact destructive cancellation between the two particle-assignment
choices (which differ by an odd permutation of indistinguishable
fermions). For an opposite-spin pair, the two assignments produce two
distinguishable target states ($\uparrow$ at $k,\downarrow$ at $l$,
or vice versa: genuinely different basis vectors, since the spin
label now marks which particle sits where); each is reached by the two
hop-orderings only, which share a sign and add constructively, giving
$(H_\infty^2)=2\lambda^2$ for each of the two targets. Summing
$(H_\infty^2)^2$ over both: same spin contributes $0$, opposite spin
contributes $2\times(2\lambda^2)^2=8\lambda^4$. The difference,
$0-8\lambda^4=-8\lambda^4=-2z\lambda^4$ at $z=4$, is the diagonal
channel's coefficient, obtained with no fitted or imported number. In
the signed closed-walk bookkeeping of Eq.~\eqref{eq:mu4decomp} the same
difference appears instead as transposition walks of sign $-1$ that
return to the reference state only for the same-spin pair: the two
bookkeepings assign the $-8\lambda^4$ to different terms but agree on
the same-spin minus opposite-spin difference.

\emph{Edge channel.} Take an adjacent pair $i,j$ and the two other
corners $m,p$ of \emph{one} plaquette containing edge $i$-$j$. Two
two-hop intermediates are relevant, $c_1=\{p,j\}$ and $c_2=\{m,i\}$
(the two ways of interleaving the $3$-hop$+1$-hop process), each
reached by exactly two paths: a genuine two-particle path (both
original particles move once) and a single-particle path (one particle
hops twice through the plaquette). For same-spin, direct sign
evaluation gives, for both $c_1$ and $c_2$, one path of each sign
($-1,+1$), so both cancel exactly: $(H_\infty^2)_{a,c_1}=(H_\infty^2)_{a,c_2}=0$.
For opposite spin, the two paths to each target are no longer the
\emph{same} target: they resolve into two distinguishable, singly-reached
configurations per intermediate ($\uparrow$/$\downarrow$ interchanged),
each with $(H_\infty^2)=\lambda^2$ (a single contributing path, sign
$+1$), contributing $\lambda^4$ each: four distinguishable targets
(two from $c_1$, two from $c_2$) give $4\lambda^4$ from this one
plaquette. Because an edge belongs to \emph{two} plaquettes on the
square lattice, the mirror plaquette (corners $m',p'$ on the other side
of edge $i$-$j$) contributes an independent, identical
$4\lambda^4$, same-spin again canceling to zero by the identical
sign argument. Summing both plaquettes: same spin $0$, opposite spin
$8\lambda^4$, difference $-8\lambda^4=-2z\lambda^4$ at $z=4$,
identical to the diagonal channel, but reaching the same total through
a structurally different route (one plaquette with two constructively-interfering
paths per target, versus two plaquettes each contributing single-path
targets).

Both channels' $-8\lambda^4$ coefficients are therefore derived by a
complete accounting of fermion signs on every contributing path, and
confirmed by exact evaluation.

\paragraph{Does the square-lattice value generalize?} Writing the
square-lattice result as $\alpha(N,2)=-16z/(N-1)$ (that is, $-64/(N-1)$
with $z=4$ restored) packages the per-pair magnitude and the pair count
together, and unlike the second-moment theorem this packaging is not
graph-general: ``$16z$'' is a square-lattice mnemonic for one specific
number, not a law in $z$. Repeating the exact sparse-trace
measurement on a $3$-dimensional simple-cubic lattice with periodic
boundaries ($z=6$) gives $\alpha(27,2)=-7.385=-96/13$, \emph{twice} the
substitution $-16z/(N-1)=-48/13\approx-3.692$. This factor of two is
\emph{not} accounted for by the edge channel alone: direct enumeration
shows \emph{both} channels independently double in three dimensions,
by two unrelated geometric mechanisms that happen to share the same
$d-1$ scaling. The edge channel doubles because each edge of the
simple-cubic lattice belongs to $4$ elementary plaquettes rather than
$2$ (one pair for each of the two directions transverse to the edge).
The diagonal channel doubles \emph{independently}, for a different
reason: a site has $2d(d-1)=12$ face-diagonal (next-nearest,
two-common-neighbor) neighbors in three dimensions, not the $z=6$ a
naive substitution would suggest: the square lattice's diagonal-type
neighbor count happening to equal its coordination number $z$ is a
$d=2$ coincidence, not a general fact. Each channel supplies an equal half of the doubled
total ($-1296\lambda^4$ of $-2592\lambda^4$ at $N=27$). Both mechanisms
are 4-cycle counts, and the general statement is Eq.~\eqref{eq:alphaN}
at $n=2$, $\alpha(N,2)=-64C_4/[N(N-1)]$: the exchange coefficient counts
4-cycles, not neighbors (S6). On the $d$-dimensional hypercubic torus with
every side $3$ or $\geq5$, the 4-cycles are the plaquettes,
$C_4=Nd(d-1)/2$, so
\begin{equation}
\alpha_d(N,2) = -\frac{32\,d(d-1)}{N-1}=-\frac{16(d-1)z}{N-1} \qquad (z=2d),
\end{equation}
reproducing $\alpha(27,2)=-96/13$ at $d=3$ (and $-48/31$ on the
$5\times5\times5$ lattice). Each side of length $4$ adds axis rings of
length four as further 4-cycles; with $m$ such sides,
$\alpha(N,2)=-16[2d(d-1)+m]/(N-1)$ ($4\times4\times4$: $-80/21$;
$3\times4$: $-80/11$; $4\times5$: $-80/19$). A lattice with the same $d$
and $z$ but a different 4-cycle structure behaves differently: the
kagome lattice, with $d=2$ and $z=4$ like the square lattice, has no
4-cycles and $\alpha(N,2)=0$ exactly.

\paragraph{The $L=4$ torus, exactly.} At $L=4$, $\alpha(16,2)=-32/5$
rather than the $-64/15$ of larger tori. In terms of
Eq.~\eqref{eq:alphaN}, the eight axis rings of length four are 4-cycles
in addition to the sixteen plaquettes, $C_4=24$, and
$-64\cdot24/(16\cdot15)=-32/5$. Pair by pair, this decomposes \emph{exactly} into
two independent, equal contributions of $-16/15$ each: (i) 16 extra
diagonal-type pairs beyond the generic 32, since sites displaced by
$(2,0)$ or $(0,2)$ acquire a spurious second common neighbor via
periodic wraparound on this specific $4\times4$ torus (generically they
share only one, insufficient for the ring-exchange mechanism); (ii) the
edge channel's per-pair magnitude is itself enhanced from $-8$ to $-12$
by an additional wraparound-enabled route unavailable at larger $L$.
Both are confirmed by direct enumeration; together,
$-64/15-16/15-16/15=-96/15=-32/5$ exactly.

\paragraph{Extension to $n>2$.} Independent pairwise blocking
(Eq.~\eqref{eq:alphaapprox} in the main text) is exactly the
transposition part of $\alpha(N,n)$; on its own it misses the slope by
$1.8\%$ ($n=3$) and $3.9\%$ ($n=4$) at $L=5$. Classifying closed walks
by hop partition ($4$, $3{+}1$, $2{+}2$, $2{+}1{+}1$, $1{+}1{+}1{+}1$)
places the four-hop 3-cycles (three particles on a 4-cycle, the empty
corner circulating once) in the $2{+}1{+}1$ class, which is where the
missing weight lies. The local values show it directly ($n=3$): a third
particle sitting on a same-spin pair's mediator site gives a clean,
spin-dependent value of $(H_\infty^4)_{aa}$ ($180$ for a same-spin and
$186$ for an opposite-spin third particle on a mediator of a diagonal
pair; $180$ and $182$ on a plaquette corner next to an edge pair, in
units of $\lambda^4$), distinct from either the blocked or fully-open
case, and \emph{$N$-independent} (identical at $L=5$ and $L=6$). Summed
over configurations with the spin-averaging weights of
Eq.~\eqref{eq:mu4decomp}, the 3-cycles give the second term of
Eq.~\eqref{eq:alphaN}, which completes $\alpha(N,n)$.

\paragraph{The hard-core-boson moment, Eq.~\eqref{eq:muHCB}.} Let $H$ be
the adjacency matrix of the $n$-token graph: its vertices are the
$n$-subsets $S$ of the sites, and $H_{S'S}=1$ when $S'=S\setminus\{u\}\cup\{w\}$
for an edge $uw$ with $u\in S$, $w\notin S$. $H$ is real symmetric, so
$(H^4)_{SS}=\sum_{S''}[(H^2)_{SS''}]^2$, and a two-hop path changes at
most two occupied sites. We sum over the three cases, averaging over
uniformly random $n$-subsets; $p_{ab}$ is the probability of a given
pattern of $a$ occupied and $b$ empty sites, and
$\binom Nn p_{11}=\binom{N-2}{n-1}$, $\binom Nn p_{22}=\binom{N-4}{n-2}$.
\begin{enumerate}
\item $S''=S$. Here $(H^2)_{SS}$ is the number of edges with exactly one
end in $S$ (the cut of $S$). In the square of this count, an edge paired
with itself ($m$ terms) contributes $2p_{11}$. An ordered pair of
distinct edges sharing a vertex ($2W$ pairs) contributes
$p_{12}+p_{21}=p_{11}$, because $(N-n-1)+(n-1)=N-2$. An ordered pair of
disjoint edges ($2D$ pairs) contributes $4p_{22}$, one term for each
choice of the occupied end of each edge. Total:
$(2m+2W)\,p_{11}+8D\,p_{22}$.
\item $S''=S\setminus\{u\}\cup\{w\}$ with $u\in S$, $w\notin S$. Each
two-hop path passes through a common neighbour $v$ of $u$ and $w$. If $v$
is empty, the particle hops $u\to v\to w$. If $v$ is occupied, the
particle at $v$ hops to $w$ and the particle at $u$ then hops to $v$. In
either case there is exactly one path per common neighbour, so
$(H^2)_{SS''}=(A_G^2)_{uw}$, whatever the occupation of $v$. With
$k=(A_G^2)_{uw}$ and $k^2=k+2\binom k2$: summed over ordered pairs
$u\neq w$, $\sum k=\sum_v d_v(d_v-1)=2W$, and $\sum\binom k2=4C_4$,
which counts the 4-cycles that have $u$ and $w$ as opposite corners.
Total: $(2W+8C_4)\,p_{11}$.
\item $S''=S\setminus\{u,v\}\cup\{w,x\}$. Each of the two hops moves a
particle directly from $\{u,v\}$ to $\{w,x\}$, in either order, so
$(H^2)_{SS''}=2(A_{uw}A_{vx}+A_{ux}A_{vw})$. With $a,b\in\{0,1\}$ the
two products, $(a+b)^2=a+b+2ab$. Summed over unordered pairs $\{u,v\}$
(occupied) and $\{w,x\}$ (empty), $\sum(a+b)=4D$: each matching is an
unordered pair of disjoint edges with a chosen occupied end on each.
Also $\sum 2ab=4C_4$: both matchings exist exactly when $u,w,v,x$ form a
4-cycle with $\{u,v\}$ as one diagonal, and each 4-cycle offers two.
Total: $4(4D+4C_4)\,p_{22}=16(D+C_4)\,p_{22}$.
\end{enumerate}
The sum is Eq.~\eqref{eq:muHCB}. At $n=1$ it reduces to the classical
count $\Tr A_G^4=2m+4W+8C_4$ of closed walks of length four (see,
e.g., Ref.~\cite{CvetkovicRowlinsonSimic2009}), divided by $N$. For $N<4$ the terms in $D$
and $C_4$ vanish (and $p_{22}$, a $0/0$ there, is set to $0$).
Multiplied by $\binom Nn$ and summed with a fugacity $\zeta^n$, the
result reads
$\Tr(\zeta^{\hat n}H^4)=(2m+4W+8C_4)\,\zeta(1+\zeta)^{N-2}+(24D+16C_4)\,\zeta^2(1+\zeta)^{N-4}$.
At $\zeta=1$ this is the unrestricted trace of the high-temperature
expansion; its coefficient of $\zeta^n$ is $\binom Nn$ times the
canonical moment.

\paragraph{The fourth moment under Peierls phases, Eq.~\eqref{eq:mu4flux}.}
Now let $H$ be $H_\infty$ with the amplitude $-e^{iA_{yx}}$ for a hop
$x\to y$ ($\lambda=1$, $A_{xy}=-A_{yx}$). It is Hermitian, so
$\Tr H^4=\sum_{c,c''}|(H^2)_{c''c}|^2$ over the basis configurations
$c,c''$ (occupied sites with their spins), and
$(H^2)_{c''c}=\sum_\pi a_\pi$ over the two-hop paths $\pi$ from $c$ to
$c''$. For two such paths, $a_\pi\overline{a_{\pi'}}$ is the product of
the four matrix elements along the closed walk $\pi$ followed by $\pi'$
reversed; the phases are scalar factors, so by the Lemma above it equals
${\rm sgn}(\pi_w)\,e^{i\Theta_w}$, with $\pi_w$ the permutation the walk
induces and $\Theta_w$ the sum of the link phases along it. Hence
$|(H^2)_{c''c}|^2=\sum_\pi1+\sum_{\pi\neq\pi'}{\rm sgn}(\pi_w)\cos\Theta_w$,
and only the second sum can depend on the phases. Classify by the change
of the occupied set, as for bosons.
\begin{enumerate}
\item $c''=c$. Every path hops out and back, $\Theta_w=0$: no phase
dependence.
\item One site replaced, $u\to w$. A path through a common neighbour $v$
of $u$ and $w$ reaches the configuration $c''$ in which the particle of
$u$ sits on $w$ if $v$ is empty (that particle hops $u\to v\to w$), or
if $v$ holds a particle of the same spin (it hops $v\to w$, then the
particle of $u$ hops to $v$). If $v$ holds the opposite spin, the path
ends in the configuration with the spins of $v$ and $w$ exchanged, which
no other path reaches. Both routes through $v$ carry the same link phase,
$A_{vu}+A_{wv}$ (the hops $u\to v\to w$, or $v\to w$ then $u\to v$), so
two paths through distinct $v,v'$ close the 4-cycle $(u,v,w,v')$ with
$\Theta_w=\pm\Theta_C$ whether or not $v$, $v'$ are occupied, and
${\rm sgn}(\pi_w)=+1$ if $v$ and $v'$ are both empty (the identity),
$-1$ if exactly one is occupied (a transposition), $+1$ if both are (a
3-cycle).
\item Two sites replaced, $\{u,v\}\to\{w,x\}$. The paths are the two
matchings $(u\to w,v\to x)$ and $(u\to x,v\to w)$, each in two orders;
the two orders of one matching close a walk with $\Theta_w=0$ and the
identity permutation. The two matchings reach the same configuration
only if the particles of $u$ and $v$ have the same spin, and then close
the 4-cycle $(u,w,v,x)$ with a transposition.
\end{enumerate}
Every phase-dependent term thus belongs to exactly one 4-cycle $C$ and
is $\pm2\cos\Theta_C$ per unordered pair of paths. Averaging over the
sector, let $s_2=P_{\rm same}/\binom n2$ and $s_3=P_3/\binom n3$ be the
probabilities that two or three given occupied sites carry the same spin.
Case~2 counts, for each of the four choices of an occupied corner $u$
and the opposite empty corner $w$ of $C$, the weight
$2[p_{13}-2s_2p_{22}+s_3p_{31}]$ (the other two corners empty; one of
them occupied with the spin of $u$; both). Case~3 counts, for each of
the two choices of the occupied diagonal, $-8s_2p_{22}$: the two
matchings, each counted twice, give $|2a_1+2a_2|^2$ with the cross term
$8\,{\rm Re}(a_1\overline{a_2})=-8\cos\Theta_C$. Every 4-cycle therefore
contributes
\begin{equation}
\bigl(8p_{13}-32s_2p_{22}+8s_3p_{31}\bigr)\cos\Theta_C,
\end{equation}
the same coefficient for every $C$, because the sector average weights
every set of four sites alike; $s_2p_{22}$ and $s_3p_{31}$ are finite
for every $n$, being $2P_{\rm same}(N-n)(N-n-1)/(N)_4$ and
$6P_3(N-n)/(N)_4$. With $p_{13}=p_{11}-2p_{22}-p_{31}$ (partitioning
$p_{11}$ by the occupation of the other two corners) and, for two spin
species, $P_3=\binom n3-\tfrac12(n-2)[\binom n2-P_{\rm same}]$
(\S\ref{sec:mu4exchange}), the coefficient is $B$ of
Eq.~\eqref{eq:Bflux}. The remaining terms do not depend on the phases,
so they equal their value at $\Theta_C\equiv0$, where the Proposition of
\S\ref{sec:mu4exchange} gives $A+\alpha P_{\rm same}-BC_4
=(2m+4W)p_{11}+24Dp_{22}$. This is Eq.~\eqref{eq:mu4flux}. $\square$

\section{Schrieffer--Wolff operator construction}
\label{app:SW}

Write $H=H_t+U\hat N_d$. From the $N_d=0$ sector $P_0$, define $T_0=P_0H_tP_0$
and $T_1=Q_1H_tP_0$, where $Q_1$ projects onto the exactly-one-doublon
sector. $T_1$ is built by, for each $P_0$ basis state $(u,d)$ (up- and
down-tuples of occupied sites), enumerating hops of a spin-$\sigma$
particle at site $s$ onto a neighbor $nb$ already occupied by the
opposite spin: the resulting state has $s$ vacated and $nb$
doubly-occupied, landing in the $Q_1$ basis. Because up and down
sectors are combined as a plain tensor product with no cross
Jordan--Wigner string (the standard Hubbard-model convention), the
fermion sign of such a hop is exactly the same sorted-tuple sign rule
used for $T_0$ (Appendix \ref{app:numerics}), applied to the hopping
species' own tuple, treating the other species as a spectator.
$T_0^{Q}=Q_1H_tQ_1$ is built the same way from each $Q_1$ basis state,
keeping the hops whose result again has exactly one doubly occupied
site (a singly occupied site's electron moving into an empty site, or a
doublon electron hopping onto an opposite-spin singly occupied site,
which moves the doublon).

The Hermitian Schrieffer--Wolff expansion~\cite{ChaoSpalekOles1977,MacDonaldGirvinYoshioka1988}
with every virtual energy denominator equal to $(0-U)^{-1}=-1/U$ (through
third order $Q_1$ is the only virtual sector, since $P_0H_tQ_2=0$ and a
path $P_0\to Q_1\to Q_2\to P_0$ needs two hops back) gives
Eq.~\eqref{eq:Heff}, with $V\equiv T_1^\dagger T_1$ and
$H^{(3)}=T_1^\dagger T_0^{Q}T_1-\tfrac12\{V,T_0\}$. Expanding $\langle(T_0+X)^2\rangle-\langle T_0+X\rangle^2=\Var(T_0)+2\langle T_0,X\rangle_{\rm HS}+\Var(X)$
(using $\langle T_0\rangle=0$) for $X=-V/U+H^{(3)}/U^2$ gives
Eq.~\eqref{eq:mu2cU}, with
$\Tr(T_0H^{(3)})=\Tr(T_0T_1^\dagger T_0^{Q}T_1)-\Tr(T_0^2V)$; all traces
are evaluated as sparse matrix products, in integers at $\lambda=1$.

\paragraph{Connection to the standard $t$-$J$/three-site expansion.}
$V=T_1^\dagger T_1$ is exactly the standard second-order strong-coupling
operator familiar from the canonical Hubbard-to-$t$-$J$
transformation~\cite{ChaoSpalekOles1977,MacDonaldGirvinYoshioka1988}: up
to the overall $1/U$ and normalization, $-V$ decomposes into (i) the
Heisenberg superexchange term $J\sum_{\langle ij\rangle}(\mathbf
S_i\cdot\mathbf S_j-\tfrac14n_in_j)$ with $J=4\lambda^2/U$, itself
splitting into a \emph{diagonal} $S_i^zS_j^z-\tfrac14n_in_j$ piece and
an off-diagonal spin-flip piece that swaps spin labels between two
already-occupied sites without changing which \emph{sites} are
occupied; and (ii) the correlated three-site hopping term, which moves
one particle between two sites both adjacent to a third (mediating,
necessarily doubly-visited) site, either keeping its spin or, in the
spin-flip form, exchanging spin labels with the electron on the
mediating site: in both forms a second-neighbor-type move on the
site-occupation graph. Both pieces are standard ingredients of
the $t/U$ expansion; what is specific to the present calculation is
what they imply for $\langle T_0,V\rangle_{\rm HS}$.

\paragraph{Locality argument for the vanishing cross term.} $T_0$'s
edges are single nearest-neighbor hops, changing the site-occupation
pattern at exactly one site pair. The diagonal spin-exchange piece
above returns to the \emph{same} configuration, contributing nothing
when multiplied by the purely off-diagonal $T_0$ and traced; the
spin-flip piece changes no \emph{site} occupation pattern at all (an
edge type disjoint from $T_0$'s, since $T_0$ always changes the site
pattern); and the three-site hop is a second-neighbor-type move, again
a different edge type from $T_0$'s nearest-neighbor hops on a
large-enough torus. All three pieces therefore contribute zero to
$\langle T_0,V\rangle_{\rm HS}$ unless two sites with a common
mediating neighbor are themselves adjacent, that is, unless the graph
has a triangle. The square-lattice torus has none for $L\geq4$; at
$L=3$ periodic wraparound creates triangles, the origin of the nonzero
cross term reported in \S\ref{sec:SW}. The machine-checked statement
(Appendix~\ref{app:lean}) is the vanishing of the trace on every
triangle-free graph; this decomposition is not used there.

\section{Numerical methods and validation}
\label{app:numerics}

\paragraph{Fermion sign rule (S8).} A $k$-fermion configuration is
represented as a sorted tuple of occupied sites; moving the fermion at
its rank-$m$ position (0-indexed) to a new site $a$ picks up sign
$(-1)^m\times(-1)^p$, where $p$ counts how many of the remaining
occupied sites are less than $a$: the standard sorted-Slater-determinant
sign rule for any $k$. Verified Hermitian for $k=2$ (direct
anticommutation derivation); cross-validated against an independent
Jordan--Wigner bit-string implementation on 4000 random moves, for
$k=1$--$8$ on 12 sites (zero mismatches); and, for a single species
with $k=4$ at $L=4$, the moments $\mu_2,\mu_4$ of the operator built
with this rule agree exactly, in rational arithmetic, with the
free-fermion values obtained by choosing 4 distinct single-particle
energies and summing. In the Lean
formalization (Appendix~\ref{app:lean}) the rule is derived from
creation and annihilation operators satisfying the canonical
anticommutation relations, for an arbitrary ordering of the orbitals.

\paragraph{Sparse-trace evaluation.} All $U=\infty$ and Schrieffer--Wolff
results (\S\ref{sec:Uinfinity}--\ref{sec:SW}) are computed via exact
sparse matrix products, never diagonalization: $\Tr(H^2)$ from the
diagonal of $H@H$, and $\Tr(H^4)=\sum_{ab}|(H^2)_{ab}|^2$ for Hermitian
$H$. At zero flux the traces are integers at $\lambda=1$, so the
moments are exact rationals; under flux they are evaluated in floating
point, to about $10^{-12}$. The method is tractable for every sector
reported here (up to $D_0\approx4\times10^6$).

\begin{figure}[htbp]
\centering
\includegraphics[width=0.98\textwidth]{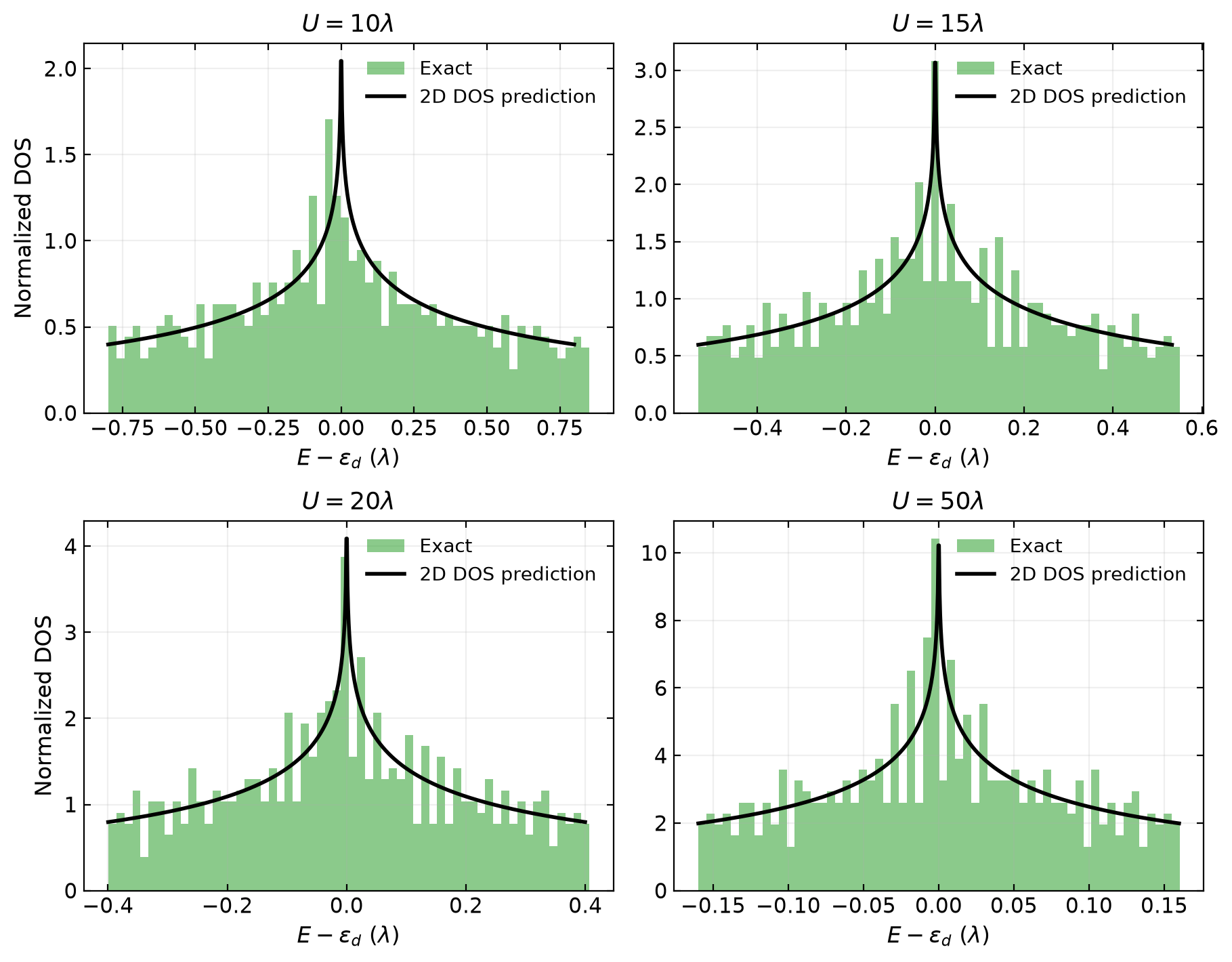}
\caption{Two-particle benchmark: the exact two-electron U-band, computed
by a direct relative-coordinate K-grid scan ($L_{\rm rel}=21$,
$48\times48$ K-points), against the parameter-free second-order
prediction $\rho_{2D}(E-\varepsilon_d;t_d)$ of Eq.~(\ref{eq:doublondisp}), at
$U=10,15,20,50\lambda$: moderate coupling through the asymptotic
doublon regime, no fitted parameters. This is the same benchmark
underlying the collapse of Fig.~\ref{fig:ubandcollapse}, shown here at
the level of the individual $U$ values rather than the rescaled
universal curve.}
\label{fig:kgridUband}
\end{figure}

\paragraph{Relative-coordinate scan (Figs.~\ref{fig:ubandcollapse} and
\ref{fig:kgridUband}).} At fixed pair momentum $K$, the two-electron
problem reduces exactly to a single particle on the lattice of relative
coordinates, with $K$-dependent hopping amplitudes along the two axes
and an on-site defect $U$ at the origin (the Koster--Slater problem of
\S\ref{sec:benchmark}). We take a relative lattice of $L_{\rm rel}=21$
sites per side and scan $K$ over a $48\times48$ grid, which acts as a
fine set of twisted boundary conditions for the centre-of-mass motion;
at each $K$ the U-band level is the eigenvalue split off above the
continuum by the repulsive defect. Each relative-coordinate problem has
only $L_{\rm rel}^2$ states, so the scan is far cheaper than diagonalizing
the full two-electron Hamiltonian at many twists.

\begin{figure}[htbp]
\centering
\includegraphics[width=0.55\textwidth]{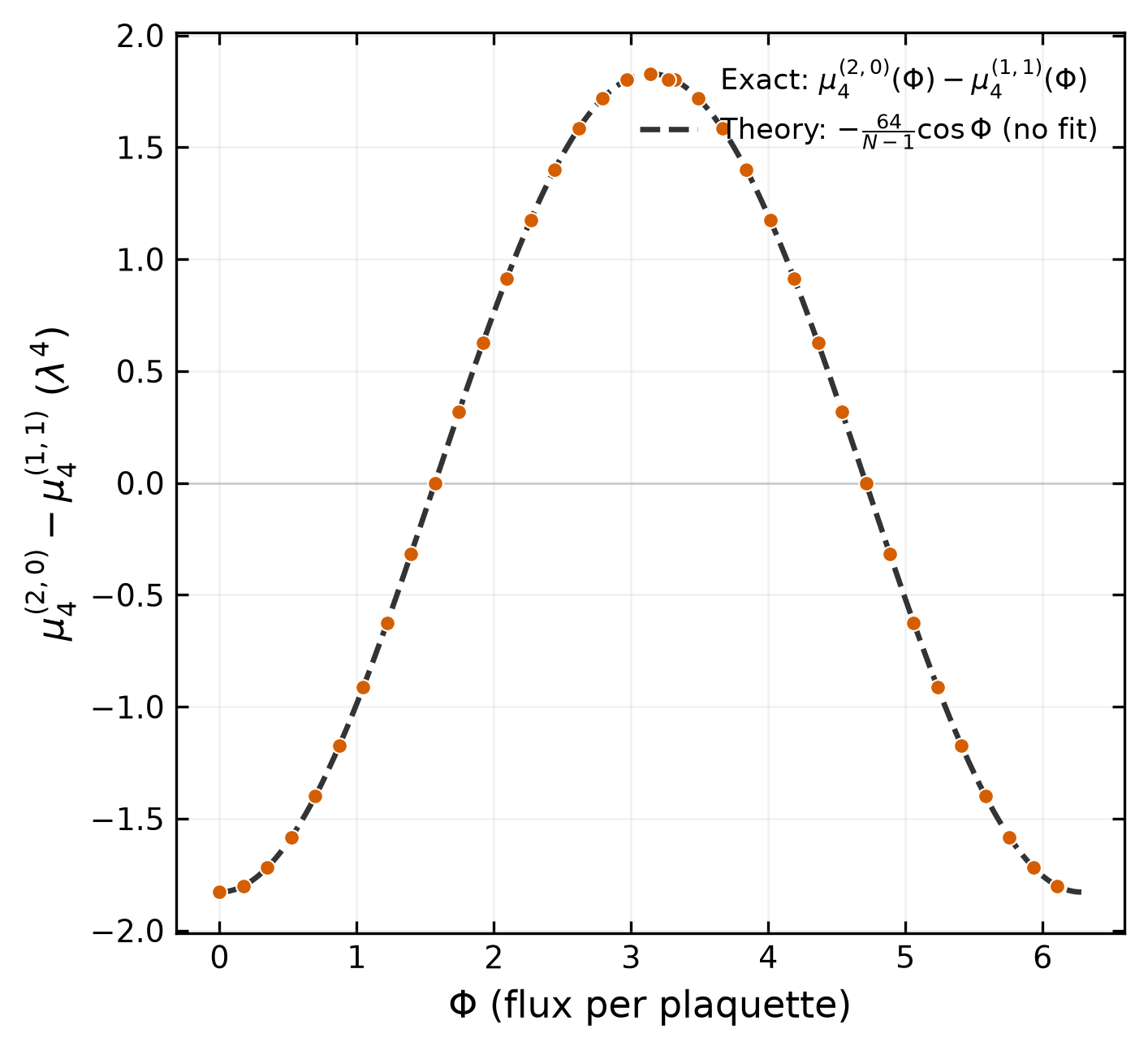}
\caption{Exchange contribution to the fourth moment: the same-spin-minus-opposite-spin
difference $\mu_4^{(2,0)}(\Phi)-\mu_4^{(1,1)}(\Phi)$ against the
closed-form prediction $-64/(N-1)\cos\Phi_\square$ (\S\ref{sec:futurework}),
$n=2$, $N=36$ ($L=6$; the horizontal axis $\Phi$ is the plaquette flux
$\Phi_\square$), exact sparse-trace evaluation with a genuine
Peierls Hamiltonian, no fitted parameters. Taking this spin-sector
difference, rather than the raw fourth moment, cancels the spin-blind
orbital plaquette-loop contribution exactly, isolating the exchange
mechanism alone.}
\label{fig:mu4exchangeflux}
\end{figure}

\begin{figure}[htbp]
\centering
\includegraphics[width=0.55\textwidth]{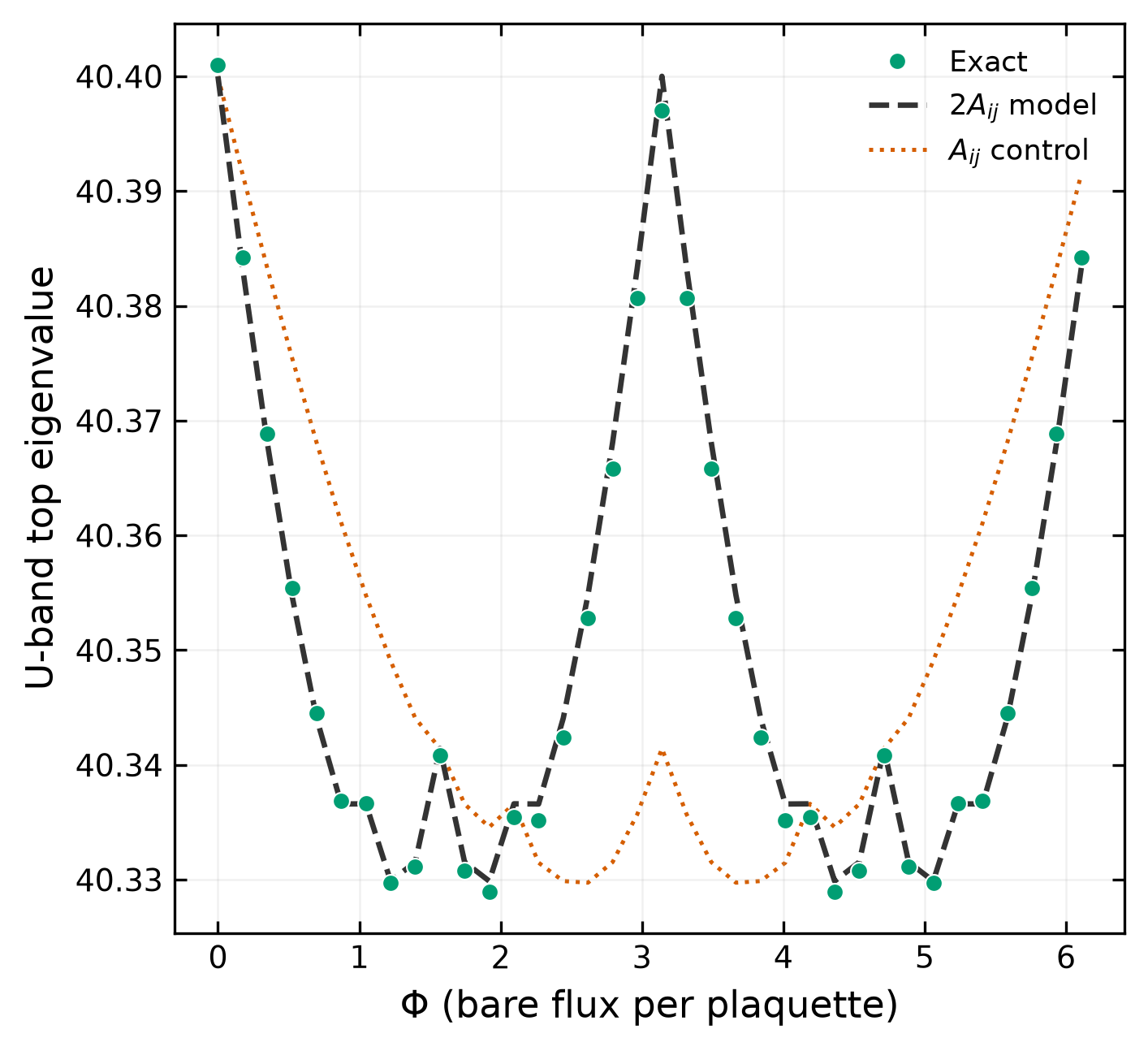}
\caption{Doublon flux response (\S\ref{sec:futurework}; energies in
units of $\lambda$): the exact flux-threaded two-electron U-band's
top eigenvalue against the effective doublon model with doubled
Peierls phase $t_de^{2iA_{ij}}$, compared with an undoubled control
$t_de^{iA_{ij}}$, $N=36$ ($L=6$), $U=40\lambda$, no fitted parameters.
This is a standard consequence of the effective doublon Hamiltonian's
charge-$2e$ gauge coupling rather than a distinctive result of this
paper on its own; it is included because it makes the bridge to the
gauge-response discussion of \S\ref{sec:futurework} concrete at the
two-particle level, not as evidence of collective or superconducting
behavior.}
\label{fig:doublonflux}
\end{figure}

\paragraph{Additional numerical validation figures.} Three further
figures close specific analytic claims made in the main text: the
two-particle U-band benchmark (Fig.~\ref{fig:kgridUband}), the exchange
contribution to $\mu_4$ under flux (Fig.~\ref{fig:mu4exchangeflux}), and
the doublon's doubled Peierls phase (Fig.~\ref{fig:doublonflux}), in
addition to the density-of-states validation of
Fig.~\ref{fig:dosfamily} above.

\section{Machine-checked results}
\label{app:lean}

The exact statements listed in Table~\ref{tab:lean} are formalized and
proved in Lean~4~\cite{deMouraUllrich2021} with
Mathlib~\cite{Mathlib2020} (Lean v4.35.0, Mathlib release v4.35.0). Each
statement is written as a proposition in one file that uses only Mathlib
and the definitions of the formalization (sector, torus, configuration,
fermion operators, $H_t$, $H_\infty$, moments), so that what is proved
can be read without reading the proofs. Every headline theorem depends
only on Lean's three standard axioms (propositional extensionality,
choice, quotient soundness): no unproved step (\texttt{sorry}) and no
trusted compiled evaluation enters. Negative controls (a wrong sector
dimension, a wrong multiplicity in the finite-population $\mu_4$, the
single-phase doublon amplitude, and sign-free annihilation operators)
were checked by hand to be rejected or refuted (recorded with the
formalization). Independently, Lean's comparator checks that every
headline theorem proves exactly its stated proposition with only the
three permitted axioms, re-checking the proofs with six kernels (Lean's
default and paranoid checkers, lean4lean, nanoda, con-leche and
con-ron), and it rejects four automated negative controls (a weakened
statement, an added axiom, a changed definition, an unproved step);
the record of this run is kept with the formalization.

\begin{table}[htbp]
\centering
\small
\begin{tabular}{@{}p{0.55\textwidth}p{0.38\textwidth}@{}}
\toprule
statement & Lean theorem \\
\midrule
sector dimensions $\binom{N}{a}\binom{a}{k}\binom{N-a}{b-k}$ ($a=N_\uparrow$, $b=N_\downarrow$, $k$ doublons), which fix the sector populations of Fig.~\ref{fig:sectors} (\S\ref{sec:fewbody})
  & \texttt{sectorDimensions}, \texttt{sectorDimensionsQuoted} \\
$\mu_2=\lambda^2\,2|E|\,n(N-n)/[N(N-1)]$ on any finite simple graph with $N\geq2$, any spin composition; $z$-regular and torus forms (\S\ref{sec:mu2theorem})
  & \texttt{secondMomentTheorem}, \texttt{secondMomentRegular}, \texttt{secondMomentTorus} \\
the same for $H_\infty=P_0H_tP_0$ built from fermion operators, any spin-independent Peierls phases, any ordering of the orbitals
  & \texttt{secondMomentHubbard}, \texttt{secondMomentHubbardTorus} \\
weighted form $\mu_2=\frac{n(N-n)}{N(N-1)}\sum_{i\ne j}|t_{ij}|^2$ for any spin-independent Hermitian amplitudes $t_{ji}=t_{ij}^*$ (\S\ref{sec:mu2theorem})
  & \texttt{secondMomentWeightedTheorem}, \texttt{secondMomentHubbardWeighted} \\
canonical anticommutation relations; sorted-tuple sign rule (App.~\ref{app:numerics}); $H_\infty$ is Hermitian with single hops of modulus $\lambda$
  & \texttt{canonicalAnticommutation}, \texttt{sortedTupleSignRule}, \texttt{projectedHubbardIsProjectedHopping} \\
finite-population variance and fourth moment, additivity; Table~\ref{tab:fpcexact} (\S\ref{sec:fermistats}, App.~\ref{app:mu4fpc})
  & \texttt{finitePopulationVariance}, \texttt{finitePopulationFourthMoment}, \texttt{independentSumFourthMoment}, \texttt{finitePopulationTable} \\
one spin species on the torus: the variance of a sum of the band energies of $k$ distinct single-particle states equals $\mu_2(U{\to}\infty)$ (\S\ref{sec:fermistats}; that these sums are the $U{=}0$ spectrum is not formalized)
  & \texttt{freeFermionSecondMomentTorus}, \texttt{freeAndProjectedSecondMomentsAgree} \\
four-hop closed walks induce only $e$, transpositions and 3-cycles; both occur (\S\ref{sec:mu4exchange}, App.~\ref{app:mu4walks})
  & \texttt{closedWalkFewMoved}, \texttt{fourHopPermutations}, \texttt{fourHopExchangesOccur} \\
$P_3$ affine in $P_{\rm same}$ at fixed $n$ (\S\ref{sec:mu4exchange})
  & \texttt{threeCycleCountAffine} \\
fourth moment of $n$ hard-core bosons, Eq.~\eqref{eq:muHCB}, on any finite simple graph with $N\geq4$; the counts $D=\binom m2-W$ and $8C_4$ (sequences) behind it (App.~\ref{app:mu4walks})
  & \texttt{hardCoreBosonFourthMoment} \\
the Proposition of \S\ref{sec:mu4exchange}: $\mu_4(\infty)=A(N,n)+\alpha(N,n)P_{\rm same}$ with Eqs.~\eqref{eq:Aclosed} and \eqref{eq:alphaN}, for $H_\infty$ built from fermion operators, any finite simple graph with $N\geq4$, any $n\leq N$, any ordering of the orbitals, no Peierls phases
  & \texttt{fourthMomentHubbard} \\
its square-lattice form ($L=3$ or $L\geq5$), the torus counts $m=2N$, $W=6N$, $C_4=N$ ($C_4=24$ at $L=4$), and the seven $\mu_4$ values of Table~\ref{tab:mu4exact}
  & \texttt{fourthMomentHubbardTorus}, \texttt{torusGraphCounts}, \texttt{fourthMomentTableL4} \\
$\Tr(T_0V)=0$ on triangle-free graphs, so the variance of $T_0-V/U$ has no $1/U$ term; the torus ($L\geq3$) is triangle-free iff $L\neq3$ (\S\ref{sec:SW}; the expansion that links this to $\mu_{2,c}$ is not formalized)
  & \texttt{schriefferWolffCrossTerm}, \texttt{schriefferWolffSecondMoment}, \texttt{schriefferWolffHubbard}, \texttt{torusTriangleFree}, \texttt{crossTermNeedsTriangleFree} \\
two-electron problem: doubled Peierls phase $t_de^{2iA_{ji}}$, nearest-neighbor range, flux-blind on-site term (\S\ref{sec:futurework})
  & \texttt{doubledPeierlsPhase}, \texttt{doublonHoppingRange}, \texttt{doublonOnSiteShift} \\
\bottomrule
\end{tabular}
\caption{Machine-checked statements and the Lean theorems that prove them.}
\label{tab:lean}
\end{table}

\paragraph{Scope.} The following are \emph{not} formalized, and the
corresponding statements of the paper rest on the derivations given in
the text and the Supplemental Material: the Schrieffer--Wolff expansion
itself, at second and third order (the formalization evaluates the
second-order terms and the operators $T_0$, $V$ it produces), and hence
the $U^{-2}$ coefficient $c_2$; the identification of the $U=0$
many-body energies with sums of the energies of distinct single-particle states (Slater
determinants) and of the tight-binding band with plane waves; the
fourth moment under Peierls phases, Eq.~\eqref{eq:mu4flux}, and the
flux-resolved exchange coefficient (the Lean statements of the fourth
moment have no Peierls phases); the fourth moment for more than two spin
species, on graphs with self-loops or multiple edges, and for
bond-dependent amplitudes; the hypercubic form $\alpha_d(N,2)$ beyond the
square lattice; the walk-by-walk classification of
Appendix~\ref{app:mu4walks} (the Lean proof reaches the same coefficients
by its own count of two-hop paths); and the physical identification of the formalized
fermion operators: they satisfy the canonical anticommutation relations
for an arbitrary ordering of the orbitals, and any other convention
differs from them by phases of the basis states, a standard fact not
proved there. All spectral, density-of-states
and finite-$U$ numerics are independent of the formalization. The Lean
text of every statement, with the definitions it uses, is reproduced
verbatim in S12; the proofs are in the accompanying repository.

\paragraph{Reproduction.} From the \texttt{lean/} directory of the
accompanying repository: \texttt{lake exe cache get} (prebuilt
Mathlib), \texttt{lake build} (must finish without errors or warnings),
and \texttt{lake env lean AxiomAudit.lean} (prints the axioms of every
headline theorem); on Linux, \texttt{lake comparator --paranoid} and
\texttt{lake check} run the independent checks above.

\paragraph{Code and data availability.} All exact-diagonalization and
sparse-trace codes used to produce the numerical results in this paper,
covering the $n=3,4$ exact diagonalization, the general
$k$-fermion sector construction, the $U=\infty$ sparse-trace
evaluation, the Schrieffer--Wolff operator construction and traces, and
the flux-threaded Peierls-Hamiltonian validation of \S\ref{sec:futurework},
are public, together with the scripts that regenerate every figure and
table, the records (command, environment and output) of the long runs,
and the Lean~4 formalization of Appendix~\ref{app:lean} (its
\texttt{lean/} directory), at
\url{https://github.com/zhineos/hubbard-spectral-moments}, release
v1.1.0 (commit \texttt{68828a3}), archived at Zenodo as
\href{https://doi.org/10.5281/zenodo.24392999}{doi:10.5281/zenodo.24392999}
(all versions:
\href{https://doi.org/10.5281/zenodo.24244236}{doi:10.5281/zenodo.24244236}).
Notes S1--S12 form the Supplemental Material.

\end{document}

%% file: fig_walktypes.tex
\providecolor{ConnBlue}{HTML}{0072B2}
\providecolor{ExchVermillion}{HTML}{D55E00}
\begin{tikzpicture}[
    occ/.style={circle,fill=black,inner sep=0pt,minimum size=5.5pt},
    emp/.style={circle,draw=black,fill=white,inner sep=0pt,minimum size=5.5pt},
    bond/.style={gray!60,line width=0.6pt},
    hop/.style={-{Stealth[length=4pt]},line width=0.9pt,ConnBlue},
    xhop/.style={-{Stealth[length=4pt]},line width=0.9pt,ExchVermillion},
    num/.style={font=\tiny,inner sep=1pt},
    lab/.style={font=\scriptsize},
    ttl/.style={font=\scriptsize\bfseries,anchor=south west},
    txt/.style={font=\scriptsize,text width=3.7cm,align=left,anchor=north west},
]
\def\a{1.15}
\begin{scope}[shift={(0,0)}]
  \node[ttl] at (-0.2,1.55) {(a) Type I(i)};
  \draw[bond] (0,0)--(\a,0);
  \node[occ] (A) at (0,0) {}; \node[emp] (B) at (\a,0) {};
  \draw[hop] (A) to[bend left=35] (B); \draw[hop] (B) to[bend left=35] (A);
  \node[lab] at (0.5*\a,0.5) {$\times2$};
  \node[txt] at (-0.2,-0.35) {out and back twice. One route: flux-blind. Any spins.};
\end{scope}
\begin{scope}[shift={(4.1,0)}]
  \node[ttl] at (-0.2,1.55) {(b) Types I(ii), I(iii)};
  \draw[bond] (0,0)--(2*\a,0);
  \node[occ] (A) at (0,0) {}; \node[emp] (B) at (\a,0) {}; \node[emp] (C) at (2*\a,0) {};
  \draw[hop] (A) to[bend left=30] (B); \draw[hop] (B) to[bend left=30] (C);
  \draw[hop] (C) to[bend left=30] (B); \draw[hop] (B) to[bend left=30] (A);
  \node[txt] at (-0.2,-0.35) {two neighbors in turn, or one step further and back. One route: flux-blind. Any spins.};
\end{scope}
\begin{scope}[shift={(8.2,0)}]
  \node[ttl] at (-0.2,1.55) {(c) Type I(iv)};
  \draw[bond] (0,0) rectangle (\a,\a);
  \node[occ] (A) at (0,0) {}; \node[emp] (B) at (\a,0) {}; \node[emp] (C) at (\a,\a) {}; \node[emp] (D) at (0,\a) {};
  \draw[hop] (A)--(B); \draw[hop] (B)--(C);
  \draw[hop,dashed] (A)--(D); \draw[hop,dashed] (D)--(C);
  \node[lab] at (0.5*\a,0.5*\a) {$\Phi_\square$};
  \node[txt] at (-0.2,-0.35) {plaquette loop: two routes to the far corner, constructive, $\propto1+\cos\Phi_\square$. Any spins.};
\end{scope}
\begin{scope}[shift={(12.2,0)}]
  \node[ttl] at (-0.2,1.55) {(d) Type II(a)};
  \draw[bond] (0,0)--(0,\a); \draw[bond] (1.6,0)--(1.6,\a);
  \node[occ] (A) at (0,0) {}; \node[emp] (B) at (0,\a) {};
  \node[occ] (C) at (1.6,0) {}; \node[emp] (D) at (1.6,\a) {};
  \draw[hop] (A) to[bend left=30] (B); \draw[hop] (B) to[bend left=30] (A);
  \draw[hop] (C) to[bend left=30] (D); \draw[hop] (D) to[bend left=30] (C);
  \node[txt] at (-0.2,-0.35) {two particles, each out and back. One route each: flux-blind. Any spins.};
\end{scope}
\begin{scope}[shift={(0,-4.4)}]
  \node[ttl] at (-0.2,1.55) {(e) Type II(b), diagonal};
  \draw[bond] (0,0) rectangle (\a,\a);
  \node[occ] (I) at (0,0) {}; \node[emp] (K) at (\a,0) {}; \node[occ] (J) at (\a,\a) {}; \node[emp] (L) at (0,\a) {};
  \node[lab,below left] at (I) {$i$}; \node[lab,above right] at (J) {$j$};
  \node[lab,below right] at (K) {$k$}; \node[lab,above left] at (L) {$l$};
  \draw[xhop] (I)--(K); \draw[xhop] (K)--(J);
  \draw[xhop,dashed] (J)--(L); \draw[xhop,dashed] (L)--(I);
  \node[txt] at (-0.2,-0.35) {ring exchange of $i,j$ through $k,l$: four routes to $\{k,l\}$. Same spins: they cancel at $\Phi_\square=0$; $\propto1-\cos\Phi_\square$. Opposite spins: flux-blind.};
\end{scope}
\begin{scope}[shift={(4.1,-4.4)}]
  \node[ttl] at (-0.2,1.55) {(f) Type II(b), edge};
  \draw[bond] (0,0) rectangle (\a,\a);
  \node[occ] (I) at (0,0) {}; \node[occ] (J) at (\a,0) {}; \node[emp] (P) at (\a,\a) {}; \node[emp] (M) at (0,\a) {};
  \node[lab,below left] at (I) {$i$}; \node[lab,below right] at (J) {$j$};
  \node[lab,above left] at (M) {$m$}; \node[lab,above right] at (P) {$p$};
  \draw[xhop] (I)--node[num,left]{1}(M); \draw[xhop] (M)--node[num,above]{3}(P); \draw[xhop] (P)--node[num,right]{4}(J);
  \draw[xhop,dashed] (J) to[bend left=25] node[num,below]{2} (I);
  \node[txt] at (-0.2,-0.35) {$i$ goes three hops around, $j$ one hop along the bond: two routes per intermediate, over two plaquettes. Same spins: $\propto1-\cos\Phi_\square$. Opposite spins: flux-blind.};
\end{scope}
\begin{scope}[shift={(8.2,-4.4)}]
  \node[ttl] at (-0.2,1.55) {(g) Type III, 3-cycle};
  \draw[bond] (0,0) rectangle (\a,\a);
  \node[occ] (I) at (0,0) {}; \node[occ] (J) at (\a,0) {}; \node[occ] (K) at (\a,\a) {}; \node[emp] (M) at (0,\a) {};
  \node[lab,below left] at (I) {$i$}; \node[lab,below right] at (J) {$j$};
  \node[lab,above right] at (K) {$k$}; \node[lab,above left] at (M) {$m$};
  \draw[xhop] (I)--node[num,left]{1}(M); \draw[xhop] (J)--node[num,below]{2}(I);
  \draw[xhop] (K)--node[num,right]{3}(J); \draw[xhop] (M)--node[num,above]{4}(K);
  \node[txt] at (-0.2,-0.35) {the empty corner goes once around: two routes ($i\to m$ or $k\to m$ first), constructive, $\propto1+\cos\Phi_\square$. All three spins equal.};
\end{scope}
\begin{scope}[shift={(12.2,-4.4)}]
  \node[occ] at (0,1.2) {}; \node[lab,anchor=west] at (0.2,1.2) {occupied site};
  \node[emp] at (0,0.8) {}; \node[lab,anchor=west] at (0.2,0.8) {empty site};
  \draw[hop] (-0.15,0.4)--(0.35,0.4); \node[lab,anchor=west] at (0.45,0.4) {hop, spin-blind type};
  \draw[xhop] (-0.15,0.0)--(0.35,0.0); \node[lab,anchor=west] at (0.45,0.0) {hop, spin-sensitive type};
  \node[lab,anchor=west] at (-0.2,-0.4) {dashed: second route or};
  \node[lab,anchor=west] at (-0.2,-0.75) {second particle};
  \node[lab,anchor=west] at (-0.2,-1.1) {numbers: hop order};
\end{scope}
\end{tikzpicture}